\documentclass[aps,prd,twocolumns,eqsecnum,preprintnumbers,showpacs,amsmath,amssymb]
{revtex4}

\usepackage{graphicx}
\usepackage{dcolumn}
\usepackage{bm}

\begin{document}

\title{INTRINSICALLY NONPERTURBATIVE QCD \\ I. A PURE DYNAMICAL THEORY OF GLUON CONFINEMENT }

\author{V. Gogohia}
\email[]{gogohia@rmki.kfki.hu}

\affiliation{HAS, CRIP, RMKI, Depart. Theor. Phys., Budapest 114,
P.O.B. 49, H-1525, Hungary}

\date{\today}
\begin{abstract}
We establish exactly and uniquely the infrared structure of the
full gluon propagator in  QCD, not solving explicitly the
corresponding dynamical equation of motion. By construction, this
structure is an infinite sum over all severe (i.e., more singular
than $1/ q^2$) infrared singularities. It reflects the zero
momentum modes enhancement effect in the true QCD vacuum. Its
existence exhibits a characteristic mass (the so-called mass gap),
which is responsible for the scale of nonperturbative dynamics in
the QCD ground state. The theory of distributions, complemented by
the dimensional regularization method, allows one to put severe
infrared singularities under firm mathematical control. By an
infrared renormalization of a mass gap only, the infrared
structure of the full gluon propagator is exactly reduced to the
simplest severe infrared singularity, the famous $(q^2)^{-2}$. So,
the smooth in the infrared limit the full gluon propagator is to
be ruled out. Collective motion of all the purely transverse
$virtual$ gluon field configurations with low-frequency
components/large scale amplitudes is solely responsible for the
color confinement phenomenon within our approach. At the
microscopic, dynamical level these field configurations are
saturated by the nonlinear fundamental four-gluon interaction. It
just makes the full gluon propagator inevitably so singular in the
infrared. The amplitudes of all the purely transverse severely
singular $actual$ gluon field configurations are totally
suppressed, leading thus to the confinement of gluons. We
formulate the gluon confinement criterion in a manifestly
gauge-invariant way, taking into account the distribution nature
of severe infrared singularities.
\end{abstract}



\maketitle

\section{Introduction}

Quantum Chromodynamics (QCD) \cite{1} is widely accepted as a
realistic, quantum field gauge theory of strong interactions not
only at the fundamental (microscopic) quark-gluon level, but at
the hadronic (macroscopic) level as well. This means that, in
principle, it should describe the properties of experimentally
observed hadrons in terms of experimentally never seen quarks and
gluons, i.e., to describe the hadronic word from first principles.
But this is a formidable task because of the color confinement
phenomenon, the dynamical mechanism of which is not yet fully
understood, and therefore the confinement problem remains unsolved
up to the present days \cite{2}. It prevents colored quarks and
gluons to be experimentally detected as asymptotic states, which
are colorless, by definition, so color confinement is permanent
and absolute \cite{1}. At present, there is no doubt left that
color confinement as well as all other dynamical effects, such as
spontaneous breakdown of chiral symmetry (SBCS), bound-state
problems, etc., are inaccessible to the perturbation theory (PT)
technics, i.e., they are very essentially nonperturbative (NP)
effects. This means that for their investigation the NP solutions,
methods and approaches are needed to be found, used and developed.
This is especially necessary taking into account that the
above-mentioned NP effects are low-energy/momentum (large
distances) phenomena, and, as it is well known, the PT methods, in
general, fail to investigate them.

 The surprising fact is that after more than thirty years
of QCD, we still don't know the interaction between quarks and
gluons. To know it means that one knows the full gluon and quark
propagators, the quark-gluon  and the pure gluon vertices. In the
weak coupling limit or in the case of heavy quarks only this
interaction is known. In the first case all the above-mentioned
lower and higher Green's functions (propagators and vertices,
respectively) become effectively free ones multiplied by the
renormalization group corresponding PT logarithm improvements. In
the case of heavy quarks all the Green's functions can be
approximated by their free counterparts from the very beginning.
In general, the Green's functions are essentially different from
their free counterparts (substantially modified) due to the
response of the highly nontrivial structure of the true QCD
vacuum. It is just this response which is taken into account by
the full ("dressed") propagators and vertices (it can be neglected
in the weak coupling limit or for heavy quarks). That is the main
reason why they are still unknown, and the confinement problem is
not yet solved. In other words, it is not enough to know the
Lagrangian of the theory. In QCD it is also necessary and
important to know the true structure of its ground state (also
there might be symmetries of the Lagrangian which do not coincide
with symmetries of the vacuum). This knowledge can only come from
the investigation of a general system of the dynamical equations
of motion, the so-called Schwinger-Dyson (SD) system of equations
\cite{1,3,4}, to which all the Green's functions should satisfy.
To solve this system means to establish all the QCD Green's
functions, and thus to establish the structure of the true QCD
ground state as well. So, the SD system of equations is the only
place where the color confinement problem can generally be solved.
In this work, we will establish the interaction between quarks,
more precisely the exact IR structure of the full gluon
propagator, which is responsible for color confinement. In part II
of our general investigation of the color confinement problem, we
will establish the quark-gluon vertex needed for its solution. The
Slavnov-Taylor (ST) identities for the pure gluon vertices will be
investigated there as well (for references see below).

 The full dynamical information of any quantum
field gauge theory such as QCD is contained in the corresponding
quantum equations of motion, the above-mentioned SD system of
equations. It is a highly nonlinear, strongly coupled system of
four-dimensional integral equations. To solve this system means to
solve QCD itself and vice versa. The Bethe-Salpeter (BS) integral
equations for the bound-state amplitudes \cite{5} should be also
included into this system. The kernels and scattering amplitudes
of these integral equations are determined by an infinite series
of the corresponding multi-loop skeleton diagrams. It is a general
feature of nonlinear systems that the number of solutions (if any)
cannot be fixed $a \ priori$. Although this system of dynamical
equations can be reproduced by expansion around the free field
vacuum, the final equations make no reference to the vacuum of PT.
They are sufficiently general and should be treated beyond PT
\cite{1}. These equations should be also complemented by the
corresponding ST identities \cite{1,3,4,6,7,8}, which, in general,
relate lower and higher Green's functions to each other. These
identities are consequences of the exact gauge invariance and
therefore $"are \ exact \ constraints \ on \ any \ solution \ to \
QCD"$ \cite{1}. The low-energy/momentum region interesting for
confinement, SBCS, etc., is usually under the control of these
identities. Precisely the SD system of dynamical equations,
complemented by the ST identities, can serve as an adequate and
effective tool for the NP approach to QCD \cite{1,9,10,11,12}.

 However, two necessary requirements should be imposed
over any solutions to QCD within its dynamical equations approach.
As emphasized above, the SD system of equations is a highly
nonlinear, very complicated and strongly coupled system of
equations, so there is definitely $"no \ hope \ for \ an \ exact \
solution(s)"$ \cite{1,9}. In this connection, let us remind that
even in the case of Quantum Electrodynamics (QED) no exact
solution(s) is known, though it is a much simpler gauge theory
than QCD. Due to the above-mentioned complexity, the SD system of
equations may contain even more additional information, which
would only complicate the solution of this or that physical
problem. The SD system of equations is, in fact, an infinite chain
of the relations between different propagators, vertices and
scattering kernels. Thus, truncations (and approximations) are
inevitable in order to formulate a closed set of equations, say,
in the quark sector. Different truncations could lead to
qualitatively different solutions. QCD contains many sectors of
different nature, such as quark, ghost, Yang-Mills (YM),
Nambu-Goldstone (NG), BS, etc. However, making use of some
truncation scheme in one sector, it is necessary to be sure that
nothing is going wrong in other sectors.

$So, \ the \ first \ necessary \ requirement\ is: \ any \
truncation \ scheme \ in \ QCD \ should \ be \ self-consistent$.
The main tool to maintain the self-consistent treatment of
different sectors in QCD is, of course, the use of different ST
identities, as it has been already underlined above. The
importance of the self-consistent treatment of the SD system of
equations within any truncation scheme, in particular ladder
approximation, for the first time has been emphasized by Maskawa
and Nakajima and by Adler in their pioneering papers \cite{13,14}.
The next significant step in this direction has been done in Ref.
\cite{15}, while in our papers \cite{16} it has been finally
proven that the ladder approximation in the whole energy-momentum
range is not self-consistent.

$The \ second \ necessary \ requirement \ is \ a \ manifest \
gauge \ invariance$, i.e., to use and develop only those
self-consistent truncation schemes (approaches), which do not
depend explicitly on the gauge fixing parameter. This is crucially
important in QCD just because there is no hope for exact
solution(s). In this connection, a few remarks are in order. In
QED the explicit gauge dependence is not a problem. In order to
calculate physical observables in this theory, one needs to
multiply the corresponding elements of $S$-matrix by the conserved
currents, which immediately eliminates the dependence on the gauge
fixing parameter (i.e., the longitudinal component of the photon
propagator does not contribute to $S$-matrix). Obviously, this is
impossible in QCD due to its non-Abelian character. First of all,
the transverse and longitudinal components of the gluon propagator
do interact with each other. Secondly, the current to which the
gauge field is coupled is not conserved \cite{17,18}. In QCD to
find a calculation scheme (approach), which does not depend
explicitly on the gauge fixing parameter from the very beginning,
is crucially important, indeed.

It is worth explaining this point in more detail. Let us consider
for this purpose, for example the $\pi^+$ meson propagator, which
can be written down analytically as follows:

\begin{equation}
D_{\pi^+}(p) \sim \int d^4 q Tr \Bigl[ \gamma_5 S_u(p+q) \gamma_5
S_{\bar d}(q) \Bigr],
\end{equation}
where all numerical factors are suppressed as unimportant for our
discussion. For simplicity, the BS bound-state amplitudes for the
$\pi^+$ meson are replaced by the $\gamma_5$ matrices. Let us
emphasize now that the left-hand-side of this relation is, by
definition, a manifestly gauge-invariant quantity, since it
describes the propagation of a physical particle, the $\pi^+$
meson. However, single quark propagators, which appear in the
right-hand-side of this relation, are gauge-dependent quantities,
i.e., they formally depend explicitly on the gauge choice. First
of all, this dependence comes from the gluon propagator, which
appears in the quark self-energy. That is why a gauge-invariant
approach is necessary in order to calculate quark propagators,
i.e., only those quark propagators, calculated within a manifestly
gauge-invariant approach, are to be used to form gauge-invariant
composite propagators such as the meson propagator \cite{19}. This
guarantees that the right-hand-side of the relation (1.1) will be
manifestly gauge-invariant, as it is required by its
left-hand-side. Thus, all the solutions to the SD dynamical
equations of motion, based on the particular gauge choice (the
so-called gauge artifact solutions) should be ruled out. In other
words, such kind of the solutions for the quark propagator are not
legitimated to use in order to form gauge-invariant meson
propagator, or to calculate any other physical observables in QCD,
in particular, in low-energy QCD.
 Let us stress once more that by "gauge artifact solutions" we
mean solutions, which are due to the particular gauge choice,
i.e., they cannot be obtained by choosing some another numerical
value of the gauge fixing parameter (for more detail remarks see
Subsec. D in Sec. 2).

Our approach to low-energy QCD is formulated in the framework of
the SD system of dynamical equations mentioned above. We propose
and develop an approach which, on one hand, is self-consistent,
and, on the other hand, is manifestly gauge-invariant. It allows
one to calculate hadron properties from first principles, i.e.,
from the underlying dynamical theory of interactions between quark
and gluons only. It is based on the existence and the importance
of quantum fluctuations and excitations of the infrared (IR)
degrees of freedom in the QCD ground state. Our approach has been
recently applied to two-dimensional (2D) covariant gauge QCD,
which has been proven to confine quarks \cite{20,21}. Its axial
gauge counterpart has been already investigated in 't Hooft's
pioneering paper \cite{22} within precisely quark SD and BS
equations approach. 2D covariant gauge QCD turned out to be an
appropriate theory in order to be generalized to four-dimensional
(4D) QCD. However, there exist, at least, four important
distinctions between 2D and 4D QCD in any gauge. First of all, 2D
QCD has explicitly (in the Lagrangian) a fundamental mass scale
parameter, which is nothing else but the coupling constant. In 4D
QCD the coupling is dimensionless and Lagrangian of this theory
does not explicitly contain a fundamental scale (the mass gap).
Second, the IR singularity of the free gluon propagator in 2D QCD
is severe one (in any gauge, of course), i.e., it is NP from the
very beginning \cite{20,21}. In 4D QCD the IR singularity of the
free gluon propagator is not severe, i.e., it is PT one (for the
exact definitions of both the PT and the NP IR singularities see
below). So, there is no justification to use it for the
description of the IR region in 4D QCD ground state. The third
important distinction is that 4D QCD has a nontrivial PT phase,
while in 2D QCD it is simple. And finally the 4D QCD ground state
is much more rich and complicated than its almost trivial 2D
counterpart. Let us also point out an interesting discussion of
the Coleman theorem concerning the absence of the NG bosons in 2D
quantum field theories (the sine-Gordon and Thirring models) in
Ref. \cite{23}.

 In our general approach to 4D QCD, i.e., to QCD
itself, we have elaborated on all the above-mentioned distinctions
from 2D QCD. It will consist of a few parts. In part I, we will
formulate our approach to the YM sector in QCD in all details. We
establish uniquely and exactly the IR structure of the full gluon
propagator, fixing thus the first part of the interaction between
quarks. This automatically requires the existence of a mass gap,
which is responsible for the NP dynamics in the true QCD ground
state. Moreover, we formulate the gluon confinement criterion in a
manifestly gauge-invariant way. In part II, we establish the
second part of the above-mentioned interaction, namely the needed
quark-gluon proper vertex. This will be done with the help of the
corresponding ST identity. A closed set of equations for the quark
propagator will be derived. By the consideration of the whole
system of the SD equations and corresponding ST identities, the IR
renormalizability of QCD will be proven under some conditions. In
part III, a closed set of equations for the quark propagator,
derived in part II, will be solved. We will show explicitly that
our approach to low-energy QCD implies quark confinement and SBCS
without involving any extra degrees of freedom. We need only the
underlying dynamics of QCD -- the self-interaction of massless
gluons. In part IV, the NG sector of QCD will be numerically
evaluated. Developing the so-called chiral PT at the fundamental
quark level, we will be able to fix the BS bound-state amplitude
in the chiral limit for the NG particles. It will also allow one
to derive a new formula for the pion decay constant. The part V
will be devoted to the investigation of the QCD ground state
itself. We will show how to calculate correctly the truly NP
vacuum energy density. It is the most important characteristics of
the QCD ground state. The BS sector (i.e., the bound-state
problem) will be investigated in the next part VI. We will show
that the bound-state problem will be always reduced to an
algebraic problem, i.e., it becomes tractable within our approach.

This paper (hereafter the above-mentioned part I, which plays a
central role in our general approach to low-energy QCD) is
organized as follows. In Sec. 2, we investigate the general
properties of the SD equation for the full gluon propagator in
QCD. It is also explicitly shown there how severe IR singularities
and a mass gap may appear in the true QCD vacuum due to the
self-interaction of massless gluons. In Sec. 3, we introduce a few
useful formulae from the distribution theory (DT) \cite{24},
complemented by the dimensional regularization (DR) method
\cite{25}. They show how severe IR singularities should be
correctly put under firm mathematical control. Sec. 4 is devoted
to establishing the general structure of the full gluon propagator
in the IR region. We will show that it is given by the special
Laurent expansion as an infinite sum over all severe power-type IR
singularities possible in nD QCD. An important case of QCD itself
is considered in Sec. 5. The general IR renormalization program is
formulated in Sec. 6. In this section, we perform the IR
renormalization of the gluon SD equation. It has been proven that
by the IR renormalization of a mass gap only, all severe IR
singularities can be removed from the full gluon propagator. In
Sec. 7, we formulate the zero momentum modes enhancement (ZMME)
model of the true QCD ground state. Within it, we thus determine a
pure gluon part of the interaction between quarks. A gluon
confinement criterion is formulated in a manifestly
gauge-invariant way. In the same way, we also define the
intrinsically NP (INP) phase in QCD. In Sec. 8, the solution of
the gluon SD equation within our approach is given. In Sec. 9, the
general discussion is present, while in Sec. 10 we summarize our
conclusions. We investigate the gluon SD equation for the PT part
of the full gluon propagator in the appendix A.

\section{Gluon propagator}

In order to investigate the problem of the true QCD ground state
structure, let us start first with one of the main objects in the
YM sector. The two-point Green's function, describing the full
gluon propagator, is (Euclidean signature here and everywhere
below)

\begin{equation}
D_{\mu\nu}(q) = i \left\{ T_{\mu\nu}(q)d(q^2, \xi) + \xi
L_{\mu\nu}(q) \right\} {1 \over q^2 },
\end{equation}
where $\xi$  is the gauge fixing parameter ($\xi = 0$ - Landau
gauge and  $\xi = 1$ - Feynman gauge) and

\begin{equation}
T_{\mu\nu}(q)=\delta_{\mu\nu}-{q_{\mu} q_{\nu} \over q^2} = \delta_{\mu\nu }
- L_{\mu\nu}(q).
\end{equation}
Evidently, $T_{\mu\nu}(q)$ is the transverse (physical) component
of the full gluon propagator, while $L_{\mu\nu}(q)$ is its
longitudinal (unphysical) one. The free gluon propagator is
obtained by setting simply the full gluon form factor $d(q^2,
\xi)=1$ in Eq. (2.1), i.e.,

\begin{equation}
D^0_{\mu\nu}(q) = i \left\{ T_{\mu\nu}(q) + \xi
L_{\mu\nu}(q) \right\} {1 \over q^2 }.
\end{equation}

 The solutions of the SD equation for the full gluon propagator (2.1) are supposed to
reflect the complexity of the quantum structure of the QCD ground
state. Just this determines one of the central roles of the full
gluon propagator in the SD system of equations \cite{26}. The SD
equation for the full gluon propagator (see Eq. (2.4)) is a highly
nonlinear system of four-dimensional integrals, containing many
different, unknown in general, propagators and vertices, which, in
their turn, satisfy too complicated integral equations, containing
different scattering amplitudes and kernels, so there is no hope
for exact solution(s). However, in any case the solutions of this
equation can be distinguished from each other by their behavior in
the IR limit, describing thus many (several) different types of
quantum excitations and fluctuations of gluon field configurations
in the QCD vacuum. Evidently, not all of them can reflect the real
structure of the QCD vacuum, for example the gauge artifact
solutions (see Subsec. E). The ultraviolet (UV) limit of these
solutions is uniquely determined by asymptotic freedom (AF)
\cite{27}.

The deep IR asymptotics of the full gluon propagator can be
generally classified into the two different types: singular, which
means that the above-mentioned ZMME effect takes place in the NP
QCD vacuum, or smooth, which means that the full gluon propagator
is IR finite or even is IR vanishing. Formally, the full gluon
propagator (2.1) has an exact power-type IR singularity, $1/q^2$,
which is due to its longitudinal component. This is the IR
singularity of the free gluon propagator, see Eq. (2.3). By the
ZMME effect we mean, in general, the IR singularities, which are
more severe than $1/q^2$ (see also Subsec. C). Evidently, the
singular asymptotics is possible at any value of the gauge fixing
parameter. At the same time, the smooth behavior of the full gluon
propagator (2.1) in the IR becomes formally possible either  by
choosing the Landau gauge $\xi = 0$ from the very beginning, or by
removing the longitudinal (unphysical) component of the full gluon
propagator with the help of ghost degrees of freedom
\cite{1,17,18} (for more detail discussion see Subsec. D).

 However, any deviation in the behavior of the full gluon propagator in the IR
domain from the free one automatically assumes its dependence on a
scale parameter (at least one) different, in general, from the QCD
asymptotic scale parameter $\Lambda_{QCD}$. It can be considered
as responsible for the NP dynamics (in the IR region) in the QCD
vacuum. If QCD itself is a confining theory, then such a
characteristic scale is very likely to exist. This is very similar
to AF, which requires the above-mentioned asymptotic scale
parameter $\Lambda_{QCD}$ associated with nontrivial PT dynamics
in the UV region (AF, scale violation, determining thus the
deviation in the behavior of the full gluon propagator from the
free one in the UV domain). In this connection it is worth
emphasizing that, being numerically a few hundred $MeV$ only, it
cannot survive in the UV limit. This means that none of the finite
scale parameters, in particular $\Lambda_{QCD}$, can be determined
by PT QCD. It should come from the IR region, so it is NP by
origin. How to establish a possible relation between these two
independent scale parameters was shown in our paper \cite{28}.
Despite the fact that the PT vacuum cannot be the true QCD ground
state \cite{29}, nevertheless, the existence of such kind of a
relation is a manifestation that "the problems encountered in
perturbation theory are not mere mathematical artifacts but rather
signify deep properties of the full theory" \cite{30}.

The message that we have tried to convey is that precisely AF
clearly indicates the existence of the NP phase with its own
characteristic scale parameter in the full QCD.

\subsection{Gluon SD equation}

The general structure of the SD equation for the full gluon
propagator can be written down symbolically as follows (for our
purposes it is more convenient to consider the SD equation for the
full gluon propagator and not for its inverse, as usual):

\begin{equation}
D(q) = D^0(q) - D^0(q)T_{gh}(q) D(q) - D^0(q)T_q(q)D(q) +
D^0(q)T_g[D](q)D(q).
\end{equation}
Here and in some places below, we omit the dependence on the Dirac
indices, for simplicity. $T_{gh}(q)$ and $T_q(q)$ describe the
ghost and quark skeleton loop contributions into the gluon
propagator. They do not contain the full gluon propagators by
themselves. A pure gluon contribution $T_g[D](q)$ is a sum of four
pure gluon skeleton loops, and consequently they explicitly
contain the full gluon propagators. Precisely this makes the gluon
SD equation highly nonlinear (NL), and this is one of the reasons
why it cannot be solved exactly. However, its linear part, which
contains only ghost and quark skeleton loops, can be summed up, so
Eq. (2.4) becomes

\begin{equation}
D(q) = \tilde{D}^0(q) + \tilde{D}^0(q)T_g[D](q)D(q) = \tilde{D}^0(q) + D^{NL}(q),
\end{equation}
with $\tilde{D}^0(q)$ being a modified free gluon propagator as
follows:

\begin{equation}
\tilde{D}^0(q) = { D^0(q)  \over 1 + [T_{gh}(q) + T_q(q)]D^0(q)},
\end{equation}
where

\begin{equation}
T_{gh}(q) =  g^2 \int {i d^4 k \over (2 \pi)^4} k_{\nu} G(k)
G(k-q)G_{\mu}(k-q, q),
\end{equation}

\begin{equation}
T_q(q) = - g^2 \int {i d^4 p \over (2 \pi)^4} Tr [\gamma_{\nu}
S(p-q) \Gamma_{\mu}(p-q, q)S(p)].
\end{equation}
Let us note in advance that, in general, these quantities can be
decomposed as follows:

\begin{equation}
T_{gh}(q) \equiv T^{gh}_{\mu\nu}(q) = \delta_{\mu\nu} q^2
T_{gh}^{(1)}(q^2) + q_{\mu} q_{\nu} T_{gh}^{(2)}(q^2),
\end{equation}

\begin{equation}
T_q(q) \equiv T^q_{\mu\nu}(q) = \delta_{\mu\nu} q^2 T_q^{(1)}(q^2) +
q_{\mu} q_{\nu} T_q^{(2)}(q^2),
\end{equation}
where all invariant functions $T_{gh}^{(n)}(q^2)$ and
$T_q^{(n)}(q^2)$ at $n=1,2$ are dimensionless with a regular
behavior at zero (they include the dependence on the coupling
constant squared $g^2$). In this connection a few remarks are in
order. Due to the definition $q_{\mu} q_{\nu} =q^2 L_{\mu\nu}$
(see relation (2.2)), instead of the independent structures
$\delta_{\mu\nu}$ and $q_{\mu} q_{\nu}$ in Eqs. (2.9) and (2.10),
one can use $T_{\mu\nu}$ and $L_{\mu\nu}$ as independent
structures with their own invariant functions. For simplicity, we
assume here and everywhere below that all integrals are finite,
and consequently all invariant functions are also finite at zero.
Anyway, how to render them finite is well known procedure (see,
for example Refs. \cite{1,17,18,31}).

From a technical point of view it is convenient to use the free
gluon propagator (2.3) in the Feynman gauge ($\xi=1$), i.e,
$D^0_{\mu\nu}(q) = \delta_{\mu\nu} (i /q^2)$. Then from Eq. (2.6)
it follows

\begin{equation}
\tilde{D}^0(q) =  D^0(q) A(q^2),
\end{equation}
where

\begin{equation}
 A(q^2)= {1  \over 1 + T(q^2)},
\end{equation}
and $T(q^2)$ is regular at zero. Obviously, it is a combination of
the previous ghost $T_{gh}^{(n)}(q^2)$ and quark  $T_q^{(n)}(q^2)$
at $n=1,2$ invariant dimensionless functions (it includes the
dependence on the coupling constant squared again and the gauge
fixing parameter as well in the general case (i.e., when $D^0(q)$
is given by Eq. (2.3)). Since $A(q^2)$ is finite at zero, the IR
singularity of the linear part of the full gluon propagator is
completely determined by the power-type IR singularity of the free
gluon propagator, as it follows from Eq. (2.11), i.e.,
$\tilde{D}^0(q) = A(0) D^0(q), \quad q^2 \rightarrow 0$. We are
especially interested in the structure of the full gluon
propagator in the IR region, so the relation (2.11) will be used
as an input in the direct iteration solution of the gluon SD
equation (2.5). Evidently, this form of the gluon SD equation
makes it possible to take into account automatically ghost and
quark degrees of freedom in all orders of linear PT. On the other
hand, it emphasizes the important role of the pure gluon
contribution (i.e., YM one), which forms its NL part.

Let us present now the NL pure gluon part, which was symbolically
denoted as $T_g[D](q)$ in the gluon SD Eq. (2.5). As mentioned
above, it is a sum of four terms, namely

\begin{equation}
T_g[D](q)  = {1 \over 2} T_t + {1 \over 2} T_1(q) + {1 \over 2}
T_2(q) + {1 \over 6} T_2'(q),
\end{equation}
where the corresponding quantities are given explicitly below

\begin{equation}
T_t =  g^2 \int {i d^4 q_1 \over (2 \pi)^4} T^0_4 D(q_1),
\end{equation}

\begin{equation}
T_1(q) =  g^2 \int {i d^4 q_1 \over (2 \pi)^4} T^0_3 (q, -q_1,
q_1-q) T_3 (-q, q_1, q -q_1) D(q_1) D(q -q_1),
\end{equation}

\begin{equation}
T_2(q) =  g^4 \int {i d^4 q_1 \over (2 \pi)^4} \int {i d^n q_2
\over (2 \pi)^4} T^0_4 T_3 (-q_2, q_3, q_2 -q_3) T_3(-q, q_1,
q_3-q_2) D(q_1) D(-q_2)D(q_3) D(q_3 -q_2),
\end{equation}

\begin{equation}
T_2'(q) =  g^4 \int {i d^4 q_1 \over (2 \pi)^4} \int {i d^4 q_2
\over (2 \pi)^4} T^0_4 T_4 (-q, q_1, -q_2, q_3) D(q_1)
D(-q_2)D(q_3).
\end{equation}
In the last two equations $q-q_1 +q_2-q_3=0$ is assumed as usual.
The $T_t$ term, which is given in Eq. (2.14), is the so-called
tadpole term constant contribution into the gluon propagator
(gluon self-energy). The $T_1(q)$ term describes the one-loop
skeleton contribution, depending on the three-gluon vertices only.
The $T_2(q)$ term describes the two-loop skeleton contribution,
depending on the three- and four-gluon vertices, while the
$T_2'(q)$ term describes the two-loop skeleton contribution,
depending on the four-gluon vertices only.

The formal iteration solution of Eq. (2.5) looks like

\begin{equation}
D(q) = D^{(0)}(q)  + \sum_{k=1}^{\infty}D^{(k)}(q) = D^{(0)}(q) +
\sum_{k=1}^{\infty} \Bigl[ D^{(0)}(q) T_g \Bigl[
\sum_{m=0}^{k-1}D^{(m)} \Bigr] (q) \Bigl( \sum_{m=0}^{k-1}D^{(m)}
(q)\Bigr) - \sum_{m=1}^{k-1}D^{(m)}(q) \Bigr],
\end{equation}
where, for example explicitly the first four terms are:

\begin{eqnarray}
D^{(0)}(q) &=& \tilde{D}^0(q), \nonumber\\
D^{(1)}(q) &=& \tilde{D}^0(q)T_g[\tilde{D}^0] (q) \tilde{D}^0(q), \nonumber\\
D^{(2)}(q) &=& \tilde{D}^0(q)T_g[\tilde{D}^0 +  D^{(1)}] (q)
(\tilde{D}^0(q) + D^{(1)}(q)) - D^{(1)}(q), \nonumber\\
D^{(3)}(q) &=& \tilde{D}^0(q)T_g[\tilde{D}^0 + D^{(1)} + D^{(2)}](q)
(\tilde{D}^0(q) + D^{(1)}(q) + D^{(2)}(q))
 - D^{(1)}(q) - D^{(2)}(q),
\end{eqnarray}
and so on. It is worth mentioning that the order of iteration does
not coincide with the order of PT in the coupling constant
squared. For example, any iteration (even zero) in Eq. (2.18)
contains ghost and quark degrees of freedom in all orders of PT,
as underlined above. In other words, the iteration solution (2.18)
is a general one, since the skeleton loop contributions (skeleton
diagrams) are to be iterated (the so-called general iteration
solution). In principle, it should be distinguished from the pure
PT iteration solution, i.e., from the expansion in powers of the
coupling constant squared. In this case the pure PT diagrams (with
free propagators and point-like vertices) are to be iterated.
 Of course, there is no hope to find solution in a
closed form, i.e., to sum up an infinite series presented in Eq.
(2.18). However, for future purpose it is useful to note its most
important algebraic feature. As it follows from Eqs. (2.19) each
subsequent iteration contains all the preceding ones. Below we
will establish the dynamical context of the general iteration
solution (2.18).

\subsection{The deep IR structure of the gluon propagator}

In order to investigate the IR structure of the full gluon
propagator it is instructive to start from the investigation of
the linear part of the gluon SD equation in the straightforward
$q=0$ limit. Since the full gluon propagator has, in general, the
IR singularity $1/q^2$ (see Eq. (2.1)), the replacement $1/q^2
\rightarrow 1/(q^2 + i\epsilon)$ is understood in order to assign
a mathematical meaning to this limit (this $\epsilon$ is not to be
mixed up with the IR regularization parameter introduced in Sec.
3). The corresponding skeleton loop integral, which contains quark
degrees of freedom (2.8) in this limit becomes

\begin{equation}
T_q(0) = - g^2 \int {i d^4 p \over (2 \pi)^4} Tr [\gamma_{\nu}
S(p) \Gamma_{\mu}(p, 0)S(p)].
\end{equation}
It is easy to see that this integral does not exhibit any
singularities in the integrand at very small values of the
skeleton loop variable. This number may be infinite in the UV
limit, and its removal is the subject to the corresponding UV
renormalization procedure, as mentioned above.

 In the same way the ghost skeleton loop integral (2.7) becomes

\begin{equation}
T_{gh}(0) =  g^2 \int {i d^4 k \over (2 \pi)^4} k_{\nu} G(k)
G(k)G_{\mu}(k, 0).
\end{equation}
At first sight an additional singularity at very small values of
the skeleton loop variable will appear because of the second ghost
propagator. However, this is not the case. The ghost-gluon vertex
$G_{\mu}(k, 0)$ is the linear function of its argument and the
combination $k_{\nu}k_{\mu}$ will cancel this additional
singularity. So, it is finite in the IR region and its removal is
again the subject to the UV renormalization program.

Since the NL part of the gluon SD equation starts from the tadpole
term (2.14), which does depend on the external gluon momentum $q$
at all, let us consider the skeleton loop integral (2.15) first.
In the exact $q=0$ limit it is

\begin{equation}
T_1(0) =  g^2 \int {i d^4 q_1 \over (2 \pi)^4} T^0_3 (0, -q_1,
q_1) T_3 (0, q_1, -q_1) D(q_1) D(-q_1).
\end{equation}
An additional singularity due to $D(-q_1)= D(q_1)$ will appear.
However, the three-gluon vertices which are present in the
nominator being the linear functions of their arguments will
cancel this additional IR singularity. Thus, this skeleton loop
integral is finite in the deep IR region.

The two-loop skeleton integral (2.16) in the exact $q=0$ limit
becomes

\begin{equation}
T_2(0) =  g^4 \int {i d^4 q_1 \over (2 \pi)^4} \int {i d^n q_2
\over (2 \pi)^4} T^0_4 T_3 (-q_2, -q_1 +q_2, q_1) T_3(0, q_1,
-q_1) D(q_1) D(-q_2)D(-q_1+q_2) D(-q_1).
\end{equation}
Again an additional singularities in the integration over the very
small values of the loop variables $q_1$ and $q_2$ due to the full
gluon propagators $D(-q_1+q_2)$ and $D(-q_1)$ are to appear.
However, contrary to the  previous case these singularities might
not be cancelled by terms which come from the corresponding
three-gluon vertices at the very small values of all the gluon
momenta involved (the number of the three-gluon vertices might not
be enough). So, this skeleton loop integral can be source of the
additional IR singularities with respect to the small values of
the external gluon momentum $q$ (however, see remarks below in
Subsec. E).

The most strict evidence of unavoidable occurrence of the
additional IR singularities in the full gluon propagator is given
by the last two-loop skeleton integral (2.17). In the exact $q=0$
limit it is

\begin{equation}
T_2'(0) =  g^4 \int {i d^4 q_1 \over (2 \pi)^4} \int {i d^4 q_2
\over (2 \pi)^4} T^0_4 T_4 (0, q_1, -q_2, -q_1+q_2) D(q_1)
D(-q_2)D(-q_1+q_2).
\end{equation}
Again an additional singularities will appear due to $D(-q_1+q_2)$
in the integration over the very small values of the loop
variables $q_1$ and $q_2$. {\bf The important observation,
however, is that they cannot be cancelled by the corresponding
terms from the nominator, for sure, since the full four-gluon
vertex, when all the gluon momenta involved go to zero, will be
reduced to the corresponding point-like four-gluon vertex, which
does not depend on the gluon momenta involved at all.} Thus, the
straightforward $q=0$ limit is certainly dangerous in this case,
and more sophisticated method is needed to investigate the region
of all the small gluon momenta involved (see below). The IR
singular structure of this two-loop skeleton integral is the
principal distinction from all other skeleton loop integrals
considered above. Let us also make once thing perfectly clear. Due
to the above-mentioned singular structure this integral implicitly
contains the corresponding mass scale parameter (the mass gap). In
other words, the mass gap is hidden in the initial skeleton
integral (2.17) at nonzero $q$. They (the mass gap and an
additional singularities) will show up when $q$ goes to zero. How
to extract them explicitly from the corresponding Feynman diagram
see next Subsec.

\subsection{The explicit functional estimate}

Let us now establish a type of a possible functional dependence of
the full gluon propagator in the IR region. For this purpose it is
convenient to start with the gluon SD equation (2.5). Up to the
first iteration it becomes

\begin{equation}
D(q) = \tilde{D}^0(q) + \tilde{D}^0(q)T_g[D](q)D(q) =
        \tilde{D}^0(q) +
        \tilde{D}^0(q)T_g[\tilde{D}^0](q)\tilde{D}^0(q)+ .....,
\end{equation}
where we will use Eq. (2.11) for the modified free gluon
propagator in the Feynman gauge in what follows. In order to
clearly separate an additional IR singularities in the skeleton
loop integral (2.17) it is sufficient to explicitly consider it at
the order $g^4$, for which we should put $T_4 = T^0_4$. To this
order the two-loop contribution (2.17) then becomes

\begin{equation}
T'_2(q) \equiv T'^{(2)}_{\nu_1\mu_1}(q) = g^4 \int {i d^4 q_1
\over (2 \pi)^4} \int {i d^4 q_2 \over (2 \pi)^4}
T^0_{\nu_1\rho\lambda\sigma} T^0_{\mu_1\rho_1\lambda_1\sigma_1}
\tilde{D}^0_{\rho\rho_1}(q_1) \tilde{D}^0_{\lambda\lambda_1}
(-q_2) \tilde{D}^0_{\sigma\sigma_1} (q-q_1+q_2),
\end{equation}
where it is assumed that the summation over color group factors
has been already done and is included into the coupling constant
(as well as some other finite numerical factors, which can appear
as results of the integration, see below), since these numbers are
not important. The summation over Dirac indices then yields

\begin{equation}
T'^{(2)}_{\nu_1\mu_1}(q) = - i \delta_{\nu_1\mu_1} g^4 \int
id^4q_1 \int id^4q_2 { A(q_1^2) A(q_2^2) A((q-q_1+q_2)^2)  \over
q_1^2 q_2^2 (q-q_1+q_2)^2 } = - i \delta_{\nu_1\mu_1} g^4
F'_2(q^2),
\end{equation}
where we introduce

\begin{equation}
F'_2(q^2) = \int id^4q_1 \int id^4q_2 { A(q_1^2) A(q_2^2)
A((q-q_1+q_2)^2) \over q_1^2 q_2^2 (q-q_1+q_2)^2 }.
\end{equation}
As emphasized above, this integral possesses very distinctive and
important feature, namely it exhibits an additional singularities
in the integration over the very small values of the loop
variables $q_1$ and $q_2$ when the straightforward $q=0$ limit for
the external gluon momentum is undertaken, so it should not be
finite in this limit. In this connection, let us remind that the
$A$-function is a regular at zero. Thus, the exact $q=0$ limit is
dangerous, since this is the IR singularity of the full gluon
propagator as well (see Eq. (2.1)). Let us also emphasize once
more that the existence of the additional IR singularities assumes
the existence of the corresponding mass scale parameter, at least
one, (the mass gap), which is "hidden" in this integral. As
mentioned above, more sophisticated method is needed to detect the
above-mentioned additional IR singularities, and hence the
existence of the corresponding mass gap in the deep IR structure
of the full gluon propagator.

 For this purpose and in order to introduce a mass gap, which determines
the deviation of the full gluon propagator from the free one in
the IR region (due to the above-mentioned additional IR
singularities) at the level of the separate diagram
(contribution), let us present the last integral as a sum of four
terms, namely

\begin{equation}
F'_2(q^2) = \sum_{n=1}^{n=4} F'^{(n)}_2(q^2),
\end{equation}
where

\begin{equation}
F'^{(1)}_2(q^2) = \int_0^{\Delta^2} id^4q_1 \int_0^{\Delta^2}
id^4q_2 {  A(q_1^2) A(q_2^2) A((q-q_1+q_2)^2) \over q_1^2 q_2^2
(q-q_1+q_2)^2 },
\end{equation}

\begin{equation}
F'^{(2)}_2(q^2) = \int_{\Delta^2}^{\infty} id^4q_1
\int_0^{\Delta^2} id^4q_2 {  A(q_1^2) A(q_2^2) A((q-q_1+q_2)^2)
\over q_1^2 q_2^2 (q-q_1+q_2)^2 },
\end{equation}

\begin{equation}
F'^{(3)}_3(q^2) = \int_0^{\Delta^2} id^4q_1
\int_{\Delta^2}^{\infty} id^4q_2 {  A(q_1^2) A(q_2^2)
A((q-q_1+q_2)^2) \over q_1^2 q_2^2 (q-q_1+q_2)^2 },
\end{equation}

\begin{equation}
F'^{(4)}_2(q^2) = \int_{\Delta^2}^{\infty} id^4q_1
\int_{\Delta^2}^{\infty} id^4q_2 {  A(q_1^2) A(q_2^2)
A((q-q_1+q_2)^2) \over q_1^2 q_2^2 (q-q_1+q_2)^2 },
\end{equation}
where not loosing generality we introduced the common mass gap
squared $\Delta^2$ for both loop variables $q_1^2$ and $q_2^2$.
The integration over angular variables is assumed.

 As mentioned above, we are especially interested in the region of
all the small gluon momenta involved, i.e., $q \approx q_1 \approx
q_2 \approx 0$. However, in Eq. (2.30) we can formally consider
the variables $q_1$ and $q_2$ as much smaller than the small gluon
momentum $q$, i.e., to approximate $q_1 \approx \delta_1q, \ q_2
\approx \delta_2q$, so that $q-q_1+q_2 \approx q(1+ \delta)$,
where $\delta = \delta_2 - \delta_1$. To leading order in
$\delta$, one obtains

\begin{equation}
F'^{(1)}_2(q^2) = - {A(q^2) \over q^2} \int_0^{\Delta^2} dq^2_1
\int_0^{\Delta^2} dq^2_2 A(q_1^2) A(q_2^2),
\end{equation}
where all the finite numbers after the trivial integration over
angular variables will be included into the numerical factors
below, for simplicity. Since $q^2$ is small, we can replace the
dimensionless function $A(q^2)$ by its Taylor expansion as
follows: $A(q^2) = A(0) + a_1 (q^2 / \Delta^2) + O(q^4)$.
Introducing further dimensionless variables $q_1^2 = x_1 \Delta^2$
and $q_2^2 = x_2 \Delta^2$, one finally obtains

\begin{equation}
F'^{(1)}_2(q^2) = - {\Delta^4 \over q^2} c_1  - \Delta^2 c'_1  +
O(q^2),
\end{equation}
where

\begin{equation}
c_1 = A(0) \int_0^1 dx_1 A(x_1) \int_0^1 dx_2 A(x_2),
\end{equation}
and $c'_1 = a_1 (c_1 /A(0))$. The both numbers are obviously
finite.

In Eq. (2.31) it makes sense to approximate $q_2 \approx
\delta_3q_1, \ q \approx \delta_4q_1$, so that $q-q_1+q_2 \approx
q_1(1+ \tilde{\delta})$, where $\tilde{\delta} = \delta_4 -
\delta_3$. To leading order in $\tilde{\delta}$ and omitting some
algebra, one finally obtains

\begin{equation}
F'^{(2)}_2(q^2) = - \Delta^2 c_2(\nu) + O(q^2),
\end{equation}
where

\begin{equation}
c_2(\nu) = \int_1^{\nu} {dx_1 \over x_1} A^2(x_1) \int_0^1 dx_2
A(x_2),
\end{equation}
and $\nu$ is the dimensionless auxiliary UV cut-off.

In Eq. (2.32) it makes sense to approximate $q_1 \approx
\delta_5q_2, \ q \approx \delta_6q_2$, so that $q-q_1+q_2 \approx
q_2(1+ \bar \delta)$, where $\bar \delta = \delta_5 + \delta_6$.
To leading order in $\bar \delta$ and similar to the previous
case, one obtains

\begin{equation}
F'^{(3)}_2(q^2) = - \Delta^2 c_3(\nu) + O(q^2),
\end{equation}
where

\begin{equation}
c_3(\nu) = \int_1^{\nu} {dx_2 \over x_2} A^2(x_2) \int_0^1 dx_1
A(x_1).
\end{equation}
The last term (2.33) is left unchanged, since all loop variables
are big. Conventionally, we will call it as the PT part of the
contribution (diagram), i.e., denoting $F'^{(4)}_2(q^2)$ as
$F'^{PT}_2(q^2)$. Since $A(x)$ is regular at zero, the both
integrals in Eqs. (2.38) and (2.40) are logarithmically divergent.
Also if one wants to neglect ghost and quark degrees of freedom
one only needs to replace everywhere the $A$-function by unity
(see Eq. (2.12)).

 Summing up all terms, one obtains

\begin{equation}
T'_2(q) \equiv T'^{(2)}_{\nu_1\mu_1}(q) = i \delta_{\nu_1\mu_1}
\Bigr[ {\Delta^4 \over q^2} c_1 + \Delta^2(c_2(\nu) + c_3 (\nu))
\Bigl] g^4 + O(q^2).
\end{equation}
The term $F'^{PT}_2(q^2)$ is hidden in terms $O(q^2)$. Here the
characteristic mass scale parameter $\Delta^2$ is responsible for
the nontrivial dynamics in the IR domain. Let us also emphasize
that the limit $\nu \rightarrow \infty$ should be taken at the
final stage. Anyway, the finite constant $c'_1$ from the expansion
(2.35) has been already suppressed in comparison with the
logarithmically divergent integrals $c_2(\nu)$ and $c_3(\nu)$ even
at this stage. So, the integral (2.28) and hence the original
integral (2.26) is divergent in the exact $q=0$ limit, indeed. In
other words, these singularities with respect to the external
gluon momentum $q$ will show explicitly up if and only if it goes
to zero. Evidently, the integral (2.30) is an example of
overlapping IR divergences, nevertheless, it plays no important
role in the deep IR structure of the gluon propagator. First of
all, the finite constant contribution is to be neglected as
mentioned above. From the IR renormalization procedure it will
follow (see Secs. VI and VII) that the terms of the order
$O(\Delta^4)$ will be suppressed.

 The constant tadpole term (2.14) produces the
contribution as follows: $T_t \equiv T^t_{\nu_1\mu_1} = - i
\delta_{\nu_1\mu_1} \Delta^2 c_t(\nu) g^2$, where $c_t(\nu)=
\int_0^{\nu} dx_1 A(x_1)$. Formally it can be identically
decomposed into the two parts by introducing $c_t(\nu) =c_t(\nu) +
c_t - c_t= c_t + c'_t(\nu)$. The finite part depending on $c_t$
can be included into the INP part of the full gluon propagator
(see below), while leaving other infinite part for its PT part.
Let us note, however, that in dimensional regularization this term
in the pure PT iteration solution (which means $D =D^0 + ...$) of
the gluon SD equation can be generally discarded \cite{17}. Thus,
this term itself is not important at all.

 Evidently, such kind of the auxiliary (but important) procedure,
described in this Subsec., is appropriate only for the
establishing the most singular (leading and next-to-leading) terms
in the deep IR structure of the full gluon propagator. On the
other hand, it makes the explicit dependence of this structure on
the mass gap perfectly clear.

\subsection{Severe IR structure of the QCD vacuum }

At the NL two-loop level, i.e., at the order $g^4$, there is a
number of the additional diagrams, which, however, contain the
three-gluon vertices (plus the two-tadpole diagram) along with the
four-gluon ones. Their possible contributions into the deep IR
structure of the gluon propagator are given by the estimates
similar to the estimate (2.41) with different coefficients, of
course. Omitting some really tedious algebra, the full gluon
propagator (2.25) up to the first iteration can be written as

\begin{equation}
D_{\mu\nu}(q) = i \delta_{\mu\nu} \Bigl[ {\Delta^2 \over (q^2)^2}
a_1 + {\Delta^4 \over (q^2)^3} a_2 + ... \Bigr] +
D^{PT}_{\mu\nu}(q)= D^{INP}_{\mu\nu}(q)+ D^{PT}_{\mu\nu}(q),
\end{equation}
where $a_1, \ a_2$ are, in general, the short-hand notations for a
sums of the different coefficients, which include the coupling
constant squared in the corresponding powers. Moreover, some of
these coefficients contain the divergent integrals (see, for
example Eqs. (2.38) and (2.40)). Here $D^{PT}_{\mu\nu}(q)$ denotes
the contribution from the PT part of the full gluon propagator,
since it is of the order $O(q^{-2})$ as $q^2 \rightarrow 0$. The
superscript "INP" stands for the intrinsically NP part of the full
gluon propagator (for the exact definition see Secs. 4 and 7
below). Due to the distinction between the behavior of the tree-
and four-gluon vertices in the deep IR domain (i.e, when all the
gluon momenta involved go to zero, see discussion in Subsec. E
below), the coefficients $a_1, \ a_2$ are, in general, not zero.
In other words, there is no way to cancel $D^{INP}_{\mu\nu}(q)$ by
performing the functional estimate at every order of the QCD
coupling constant squared. In the deep IR region the quark and
ghost degrees of freedom are taken into account in all orders of
linear PT numerically, i.e., they are simply numbers. As functions
they can contribute into the PT part only of the full gluon
propagator. So, in the first approximation the gluon propagates
like Eq. (2.42) and not like the modified free one (2.11), though
we just started from it.

 The true QCD vacuum is really beset with severe (i.e., more singular
than $1/q^2$ as $q^2 \rightarrow 0$) IR singularities if standard
PT is applied. Moreover, each severe IR singularity is to be
accompanied by the corresponding powers of the mass gap,
responsible for the NP dynamics in the IR region. In more
complicated cases of the multi-loop diagrams (i.e., the next
iterations in Eq. (2.25)) more severe IR divergences will appear.
The coefficients at each severe IR singularity become by
themselves an infinite series in the coupling constant squared,
and the coefficients of these expansions may depend on the gauge
fixing parameter as well \cite{17}. These coefficients include
numerically the information about quark and ghost degrees of
freedom in all orders of linear PT, as underlined above.

It is worth emphasizing, however, that the ZMME effect in the QCD
vacuum, which is explicitly shown in Eq. (2.42) in the Feynman
gauge, can be demonstrated in any covariant gauge, for example in
the Landau one $\xi=0$. In other words, this effect itself is
gauge-invariant, though the finite sum of all the relevant
diagrams in the deep IR region at the same order of the coupling
constant squared may be not. Let us also remind that this effect
is not something new. It has been well known for a long time from
the very beginning of QCD, and it was the basis for the proposed
then IR slavery (IRS) mechanism of quark confinement
\cite{1,32,33,34,35,36}. Just this IR violent behavior makes QCD
as a whole an IR unstable theory, and therefore it has no IR
stable fixed point, indeed \cite{1,32}. {\bf Moreover, we
unambiguously identify the main source of the IR instability of
QCD, namely the four-gluon vertex at the Lagrangian level and the
two-loop skeleton term, which contains only the four-gluon
vertices, at the level of the gluon SD equation. Precisely this
interaction exhibits an additional IR singularities in the
corresponding loop integrals when all the gluon momenta involved
go to zero.}

The existence of a severe IR singularities automatically requires
an introduction of a mass gap, responsible for the nontrivial
dynamics in the IR region. This is important, since there is none
explicitly present in the QCD Lagrangian (the current quark mass
cannot be considered as a mass gap, since it is not
renormalization group invariant). It precisely determines to what
extent the full gluon propagator effectively changes its behavior
from the behavior of the free one in the IR domain. The phenomenon
of ''dimensional transmutation'' \cite{1,37} only supports our
general conclusion that QCD may exhibit a mass, determining the
characteristic scale of the NP dynamics in its ground state. Of
course, such gluon field configurations, which are to be described
by severely IR structure of the full gluon propagator, can be only
of dynamical origin. The only dynamical mechanism in QCD which can
produce such configurations in the vacuum, is the self-interaction
of massless gluons -- the main dynamical NL effect in QCD. Hence,
the above-mentioned mass gap appears on dynamical ground. Let us
remind that precisely this self-interaction in the UV limit leads
to AF.

We have explicitly shown that the low-frequency components of the
virtual fields in the true vacuum should have larger amplitudes
than those of a PT ("bare") vacuum \cite{10}, indeed. "But it is
to just this violent IR behavior that we must look for the key to
the low energy and large distance hadron phenomena. In particular,
the absence of quarks and other colored objects can only be
understood in terms of the IR divergences in the self-energy of a
color bearing objects" \cite{33}. So, let us introduce the
following definitions:

(i). The power-type IR singularity which is more severe than the
exact power-type IR singularity of the free gluon propagator will
be called a severe (or equivalently NP IR) singularity. In other
words, the NP IR singularity is more severe than $1/ q^2$ at $q^2
\rightarrow 0$.

(ii). At the same time, the IR singularity which is as much
singular as the exact power-type IR singularity of the free gluon
propagator, i.e., as much singular as $1/ q^2$ at $q^2 \rightarrow
0$, will be called PT IR singularity.

It makes worth emphasizing in advance that the decomposition of
the full gluon propagator into the INP and PT parts (2.42) can be
made exact (see below). From the distribution theory point of view
the NP IR singularities defined above present a rather broad and
important class of functions with algebraic singularities
\cite{24}. This explicit derivation shows how precisely the NP IR
singularities, accompanied with a mass gap in the corresponding
powers, may appear in the vacuum of QCD. Thus, the NP IR
singularities should be summarized (accumulated) into the full
gluon propagator and effectively correctly described by its
structure in the deep IR domain, presented by its INP part. The
second step is, of course, to assign a mathematical meaning to the
integrals, where such kind of the NP IR singularities will
explicitly appear, i.e., to define them correctly in the IR region
(see the next Sec.).

{\bf One important thing should be made perfectly clear. In the
exact calculation of a separate diagram the dependence on the
characteristic masses (determining the deviation of the full gluon
propagator from the free one in the IR and UV regions) is hidden.
In other words, these masses cannot be "seen" by the calculation
of the finite number of diagrams, which may be not even
gauge-invariant. An infinite number of the corresponding diagrams
should be summed up in order to trace such NP masses (i.e., to go
beyond PT). The final result of such summation should, in
principle, be gauge-invariant. In the weak coupling regime we know
how to do this with the help of the renormalization group
equations. As a result, the dependence on $\Lambda_{QCD} \equiv
\Lambda_{PT}$ will finally appear. At the same time, we do not
know how to solve these equations in the strong coupling regime.
So, in order to avoid this problem, we decided to show the
existence of a mass gap explicitly, by extracting the deep IR
asymptotics of the gluon propagator within the separate relevant
diagrams. The rest of the problem is to sum up an infinite number
of the most singular (leading and next-to-leading) contributions
in order to see whether or not a mass gap will finally survive.
Precisely this program will be carried out in what follows. We
will show that a mass gap remains, indeed, and consequently the
full gluon propagator becomes unavoidably severely singular in the
deep IR domain.}

\subsection{Discussion}

\subsubsection{A necessary generalization}

 The three-gluon proper vertex vanishes when all the gluon
independent momenta involved go to zero, i.e., $T_3(0,0)
\rightarrow T_3^0(0,0) =0$. This is also true for the ghost-gluon
proper vertex when all the momenta involved go to zero, namely
$G_{\mu}(0,0)=0$. Because of this behavior, the three-gluon and
ghost-gluon vertices should not play any noticeable role in the
deep IR structure of the full gluon propagator. Though separate
terms of an infinite series presented by the corresponding
skeleton loop integrals (2.7) (this loop integral does not exhibit
any IR singularities at all), (2.8), (2.15) and (2.16) may be
formally singular (and hence depend on the mass gap), their
general tensor decompositions are to be present by the
decompositions (2.9) and (2.10) but with different invariant
functions for the skeleton loop integrals (2.15) and (2.16), of
course. In other words, these terms will not survive after summing
up an infinite number of the corresponding contributions. At the
same time the four-gluon proper vertex is not zero when all the
gluon momenta involved go to zero, i.e., $T_4(0,0,0) \rightarrow
T_4^0(0,0,0) \neq 0$. This is the main dynamical source of the
additional IR singularities (and hence of the mass gap), which are
hidden in the corresponding two-loop skeleton integral (2.17).
That is why it plays so important role in the IR structure of the
full gluon propagator.

One can conclude that for all the skeleton loop integrals
mentioned above the exact $q=0$ limit is smooth. Thus, not loosing
generality, they produce the contributions which are of the order
$O(q^2)$ always. However, this is not the case for the two-loop
skeleton integral (2.17), which contains the four-gluon vertices.
In this case the exact $q=0$ limit is singular, and the
next-to-leading constant contributions are multiplied by the
divergent quantities. This everything means that the tensor
decomposition of the NL part $T_g[D](q) \equiv T^g_{\mu\nu}[D](q)$
is necessarily to be generalized as follows:

\begin{equation}
T^g_{\mu\nu}[D](q) =  \delta_{\mu\nu} \Bigl[ {\Delta^4 \over q^2}
L_g^{(1)}(q^2) + \Delta^2 L_g^{(2)}(q^2) +
 q^2 T_g^{(3)}(q^2) \Bigr]
 +  q_{\mu} q_{\nu} \Bigl[ {\Delta^2
\over q^2} L_g^{(4)} (q^2) + T_g^{(5)} (q^2)\Bigr],
\end{equation}
where $T_g^{(n)}(q^2)$ at $n=3,5$ are invariant dimensionless
functions. They are regular functions of $q^2$, i.e., they can be
present by the corresponding Taylor expansions, but possessing AF
at infinity, and depending thus on $\Lambda_{QCD}$ in this limit.
They are saturated by the skeleton loop integrals (2.15), (2.16)
and (2.17). At the same time, the invariant dimensionless
functions $L_g^{(n)}(q^2)$ at $n=1,2,4$ are to be present by the
corresponding Laurent expansions, namely

\begin{equation}
L_g^{(1,2,4)}(q^2) \equiv L_g^{(1,2,4)}(q^2, \Delta^2) =
\sum_{k=0}^{\infty} (\Delta^2 / q^2)^k a_k^{(1,2,4)},
\end{equation}
where the numbers $a_k^{(1,2,4)}$ by themselves are expansions in
the coupling constant squared (see below). These invariant
functions are to be saturated by the skeleton loop integral (2.17)
only. Let us emphasize the inevitable appearance of the mass gap
$\Delta^2$. It characterizes the nontrivial dynamics in the IR
region. This precisely makes the difference between the linear and
NL insertions into the gluon self-energy. Evidently, this
difference is due to different dynamics: in the linear part there
is no explicit direct interaction between massless gluons, while
in the NL part there is. When the mass gap is zero then this
decomposition takes the standard form. So, the generalization
(2.43) makes the explicit dependence on the mass gap of the full
gluon propagator perfectly clear. Let us emphasize once more that
ghost and quark degrees of freedom contribute into the INP part of
the full gluon propagator numerically only, and as a functions
they contribute into its PT part. Neither ghost nor quark skeleton
loops (2.7) and (2.8), which appear in Eq. (2.4), can cancel its
severely singular behavior in the IR, which was demonstrated
above.  It was due to the pure YM part, i.e., to its NL part of
the gluon SD equation (2.5). More precisely mainly to its two-loop
skeleton integral, which contain the four-gluon vertices only.

\subsubsection{The role of ghosts}

It is well known that in order to maintain the unitarity of
$S$-matrix in QCD the ghosts have to cancel unphysical degrees of
freedom (longitudinal ones) of the gauge bosons \cite{1,7,17,18}.
Evidently, this is due to the general decomposition of the ghost
skeleton loop (2.9), which shows that it always gives the
contribution of the order $q^2$. In the iteration solution for the
gluon propagator (2.4), $D(q) = D^0(q) - D^0(q)T_{gh}(q)D^0(q)
+D^0(q)T_{gh}(q)D^0(q)T_{gh}(q)D^0(q) + ....$, it cancels one of
$q^2$ in the denominator, which comes from the free gluon
propagator. Thus, each term in this expansion becomes always as
singular as $1/ q^2$. Precisely this makes it possible for ghosts
to cancel, in general, the longitudinal component of the full
gluon propagator, which is, by definition, as singular as $1/
q^2$. From a technical point of view the cancellation can be
explicitly demonstrated in the lowest orders of PT in powers of
the coupling constant squared (see, for example Ref. \cite{17}).
However, this is valid term by term in PT. In other words, in
every order of PT the ghosts will cancel unphysical degrees of
freedom of gauge bosons, making thus them always transverse. The
above-mentioned cancellation in all orders of PT means that it
goes beyond PT. It is a general feature, i.e., it does not depend
on whether the solution, for example to the gluon SD equation is
PT or NP, singular or regular at origin, etc. In other words, the
general role of ghosts should not be spoiled by any truncation
scheme (approach).

On the other hand, the general decomposition (2.9) of the ghost
skeleton loop (2.7) takes place if and only if (iff) the full
ghost propagator (Euclidean signature) $G(k) = - (i / k^2[1 +
b(k^2)])$, where $b(k^2)$ is the ghost self-energy, is as singular
as $1/k^2$ at $k^2 \rightarrow 0$. When the ghost self-energy is
zero, i.e., $b(k^2)=0$., then the full gluon propagator becomes
the free one, i.e., $G(k) \rightarrow G_0(k) = - (i / k^2)$. Thus
the IR singularity of the full ghost propagator cannot be more
severe than the exact IR singularity of the free ghost propagator
in order to maintain the cancellation role of unphysical degrees
of freedom of gauge bosons by ghosts at any nonzero covariant
gauge in all sectors of QCD. There is no way for ghosts to cancel
severe IR singularities, which are of dynamical origin due to the
self-interaction of massless gluons in the true QCD vacuum. There
is no doubt left that the full gluon propagator is essentially
severely modified in the IR because of the response of the NP QCD
vacuum, which is not provided by the PT vacuum.

 However, there
exists one gap in these arguments. If one chooses by hand the
Landau gauge $\xi=0$ from the very beginning, then the unphysical
longitudinal component of the full gluon propagator vanishes. Only
the physical transverse component will contribute to the full
gluon propagator, and it may become regular at zero in this case,
indeed. Otherwise, it is always singular at the origin because the
existence of the longitudinal component always produces, at least,
the IR singularity $1/q^2$ (see Eq. (2.1)). In this case there is
no restriction on the behavior of the ghost propagator in the IR,
and it may become (depending on the truncation scheme) more
singular in the IR than its free counterpart. In Ref. \cite{38}
(and references therein) precisely this type of the solution
(regular gluon propagator and more singular than the free one
ghost propagator) to the system of the SD equations in the Landau
gauge has been found. However, this solution is due to the choice
of the special Landau gauge, so it is a gauge artifact solution.
Being thus a gauge artifact, it can be related to none of the
physical phenomena such as quark and gluon confinement, SBCS, etc,
which are, by definition, manifestly gauge-invariant. At the same
time, gauge artifact solutions may exist as formal solutions to
the SD system of equations. If a regular at zero gluon propagator
will be found in a manifestly gauge-invariant way (i.e., in the
way which does not explicitly depend on the particular covariant
or non-covariant gauge choice), only then it should be taken
seriously into the consideration. To our present knowledge a
manifestly gauge-invariant solution for the smooth gluon
propagator, which will not compromise the general role of ghosts,
is not yet found. Moreover, there exists a serious doubt, in our
opinion, that such kind of the solution can be found at all. Thus,
we are left with singular at the origin gluon propagator, which is
possible in any covariant gauge. In principle, the free gluon
propagator can be also used in any gauge. The Feynman gauge free
gluon propagator in the IR has been used by Gribov \cite{39} in
order to investigate the quark confinement problem within
precisely the SD system of dynamical equations approach.

Evidently, the iteration solution (2.38) does not reproduce all
aspects of the general iteration solution (2.18). The pure gluon
loops are reproduced up to $g^2$ and $g^4$ orders, while the ghost
and quark loops can be taken into account to all orders in $g^2$.
Moreover, the most important general feature of the iteration
solution, mentioned above is not seen clearly. The main purpose of
this paper is to establish the deep IR structure of the full gluon
propagator, not solving the gluon SD equation directly, which is a
formidable task, anyway. At the same time, we will show the way
how to reconstruct the structure of the full gluon propagator in
the IR domain in complete agreement with the structure of its
general iteration solution (2.18). However, it is convenient first
in the next Sec. to emphasize the distribution nature of severe IR
singularities.

 In summary, we have discussed the general
properties of the gluon SD equation and its formal iteration
solution. We have explicitly shown how the NP IR singularities
inevitably appear in the QCD vacuum. We have discussed the role of
ghosts, and it has been also explained why the smooth in the IR
gluon propagator is a gauge artifact.

\section{ IR dimensional regularization within the distribution theory}

In general, all the Green's functions in QCD are generalized
functions, i.e., they are distributions. This is true especially
for the NP IR singularities due to the self-interaction of
massless gluons in the QCD vacuum. They present a rather broad and
important class of functions with algebraic singularities, i.e.,
functions with nonsummable singularities at isolated points
\cite{24} (at zero in our case). Roughly speaking, this means that
all relations involving distributions should be considered under
corresponding integrals, taking into account the smoothness
properties of the corresponding class of test functions (for
example, $\varphi(q)$ below. Let us note in advance that in part
III we will establish the class of test functions). In principle,
any regularization scheme (i.e., how to parameterize severe IR
singularities and thereby to put them under control) can be used;
it should, however, be compatible with DT \cite{24}.

Let us consider the positively definite ($P>0$) squared
(quadratic) Euclidean form $P(q) = q_0^2 +  q_1^2 + q_2^2 + ... +
q_{n-1}^2 = q^2$, where $n$ is the number of the components. The
generalized function (distribution) $P^{\lambda}(q)$, where
$\lambda$ is, in general, an arbitrary complex number, is defined
as $(P^{\lambda}, \varphi) = \int_{P>0}P^{\lambda}(q) \varphi(q)
d^nq$. At $Re \lambda \geq 0$ this integral is convergent and is
an analytic function of $\lambda$. Analytical continuation to the
region $Re \lambda < 0$ shows that it has a simple pole at points
\cite{24}

\begin{equation}
\lambda = - {n \over 2} - k, \quad k=0, 1, 2 ,3...
\end{equation}

In order to actually define the system of the SD equations in the
deep IR domain, it is necessary to introduce the IR regularization
parameter $\epsilon$, defined as $D = n + 2 \epsilon, \ \epsilon
\rightarrow 0^+$ within a gauge-invariant DR method \cite{25}. As
a result, all the Green's functions and "bare" parameters should
be regularized with respect to $\epsilon$ (see below), which is to
be set to zero at the end of the computations. The structure of
the NP IR singularities is then determined (when $n$ is even
number) as follows \cite{24}:

\begin{equation}
(q^2)^{\lambda} = { C_{-1}^{(k)} \over \lambda +(D/2) + k} +
finite \ terms,
\end{equation}
where the residue is

\begin{equation}
 C_{-1}^{(k)} = { \pi^{n/2} \over 2^{2k} k! \Gamma ((n/2) + k) } \times
L^k \delta^n (q)
\end{equation}
with $L = (\partial^2 / \partial q^2_0) + (\partial^2 /
\partial q^2_1) + ... + (\partial^2 / \partial q^2_{n-1})$.

Thus, the regularization of the NP IR singularities (3.2) is
nothing but the so-called Laurent expansion.
 Let us underline its most remarkable feature.
The order of singularity does not depend on $\lambda$, $n$ and
$k$. In terms of the IR regularization parameter $\epsilon$, it is
always a simple pole $1/ \epsilon$. This means that all power
terms in Eq. (3.2) will have the same singularity, i.e.,

\begin{equation}
(q^2)^{- {n \over 2} - k } = { 1 \over \epsilon} C_{-1}^{(k)} +
finite \ terms, \quad \epsilon \rightarrow 0^+,
\end{equation}
where we can put $D=n$ now (i.e., after introducing this
expansion). By "$finite \ terms$" here and everywhere a number of
necessary subtractions under corresponding integrals is understood
\cite{24}. However, the residue at a pole will be drastically
changed from one power singularity to another. This means
different solutions to the whole system of the SD equations for
different set of numbers $\lambda$ and $k$. Different solutions
mean, in their turn, different vacua. In this picture different
vacua are to be labelled by two independent numbers: the exponent
$\lambda$ and $k$. At a given number of $D(=n)$ the exponent
$\lambda$ is always negative being integer if $D(=n)$ is an even
number or fractional if $D(=n)$ is an odd number. The number $k$
is always integer and positive and precisely it determines the
corresponding residue at the pole, see Eq. (3.3). It would not be
surprising if these numbers were somehow related to the nontrivial
topology of the nD QCD vacuum.

Concluding, let us note that the structure of severe IR
singularities in Euclidean space is much simpler than in Minkowski
space, where kinematical (unphysical) singularities due to the
light cone also exist \cite{1,24,40}. In this case it is rather
difficult to untangle them correctly from the dynamical
singularities, the only ones which are important for the
calculation of any physical observable. Also the consideration is
much more complicated in the configuration space \cite{24}. That
is why we always prefer to work in the momentum space (where
propagators do not depend explicitly on the number of dimensions)
with Euclidean signature. We also prefer to work in the covariant
gauges in order to avoid peculiarities of the noncovariant gauges
\cite{1,41}, for example how to untangle the gauge pole from the
dynamical one.

In summary, first we have emphasized the distribution nature of
the NP IR singularities. Secondly, we have explicitly shown how
the DR method should be correctly implemented into DT. This makes
it possible to put severe IR singularities under firm mathematical
control provided by DT itself, complemented by the DR method.

\section{The general structure of the full gluon propagator}

To say today that QCD is a NP theory is almost a tautology. The
problem is how to define it exactly, since we know for sure that
QCD has a PT phase as well because of AF. In order to investigate
this problem, namely, how to define the NP phase in QCD, it is
convenient to begin with the algebraic decomposition of the full
gluon form factor in Eq. (2.1) as follows:

\begin{equation}
d(q^2) = d(q^2) - d^{PT}(q^2) + d^{PT}(q^2) = d^{NP}(q^2) +
d^{PT}(q^2),
\end{equation}
where, for simplicity, the dependence on the gauge fixing
parameter is omitted. In fact, this formal equation represents one
unknown function (the full gluon form factor) as an exact sum of
the two other unknown functions, which can be always done. So, at
this stage there is no approximation made. We would like to let
the PT part of this decomposition to be responsible for the known
UV asymptotics (since it is fixed by AF) of the full gluon
propagator, while the NP part is chosen to be responsible for its
unknown yet IR asymptotics. It is worth emphasizing that in
realistic models of the full gluon propagator, the NP part
reproduces usually correctly its deep IR asymptotics, determining
thus the strong intrinsic influence of the IR properties of the
theory on its NP dynamics. Evidently, the decomposition (4.1)
represents an exact subtraction of the PT contribution at the
fundamental gluon level, and consequently both terms in the
right-hand-side of Eq. (4.1) are formally determined in the whole
momentum range $[0, \infty)$. Let us emphasize that the full gluon
form factor $d(q^2)$ being also NP, nevertheless, is
"contaminated" by the PT contributions, while $d^{NP}(q^2)$ due to
the subtraction (4.1) is free of them, i.e., it is truly NP.

Substituting the decomposition (4.1) into the full gluon
propagator (2.1), one obtains

\begin{equation}
D_{\mu\nu}(q) = D^{INP}_{\mu\nu}(q) + D^{PT}_{\mu\nu}(q),
\end{equation}
where

\begin{equation}
D^{INP}_{\mu\nu}(q) = i T_{\mu\nu}(q) d^{NP}(q^2){ 1 \over q^2} =
i T_{\mu\nu}(q) d^{INP}(q^2),
\end{equation}

\begin{equation}
D^{PT}_{\mu\nu}(q) = i \{ T_{\mu\nu}(q)  d^{PT}(q^2) + \xi
L_{\mu\nu}(q) \} { 1 \over q^2}.
\end{equation}
Let us remind that the superscript "INP" is the short-hand
notation for the intrinsically NP phase in QCD. Its definition
will be given below in Sec. 7. The exact decomposition (4.2) has a
remarkable feature. The explicit gauge dependence of the full
gluon propagator is shifted from its INP part to its PT part. In
other words, we want the INP part to be always transverse, while
leaving the PT part to be of arbitrary gauge. This exact
separation will have also a dynamical ground. It is clear also
that the PT part of the full gluon propagator is, by definition,
as much singular as the free gluon propagator's power-type IR
singularity. This is the first reason why the longitudinal part of
the full gluon propagator has been shifted to its PT part (the
longitudinal part has the same IR singularity as the free gluon
propagator). At the same time, the PT part of the full gluon
propagator otherwise remains arbitrary, but preserving AF. Let us
also emphasize that ghost and quark degrees of freedom will
contribute into its PT part only. As explained above, the
corresponding skeleton loop terms in the gluon SD equation produce
only $1/q^2$-type IR singularities in the gluon self-energy.

 We want the INP gluon form factor
$d^{INP}(q^2)$ to be responsible for the deep IR structure of the
full gluon propagator, which is saturated by severe IR
singularities. For this aim, it is convenient to introduce the
auxiliary INP gluon form factor as follows:

\begin{equation}
d^{INP}_{\lambda_k}(q^2,
 \Delta^2)= (\Delta^2)^{-\lambda_k - 1} (q^2)^{\lambda_k}
 f_{\lambda_k}(q^2),
\end{equation}
where the exponent $\lambda_k$ is, in general, an arbitrary
complex number with $Re \lambda_k < 0$ (see below). The mass
squared parameter $\Delta^2$ (the above-mentioned mass gap) is
responsible for the scale of NP dynamics in the IR region in our
approach. The functions $f_{\lambda_k}(q^2)$ are, by definition,
dimensionless, regular at zero, and otherwise remaining arbitrary,
but preserving AF in the UV limit. And finally the number $k$ is a
positive integer, i.e., $k=0,1,2,3,....$ (see above). Evidently, a
real INP gluon form factor $d^{INP}(q^2)$, which now should depend
on the mass gap as well, i.e., $d^{INP}(q^2) \equiv d^{INP}(q^2,
\Delta^2)$, is a sum over all auxiliary $d^{INP}_{\lambda_k}(q^2,
\Delta^2)$.

However, this is not the whole story yet. Since we are especially
interested in the deep IR structure of the full gluon propagator,
the arbitrary functions $f_{\lambda_k}(q^2)$ should be also
expanded around zero in the form of the Taylor series in powers of
$q^2$, i.e,

\begin{equation}
f_{\lambda_k}(q^2) = \sum_{m=0}^{[- \lambda_k] -(n/2)} {(q^2)^m
\over m!} f^{(m)}_{\lambda_k}(0) + \sum_{m=[-\lambda_k] - (n/2)
+1}^{\infty} {(q^2)^m \over m!} f^{(m)}_{\lambda_k}(0),
\end{equation}
where $[-\lambda_k]$ denotes its integer number and $n$ is the
number of the components in the Euclidean squared form $q^2$. Also

\begin{equation}
f^{(m)}_{\lambda_k}(0) =  \Bigl( d^m f_{\lambda_k}(q^2) / d(q^2)^m
\Bigr)_{q^2=0}.
\end{equation}
As a result, we will be left with the finite sum of power terms
with an exponent decreasing by unity starting from $- \lambda_k$.
All other remaining terms from the Taylor expansion (4.6),
starting from the term having already a PT IR singularity (the
second sum in Eq. (4.6)), should be shifted to the PT part of the
full gluon propagator in Eq. (4.2). The INP part in Eq. (4.5) then
becomes

\begin{equation}
d^{INP}_{\lambda_k}(q^2, \Delta^2)= (\Delta^2)^{-\lambda_k - 1}
 (q^2)^{\lambda_k}
\sum_{m=0}^{[- \lambda_k] -(n/2)} {(q^2)^m \over m!}
f^{(m)}_{\lambda_k}(0),
\end{equation}
while the piece which is to be shifted to the PT part (4.4) of the
full gluon propagator (4.2) is as follows:

\begin{eqnarray}
d^{INP}_{(s)\lambda_k}(q^2,
 \Delta^2)   &=& (\Delta^2)^{- \lambda_k -1} (q^2)^{\lambda_k}
\sum_{m=[-\lambda_k] - (n/2) +1}^{\infty} {(q^2)^m \over m!}
f^{(m)}_{\lambda_k}(0) \nonumber\\
&=& (\Delta^2)^{-\lambda_k -1} \sum_{m=0}^{\infty} {(q^2)^m \over
(m + [-\lambda_k] - (n/2)+1)!} f^{( m + [-\lambda_k] - (n/2)+1
)}_{\lambda}(0),
\end{eqnarray}
where the subscript "(s)" means "shifted". The above-mentioned sum
over all $\lambda_k$ is also assumed. The important thing here is
that the expression (4.9) does not contain the NP (severe) IR
divergences with respect to the gluon momentum, indeed.

\section{4D QCD}

We are particulary interested in 4D QCD (i.e., $n=4$), which is a
realistic dynamical theory of strong interactions not only at the
fundamental quark-gluon level, but at the hadronic level as well
\cite{1}. Let us discuss the gluon propagator (4.2) in more detail
for QCD itself. On account of the expansion (4.8) and Eq. (3.1) at
$n=4$ with the obvious identification $\lambda_k \equiv \lambda$,
its INP part becomes

\begin{equation}
d^{INP}(q^2, \Delta^2) = \sum_{k=0}^{\infty} d^{INP}_k(q^2,
\Delta^2)= \sum_{k=0}^{\infty} (\Delta^2)^{1 + k} (q^2)^{-2 -k}
\sum_{m=0}^{k} {(q^2)^m \over m!} f^{(m)}_k(0),
\end{equation}
and $f^{(0)}_k(0) \equiv f_k(0)$.
 Obviously, in this case the subscript "$\lambda_k$" should be
replaced by the subscript "$k$", since $\lambda \equiv \lambda_k
=-2-k, \ k=0,1,2,3,...$. Thus, $d^{INP}(q^2, \Delta^2)$ describes
the true (physical) NP vacuum of QCD, while $d^{INP}_k(q^2,
\Delta^2)$ describe auxiliary ones, and the former is an infinite
sum of the latter ones. The expansion (5.1) is obviously the
Laurent expansion in the inverse powers of the gluon momentum
squared, which every term ends at the simplest NP IR singularity
$(q^2)^{-2}$. The only physical quantity (apart from the mass gap,
of course) which can appear in this expansion is the coupling
constant squared in the corresponding powers. In QCD it is
dimensionless and is evidently included into the $f_k$ functions.
However, let us note in advance that all the finite numerical
factors and constants (for example, the coupling constant) play no
independent role in the presence of a mass gap.

It is instructive to show explicitly expansions for a few first
different $d^{INP}_k(q^2, \Delta^2)$, namely

\begin{eqnarray}
d^{INP}_0(q^2, \Delta^2)&=& \Delta^2 f_0(0)(q^2)^{-2},
\nonumber\\
d^{INP}_1(q^2, \Delta^2)&=& (\Delta^2)^2 f_1(0)(q^2)^{-3} +
(\Delta^2)^2 f^{(1)}_1(0) (q^2)^{-2},  \nonumber\\
d^{INP}_2(q^2, \Delta^2)&=&(\Delta^2)^3 f_2(0)(q^2)^{-4} +
(\Delta^2)^3 f^{(1)}_2(0) (q^2)^{-3} + {1 \over 2} (\Delta^2)^3
f^{(2)}_2(0)(q^2)^{-2},
 \nonumber\\
d^{INP}_3(q^2, \Delta^2)&=& (\Delta^2)^4 f_3(0)(q^2)^{-5} +
(\Delta^2)^4 f^{(1)}_3(0)(q^2)^{-4} + {1 \over 2} (\Delta^2)^4
f^{(2)}_3(0)(q^2)^{-3} + {1 \over 6} (\Delta^2)^4
f^{(3)}_3(0)(q^2)^{-2}, \nonumber\\
\end{eqnarray}
and so on. Apparently, there is no way that such kind of an
infinite series could be summed up into the finite functions, for
example functions which could be regular at zero. That is why the
above-mentioned smooth gluon propagator is, in general, very
unlikely to exist (see remarks below as well). At the same time,
an infinite series (5.1), on account of the relations (5.2),
correctly reproduces the algebraic structure of the general
iteration solution (2.18), on account of the relations (2.19).
Each previous iteration is embodied into the subsequent one with
different coefficients. Let us remind now that because of Eq.
(4.7) all $f^{(m)}_k(0)$ have the dimensions of the inverse mass
squared in powers of $m$, i.e.,

\begin{equation}
[f^{(m)}_k(0)]= [\Delta^{-2}]^m  = [\Delta^2]^{-m},
\end{equation}
not losing generality. Let also note in advance that the simplest
NP IR singularity $(q^2)^{-2}$ is present in each iteration, which
emphasizes its special and important role (see below).

 Evidently, the expansion (5.1), on account of the relations (5.2), can be
equivalently written down as follows:

\begin{equation}
d^{INP}(q^2, \Delta^2) = \sum_{k=0}^{\infty} (q^2)^{-2 -k}
\sum_{m=0}^{\infty} {1 \over m!} (\Delta^2)^{k+m+1}
f^{(m)}_{k+m}(0) = \sum_{k=0}^{\infty} (q^2)^{-2
-k}(\Delta^2)^{k+1} \sum_{m=0}^{\infty} {1 \over m!}
\varphi_{k,m}(0),
\end{equation}
where we use the relation

\begin{equation}
f^{(m)}_{k+m}(0) = (\Delta^2)^{-m} \varphi_{k,m}(0),
\end{equation}
which obviously follows from the relation (5.3). Here
$\varphi_{k,m}(0)$ are dimensionless quantities, by definition.,
This expansion explicitly shows that the coefficient at each NP IR
singularity is an infinite series itself. It also shows that we
can analyze the IR properties of the INP part of the full gluon
form factor in terms of the mass gap $\Delta^2$ and the
dimensionless quantities $\varphi_{k,m}(0)$ only, which is very
convenient from a technical point of view (see below). This form
of the Laurent expansion shows also clearly the dynamical context
of the INP part of the full gluon propagator.

 As underlined above,
the piece of the NP part of the full gluon propagator, which does
not suffer from the NP IR singularities with respect to the gluon
momentum (the second sum in Eq. (4.6)), should be shifted to its
PT part. In QCD it comes from the expression (4.9) and looks like

\begin{equation}
d^{INP}_{(s)}(q^2, \Delta^2) = \sum_{k=0}^{\infty}
d^{INP}_{(s)k}(q^2, \Delta^2)= \sum_{k=0}^{\infty} (\Delta^2)^{1 +
k} (q^2)^{-2 -k} \sum_{m=k+1}^{\infty} {(q^2)^m \over m!}
f^{(m)}_k(0).
\end{equation}
Obviously, in the replacement

\begin{equation}
d^{PT}(q^2) \Longrightarrow d^{PT}(q^2) + \phi (q^2, \Delta^2) =
d'^{PT}(q^2),
\end{equation}
the function $\phi (q^2, \Delta^2)$ summed up over $k$ becomes

\begin{eqnarray}
\phi (q^2, \Delta^2) = \sum_{k=0}^{\infty} \phi_k (q^2, \Delta^2)
&=& \sum_{k=0}^{\infty} (\Delta^2)^{1 + k} (q^2)^{-1 -k}
\sum_{m=k+1}^{\infty} {(q^2)^m \over m!} f^{(m)}_k(0) \nonumber\\
&=& \sum_{k=0}^{\infty} (\Delta^2)^{1 + k} \sum_{m=0}^{\infty}
{(q^2)^m \over (m+k+1)!} f^{(m+k+1)}_k(0),
\end{eqnarray}
and it is free from the NP IR singularities with respect to the
gluon momentum.

 It has been already mentioned that an infinite
series (5.4) cannot be summed up into the some finite functions,
which can be regular at origin. Indeed, let us define the
coefficients $b_k$ as follows:

\begin{equation}
b_k = \sum_{m=0}^{\infty} {1 \over m!} \varphi_{k,m}(0), \quad
k=0,1,2,3... .
\end{equation}
It is also instructive to introduce new dimensionless variables as
$x_k = \Delta^2_k / q^2$, where $\Delta^2_k = c_k \Delta^2$ and
$b_k=c_k^k$. All these relations are relevant at $k=1,2,3...$. The
Laurent expansion (5.4) then become

\begin{equation}
d^{INP}(q^2, \Delta^2) = \Delta^2 (q^2)^{-2} \Bigl[ b_0 +
\sum_{k=1}^{\infty} (x_k)^k \Bigr].
\end{equation}
There is no any hope to find the sum of an infinite series over
$k$ even at all $x_k \ll 1$ (though this is not the case, since we
are interested in the IR region when $q^2$ is small). Moreover, in
order to get finite result this sum should cancel first an
arbitrary constant $b_0$ and than to cancel the simplest NP IR
singularity $(q^2)^{-2}$. Of course, this is unbelievable. In
other words, such kind of an infinite series are nonsummable in
mathematics due to the arbitrariness of $x_k$ (which, in fact, is
reduced to the arbitrariness of the coefficients $b_k$). The full
gluon propagator is therefore inevitably IR singular (no smooth in
the IR gluon propagator). The only hope to proceed further is to
get rid of this sum by carrying out an appropriate IR
renormalization program (see below).

In summary, starting from the preceding section and especially in
this section, we have established the deep IR structure of the
full gluon propagator in QCD. By construction, it is an infinite
sum over all the NP IR singularities, accompanied by the
corresponding powers of the mass gap. At the same time, its
structure in the PT regime remains arbitrary (but preserving AF).

\section{General IR renormalization program}

 The regularization of the NP IR singularities in QCD is
determined by the Laurent expansion (3.4) at $n=4$ as follows:

\begin{equation}
(q^2)^{- 2 - k } = { 1 \over \epsilon} a(k)[\delta^4(q)]^{(k)} +
f.t., \quad \epsilon \rightarrow 0^+,
\end{equation}
where $a(k)$ is a finite constant depending only on $k$ and
$[\delta^4(q)]^{(k)}$ represents the $k$th derivative of the
$\delta$-function (see Eqs. (3.2) and (3.3)). We point out that
after introducing this expansion everywhere one can fix the number
of dimensions, i.e., put $D=n=4$ for QCD without any further
problems. Indeed there will be no other severe IR singularities
with respect to $\epsilon$ as it goes to zero, but those
explicitly shown in this expansion. Let us underline that while
the initial expansion (5.4) is the Laurent expansion in the
inverse powers of the gluon momentum squared, the regularization
expansion (6.1) is the Laurent expansion in powers of $\epsilon$.
This means that its regular part is as follows: $f.t. =(q^2)^{- 2
- k }_{-} + \epsilon (q^2)^{- 2 - k }_{-} \ln q^2 +
O(\epsilon^2)$, where for the unimportant here definition of the
functional $(q^2)^{- 2 - k }_{-}$ see Ref. \cite{24}. Let us note
that in the Laurent expansion (5.4) there are no $\ln q^2$-type
terms, since they appear in the PT part. At the same time, when
the Laurent expansion (5.4) will be dimensionally regularized with
the help of the expansion (6.1) for each NP IR singularity, then
such kind of terms will appear. However, they will appear in the
next-to-leading terms, and therefore will be suppressed in the
$\epsilon \rightarrow 0^+$ limit. Thus, as it follows from the
Laurent expansion (6.1) that is dimensionally regularized, any
power-type NP IR singularity, including the simplest one, scales
as $1 /\epsilon$ as it goes to zero. Just this plays a crucial
role in the IR renormalization of the theory within our approach.

\subsection{IR renormalization of the gluon SD equation}

We are able now to consider the IR renormalization of the full
gluon SD equation. Fortunately, the gluon SD equation (2.4) does
not contain unknown scattering amplitudes, which usually are
determined by the infinite series of the multi-loop skeleton
diagrams. It is a closed system in the sense that there is a
dependence only on the pure gluon vertices, quark- and ghost-gluon
vertices and on the corresponding propagators. Its IR
renormalization can be carried out with the help of the
above-mentioned quantities only. For this purpose and on account
of Eq. (2.13), let us rewrite it in the following form, namely

\begin{eqnarray}
D(q) = D^0(q) &-& D^0(q)T_{gh}(q) D(q) - D^0(q)T_q(q)D(q)
\nonumber\\
&+& D^0(q) {1 \over 2} T_t(q)D(q) + D^0(q){1 \over 2} T_1(q)D(q) +
D^0(q){1 \over 2} T_2(q)D(q) + D^0(q){1 \over 6} T_2'(q)D(q),
\end{eqnarray}
where all quantities have been explicitly defined in the
expressions (2.7), (2.8) and in Eqs. (2.14)-(2.17). For
simplicity, here and below we neglect Dirac indices, since they
play no any role in tracking down of the IR singularities in the
corresponding equations and expressions. The next step is to
introduce the IR regularized quantities. In the presence of such
severe IR singularities (6.1), all the quantities should, in
principle, depend on $\epsilon$ as well, i.e., they become IR
regularized. So, one has to put formally

\begin{eqnarray}
g^2 &=& X(\epsilon) \bar g^2, \quad G(k) = \tilde{Z}_2(\epsilon)
\bar G(k), \quad S(p) = Z_2(\epsilon) \bar S(p), \nonumber\\
G_{\mu}(k, q) &=& \tilde{Z}_1(\epsilon) \bar G_{\mu}(k,q), \quad
\Gamma_{\mu}(p,q) = Z_1^{-1}(\epsilon) \bar \Gamma_{\mu}(p,q),
\nonumber\\
D(q) &=& Z_d(\epsilon) \bar D(q), \nonumber\\
T_3(q,q_1) &=& Z_3(\epsilon) \bar T_3(q,q_1), \nonumber\\
T_4(q,q_1,q_2) &=& Z_4(\epsilon) \bar T_4(q,q_1,q_2).
\end{eqnarray}
In all these relations the quantities with an overbar are, by
definition, IR renormalized, i.e., they are supposed to exist as
$\epsilon$ goes to zero. In both quantities, the IR regularized
and IR renormalized ones, the explicit dependence on $\epsilon$ is
omitted, for simplicity. In the corresponding IR maltiplicative
renormalization (IRMR) constants this dependence is not omitted in
order to distinguish them clearly from the corresponding UVMR
constants. Since we are interested in the IR renormalization of
the SD equation for the full gluon propagator, it is convenient
not to distinguish between the IR renormalization of its INP and
PT parts at this stage. Substituting these relations into the
gluon SD equation (6.2), one obtains

\begin{eqnarray}
\bar D(q) = D^0(q) &-& D^0(q)\bar T_{gh}(q) \bar D(q) - D^0(q)\bar
T_q(q)\bar D(q) + D^0(q) {1
\over 2} \bar T_t(q)\bar D(q) \nonumber\\
&+& D^0(q){1 \over 2} \bar T_1(q)\bar D(q) + D^0(q){1 \over 2}
\bar T_2(q) \bar D(q) + D^0(q){1 \over 6} \bar T_2'(q) \bar D(q).
\end{eqnarray}
The gluon SD equation (6.4) will exactly reproduce the structure
of Eq. (6.2) iff the following IR convergence conditions hold:

\begin{eqnarray}
Z_d(\epsilon) &=&1, \quad X(\epsilon) Z_d^2(\epsilon)=1, \nonumber\\
X(\epsilon) Z_d^3(\epsilon) Z_3(\epsilon)&=&1, \quad X^2(\epsilon)
Z_d^5(\epsilon) Z_3^2(\epsilon)=1, \quad
X^2(\epsilon) Z_d^4(\epsilon) Z_4(\epsilon)=1, \nonumber\\
X(\epsilon) Z_2^2(\epsilon) Z_1^{-1}(\epsilon)Z_d(\epsilon)&=&1,
\quad X(\epsilon) \tilde{Z}_2^2(\epsilon)
\tilde{Z}_1(\epsilon)Z_d(\epsilon)=1.
\end{eqnarray}
In connection with this system a few remarks are in order. Here
and everywhere, one can show that all the finite but arbitrary and
different constants, which only ones can appear in the
right-hand-side of these relations, can be put to unity not losing
generality. Moreover, this is general feature of our approach. In
all the IR convergence conditions, all the finite, but arbitrary
numbers can be put to unity, by simply redefining the
corresponding IRMR constants as well as the corresponding IR
renormalized quantities. In what follows this always will be
assumed. These IR convergence conditions should be fulfilled
simultaneously and independently, of course, in order to maintain
the algebraic and tensor structure of the gluon SD equation
itself. This makes it possible not to lose even one bit of the
information on the QCD vacuum, the dynamical and topological
structures of which are supposed to be reflected by the solutions
of this equation. Let us also note in advance that two last IR
convergence conditions will be known as quark and ghost
self-energy IR convergence conditions, respectively (part II).

 Evidently, the solutions of these relations are

\begin{equation}
Z_d(\epsilon) = X(\epsilon) = 1, \quad  Z_3(\epsilon) =
Z_4(\epsilon)=1, \quad Z_2^2(\epsilon) Z_1^{-1}(\epsilon)=1, \quad
\tilde{Z}_2^2(\epsilon) \tilde{Z}_1(\epsilon)=1.
\end{equation}
Thus, the IRMR constants of quark and ghost degrees of freedom
remain undetermined at this stage. They will be determined via the
corresponding Slavnov-Taylor (ST) identities
\cite{1,2,3,4,6,7,8,9,12}, which relate them to each other. It is
worth mentioning in advance that they also are IR finite from the
very beginning, i.e., $Z_2(\epsilon) = Z_1^{-1}(\epsilon)=
\tilde{Z}_2(\epsilon) =\tilde{Z}_1(\epsilon)=1$ (see part II).
However, the most important observation here is that
$X(\epsilon)=1$, which means that the QCD coupling constant is IR
finite from the very beginning as well, i.e., $g^2 = \bar g^2$.
That is $Z_d(\epsilon)=1$ is in complete agreement with our
general result that by the IR renormalization of a mass gap only,
we are able to remove all severe IR singularities from the INP
part of the full gluon propagator (see the next Subsec.).
Moreover, from the fact that the fill gluon propagator is not IR
renormalized ($Z_d(\epsilon)=1$), it is easy to understand that
the gauge fixing parameter is also IR finite from the very
beginning, i.e., $\xi = \bar \xi$. Let us underline also that from
$Z_d(\epsilon) = X(\epsilon)=1$ follows that $Z_3(\epsilon) =
Z_4(\epsilon)=1$, and not vice versa. Evidently, one can start
from any place in the SD system of equations, for example to start
from the quark and ghost sectors, ST identities, etc. However,
finally the system of the corresponding IR convergence conditions
will have the same solutions (6.6), of course.

\subsection{IR renormalization of a  mass gap}

In the preceding Subsec. it has been proven that the QCD coupling
constant and the gauge fixing parameter are not IR renormalized.
They are IR finite from the very beginning, i.e., $g^2 = \bar g^2$
and $\xi = \bar \xi$. In this case it is convenient to rewrite the
Laurent expansion (5.4) for the INP part of the full gluon
propagator as follows:

\begin{equation}
d^{INP}(q^2, \Delta^2) = \sum_{k=0}^{\infty} (q^2)^{-2
-k}(\Delta^2)^{k+1} \sum_{m=0}^{\infty} a_{k,m}(\xi) g^{2m},
\end{equation}
i. e., to show the dependence on the coupling constant squared and
the gauge fixing parameter explicitly. The rest in the arbitrary
numbers $a_{k,m}(\xi)$ is simply the product of the numerical
factors like $\pi$'s in different powers, eigenvalues of the color
group generators (we are not considering the numbers of different
colors and flavors as free parameters of the theory), etc. Let us
also remind that these numbers contain ghost and quark degrees of
freedom in all orders of linear PT, which, nevertheless, have been
integrated out numerically, so some of these numbers are UV
divergent.

Similar to the relations (6.3), let us now introduce the IR
renormalized mass gap

\begin{equation}
\Delta^2 = X_{\Delta}(\epsilon) \bar \Delta^2,
\end{equation}
where $X_{\Delta}(\epsilon)$ is the corresponding IRMR constant.
We already know that all the NP IR singularities, which can appear
in the full gluon propagator scale as $1/ \epsilon$ with respect
to $\epsilon$ (see Eq. (6.1)). Introducing now the so-called IR
convergence conditions as follows:

\begin{equation}
X_{\Delta}^{k+1}(\epsilon) = \epsilon \bar A_k(\epsilon), \quad
k=0,1,2,3,..., \quad \epsilon \rightarrow 0^+,
\end{equation}
the cancellation of the NP IR singularities with respect to
$\epsilon$ will be guaranteed term by term (since the NP IR
singularities are completely independent distributions) in the
Laurent expansion (6.7). Here $\bar A_k(\epsilon)$ are the IR
renormalized quantities, which, by definition, exist (are not
singular) as $\epsilon$ goes to zero. In terms of the IR
renormalized quantities the INP part of the full gluon propagator
(6.7) then becomes

\begin{equation}
d^{INP}(q^2, \bar \Delta^2) = \epsilon \sum_{k=0}^{\infty} (\bar
\Delta^2)^{1 + k} (q^2)^{-2 -k} \bar B_k(\epsilon),
\end{equation}
where

\begin{equation}
\bar B_k(\epsilon) = \bar A_k(\epsilon)a_k = \bar A_k(\epsilon)
\sum_{m=0}^{\infty} a_{k,m}(\xi) g^{2m},
\end{equation}
and it is exists as $\epsilon$ goes to zero (at any
$k=0,1,2,3,...$).

 Due to the distribution nature of the NP IR singularities,
in principle, two different cases should be considered.

I. There is an explicit integration over the gluon momentum, then
the gluon form factor (6.10) becomes

\begin{equation}
d^{INP}(q^2, \bar \Delta^2) = \sum_{k=0}^{\infty} (\bar
\Delta^2)^{1 + k} a(k) [\delta^4(q)]^{(k)} \bar B_k(\epsilon),
\end{equation}
provided the INP part of the full gluon propagator to be IR finite
from the very beginning, i.e., its IRMR constant will not depend
on $\epsilon$ at all as it goes to zero. It is easy to understand
that in this case we have to replace the NP IR singularity
$(q^2)^{-2-k}$ by its Laurent expansion (6.1), which shows that
$(q^2)^{-2-k}$ always scales as $1 /\epsilon$. The the so-called
$f.t.$ shown there become terms of the order $\epsilon$ as
$\epsilon \rightarrow 0^+$, so they vanish in this limit.

II. There is no explicit integration over the gluon momentum, then
the INP part of the full gluon form factor (6.10) vanishes, i.e.,

\begin{equation}
d^{INP}(q^2, \bar \Delta^2) \sim \epsilon, \quad \epsilon
\rightarrow 0^+.
\end{equation}
In this case the NP IR singularities $(q^2)^{-2-k}$ cannot be
treated as distributions, and therefore the Laurent expansion
(6.1) is not to be used, i.e., the above-mentioned functions
$(q^2)^{-2-k}$ are to be considered as standard mathematical
functions. This behavior can be treated as gluon confinement
criterion (see below). It simply means that there are no
transverse gluons in the IR, i.e., at large distances one cannot
detect gluons as free particles.

Let us emphasize now that the IR convergence conditions (6.9)
should be valid at any $k$, in particular at $k=0$, then from Eq.
(6.9) it follows

\begin{equation}
X_{\Delta}(\epsilon) = \epsilon \bar A_0, \quad \epsilon
\rightarrow 0^+,
\end{equation}
where we put $\bar A_0 \equiv \bar A_0(0)$, i .e., its value in
the $\epsilon \rightarrow 0^+$ limit. Thus the mass gap is IR
renormalized as follows:

\begin{equation}
\Delta^2 = \epsilon \bar \Delta^2, \quad \epsilon \rightarrow 0^+,
\end{equation}
where we include an arbitrary but finite constant $\bar A_0$ into
the IR renormalized mass gap $\bar \Delta^2$, and retaining, for
simplicity, the same notation. This means that in what follows we
can put it to unity, not losing generality, i.e., $\bar A_0 =1$.
From the IR convergence conditions (6.9) it follows that $\bar A_k
(\epsilon) \sim \epsilon^k$, which through the relation (6.11)
yields $\bar B_k (\epsilon) \sim \epsilon^k$ as well. In other
words, the solution obtained at $k=0$ should be used in the IR
convergence condition at $k=1$ and so on.

It is instructive to rewrite the Laurent expansion (6.10), which
is already IR renormalized, as follows:

\begin{equation}
d^{INP}(q^2, \bar \Delta^2) = \epsilon \bar \Delta^2 (q^2)^{-2}
\sum_{m=0}^{\infty} a_{0,m}(\xi) g^{2m} + \epsilon \bar \Delta^2
(q^2)^{-2} \sum_{k=1}^{\infty}(\bar \Delta^2 / q^2)^k \bar
B_k(\epsilon).
\end{equation}
Since we already know that the quantities $\bar B_k(\epsilon)$
scale as $\epsilon^k$, then the second sum in this decomposition
is additionally suppressed in the $\epsilon \rightarrow 0^+$ limit
(it scales as $\epsilon^2$ as $\epsilon \rightarrow 0^+$, at
least). We are thus left with the first term in this expansion,
namely

\begin{equation}
d^{INP}(q^2, \bar \Delta^2) = \epsilon \bar \Delta^2 (q^2)^{-2}
\sum_{m=0}^{\infty} a_m(\xi) g^{2m},
\end{equation}
where $a_m (\xi) \equiv a_{0,m}(\xi)$. In other words, in the
Laurent expansion (6.16) only the term which contain the simplest
NP IR singularity with respect to the gluon momentum $(q^2)^{-2}$
will survive as $\epsilon \rightarrow 0^+$.

 In summary, by the IR
renormalization of the mass gap $\Delta^2$ only, we can remove all
the NP IR singularities, parameterized in terms of the IR
regularization parameter $\epsilon$, from the INP part of the
gluon propagator. In its turn, this makes it possible to fix the
functional dependence of the INP part of the full gluon
propagator.

\section{ZMME quantum model of the true QCD ground state}

The true QCD ground state is a very complicated confining medium,
containing many types of gluon field configurations, components,
ingredients and objects of different nature
\cite{1,42,43,44,45,46}. Its dynamical and topological complexity
means that its structure can be organized at both the quantum and
classical levels. It is definitely "contaminated" by such gluon
field excitations and fluctuations, which are of PT origin, nature
and magnitude. Moreover, it may contain such extra gluon field
configurations, which cannot be detected as possible solutions to
the QCD dynamical equations of motion, either quantum or
classical, for example vortex-type ones \cite{47}. The only well
known classical component of the QCD ground state is the
topologically nontrivial instanton-antiinstanton type of
fluctuations of gluon fields, which are solutions to the Euclidean
YM classical equations of motion in the weak coupling regime
\cite{48,49}. However, they are by no means dominant but,
nevertheless, playing a special role in the QCD vacuum. In our
opinion their main task is to prevent quarks and gluons to freely
propagate in the QCD vacuum. It seems to us that this role does
not contradict their standard interpretation as an tunneling
trajectories linking vacua with different topology \cite{1, 49}
(and references therein). A highly nontrivial dynamical and
topological structure of the QCD vacuum emerges within our
approach (for more detail qualitative discussion see Ref.
\cite{12}).

Today there is no doubt left that dynamical mechanisms of the
important NP quantum phenomena such as color confinement and SBCS
are closely related to the above-mentioned complicated and
topologically nontrivial structure of the QCD vacuum. On the other
hand, it also becomes clear from what was discussed above that the
NP IR singularities play an important role in the large distances
behavior of QCD. For that very reason, any correct NP model of
color confinement and SBCS necessary turns out to be a realistic
model of the true QCD vacuum and the other way around. Evidently,
our approach to it is based on the importance of such gluon fields
which are solutions to the QCD quantum equations of motion.

Our quantum, dynamical model of the true QCD ground state is based
on the existence and the importance of such kind of the NP
excitations and fluctuations of gluon fields which are precisely
due to the self-interaction of massless gluons only without
explicitly involving some extra degrees of freedom. They are to be
summarized (accumulated) into the purely transverse part of the
full gluon propagator, and are to be effectively correctly
described by its severely singular structure in the deep IR
domain. We will call them the purely transverse singular gluon
fields, for simplicity. At the microscopic, dynamical level the
self-interaction of massless gluons is a twofold: the three- and
four-gluon vertices. We have established that just the latter ones
are behind the purely transverse singular gluon fields.

 At this stage, it is difficult to identify actually which type of gauge
field configurations can be finally formed by the purely
transverse singular gluon fields in the QCD ground state, i.e., to
identify relevant field configurations: chromomagnetic, self-dual,
stochastic, etc. However, if these gauge field configurations can
be absorbed into the gluon propagator (i.e., if they can be
considered as solutions to the corresponding SD equation), then
its severe IR singular behavior is a common feature for all of
them. Being thus a general phenomenon, the existence and the
importance of quantum excitations and fluctuations of severely IR
degrees of freedom inevitably lead to the general ZMME effect in
the QCD ground state. That is why we call our model of the QCD
ground state as the ZMME quantum model, or simply zero modes
enhancement (ZME, since we work always in the momentum space). For
preliminary investigation of this model see our papers
\cite{50,51} and references therein.

Our approach to the QCD true ground state, based on the general
ZMME phenomenon there, can be analytically formulated in terms of
the exact decomposition of the full gluon propagator as follows:

\begin{equation}
D_{\mu\nu}(q) = D^{INP}_{\mu\nu}(q, \Delta^2) +
D^{PT}_{\mu\nu}(q),
\end{equation}
where the INP part of the full gluon propagator effectively
becomes

\begin{equation}
D^{INP}_{\mu\nu}(q, \Delta^2) = i T_{\mu\nu}(q) d^{INP}(q^2,
\Delta^2) = i T_{\mu\nu}(q) \times \Delta^2 (q^2)^{-2}
\sum_{m=0}^{\infty} a_m(\xi) g^{2m},
\end{equation}
since only the first term in the expansion (6.16) will finally
survive in the $\epsilon \rightarrow 0^+$ limit, as explained
above. The PT part of the full gluon propagator
$D^{PT}_{\mu\nu}(q)$ in any case remains undetermined. Anyway, it
is not important in our approach (see below). Let us only remind
that it depends on a new gluon PT form factor $d'^{PT}(q^2)$ (see
Eq. (5.7)). At the level of a single full gluon propagator, its PT
part is defined as to be of the arbitrary covariant gauge and it
is as much singular as $1 /q^2$ in the IR.

\subsection{Confinement criterion for gluons}

Before going to the direct solution of the gluon SD equation
within our approach, it is worth discussing the properties of the
obtained solution for the full gluon propagator in more detail. We
already know that the full gluon propagator is not IR
renormalized, i. e., it is IR finite from the very beginning $D(q)
= \bar D(q)$ ($Z_d(\epsilon)=1$). As mentioned above, however, the
two different cases should be considered due to the distribution
nature of the simplest NP IR singularity$(q^2)^{-2}$, which
saturates the INP part of the full gluon propagator.

I. If there is an explicit integration over the gluon momentum
(the so-called virtual gluon due to Mandelstam \cite{10}), then
from Eq. (7.2) it follows

\begin{equation}
D^{INP}_{\mu\nu}(q, \bar \Delta^2)= iT_{\mu\nu}(q) \bar \Delta^2
\pi^2 \delta^4(q),
\end{equation}
i.e., in this case we have to replace the NP IR singularity
$(q^2)^{-2}$ in Eq. (7.2) by its $\delta$-type regularization
(6.1) at $k=0$. Also we should always take into account the
relation (6.15) for the IR renormalization of the mass gap in
order to express everything in terms of the IR renormalized
quantities. Thus, in this case the auxiliary IRMR constant for the
INP part of the full gluon propagator, defined as $D^{INP}(q) =
Z^{INP}_d(\epsilon) \bar D^{INP}(q)$, is equal to unity, i.e.,

\begin{equation}
Z_d^{INP}(\epsilon) = \epsilon / \epsilon = 1, \quad   \epsilon
\rightarrow 0^+.
\end{equation}
The $\delta$-type regularization is valid even for the multi-loop
skeleton diagrams, where the number of independent loops is equal
to the number of the gluon propagators. In the multi-loop skeleton
diagrams, where these numbers do not coincide (for example, in the
diagrams containing three or four-gluon proper vertices), the
general regularization (6.1) should be used (i.e., the derivatives
of the $\delta$-functions), and not the product of the
$\delta$-functions at the same point, which has no mathematical
meaning in the DT sense \cite{24} (for concrete examples see
appendix A). At the same time, the auxiliary IRMR constant for the
INP part (7.4) remains the same, of course. The so-called "f.t."
terms in the Laurent expansion (6.1) become terms of the order of
$\epsilon$, at least, so they vanish in the $\epsilon \rightarrow
0^+$ limit. In Eq. (7.3) an infinite series over the coupling
constant squared is included to the IR renormalized mass gap $\bar
\Delta^2$ with retaining the same notation, for simplicity, making
it thus UV renormalized as well (let us remind that some
coefficients in this expansion are UV divergent). This also makes
it possible to eliminate the explicit dependence on the gauge
fixing parameter in the gluon form factor for the INP part of the
full gluon propagator.

II. If there is no explicit integration over the gluon momentum
(the so-called actual gluon \cite{10}), then the function
$(q^2)^{-2}$ in Eq. (7.2) cannot be treated as the distribution.
The INP part of the full gluon propagator in this case disappears
as $\epsilon $ as $\epsilon \rightarrow 0^+$, namely

\begin{equation}
D^{INP}_{\mu\nu}(q, \bar \Delta^2) \sim \epsilon, \quad \epsilon
\rightarrow 0^+,
\end{equation}
i.e., the IR renormalization of the mass gap (6.15) only comes out
into the play. This means that any amplitude (more precisely its
INP part, see the next Subsec. B) for any number of soft-gluon
emissions (no integration over their momenta) will vanish in the
IR limit in our picture. In other words, there are no transverse
gluons in the IR, i.e., at large distances (small momenta) there
is no possibility to observe gluons experimentally as free
particles. Thus, the color gluons can never be isolated. This
behavior can be treated as the gluon confinement criterion (see
also Ref. \cite{9}), and it supports the consistency of the exact
solution (7.2) for the INP part of the full gluon propagator.
Evidently, this behavior does not explicitly depend on the gauge
choice in the full gluon propagator, i.e., it is a manifestly
gauge-invariant as it should be, in principle. These two
observations greatly simplify the analysis of the gluon SD
equation within our approach (see below). Thus, in this case the
auxiliary IRMR constant for the INP part of the full gluon
propagator vanishes, i.e.,

\begin{equation}
Z_d^{INP}(\epsilon) = \epsilon, \quad   \epsilon \rightarrow 0^+,
\end{equation}
so that the full gluon propagator is reduced to the PT one in the
$\epsilon \rightarrow 0^+$ limit.

It makes sense to postpone the comparison of our manifestly
gauge-invariant criterion for gluon confinement (7.5) with the
color confinement criterion due to Kugo and Ojima (KO) \cite{52},
which is based on ghost degrees of freedom, until part III of our
investigation. However, the surprising agreement with the color
confinement criterion due to Nishijima and Oehme (NO) \cite{53}
(see also Ref. \cite{54}) is to be briefly discussed here. The NO
color confinement criterion is formulated as follows:

\begin{equation}
Z_3^{-1} = 0,
\end{equation}
where $Z_3^{-1}$ is, in fact, the renormalization constant of the
transverse part of the full gluon propagator. Within our notations
it should be identified with the IR renormalization constant of
the purely transverse part (i.e., the INP part) of the full gluon
propagator. From Eq. (7.6) in the final $\epsilon \rightarrow 0^+$
limit, one then obtains

\begin{equation}
Z_d^{INP}(0) \equiv Z_d^{INP} =0,
\end{equation}
in complete agreement with the NO criterion (7.7). Behind our
criterion (7.5) is clear dynamical mechanism, while the NO
criterion (7.7) is rather kinematical (metric confinement) than
dynamical. However, in both cases the transverse (physical) gluons
are removed from the spectrum in a quite similar way. The
coincidence between the IRMR and UVMR constants of the full gluon
propagator once more may signify deep properties of the full
theory \cite{30}. We think that this coincidence is neither
accidental nor formal, and it deserves to be investigated in more
detail elsewhere.

\subsection{INP phase in QCD}

For the sake of self-consistency and transparency of our approach
to low-energy QCD, it is convenient to discuss in more detail what
we mean by the INP phase in QCD. In the decomposition (7.1)
$D^{PT}_{\mu\nu}(q)$ is given by Eq. (4.4), on account of the
replacement for the PT gluon form factor in Eq. (5.7). The INP
part of the full gluon propagator is, in general, given in Eqs.
(7.3) and (7.5) for the above-discussed two different cases. Let
us remind that within our approach all severe IR singularities of
the dynamical origin possible in QCD are to be incorporated into
the full gluon propagator and are to be effectively correctly
described by its INP part. So, all other QCD proper vertices can
be considered as regular functions with respect to all the gluon
momenta involved. In QCD there is some kind of a correspondence
between the pure gluon proper vertices in the deep IR region
(i.e., when all the gluon momenta involved go to zero) and their
point-like counterparts, namely $T_4(0,0,0) \rightarrow T^0_4 \neq
0$, while $T_3(0,0) \rightarrow T^0_3(0,0) = 0$. If, nevertheless,
they might be singular, then this correspondence is violated and
it would require a completely different investigation, anyway.

 In principle, all other fundamental quantities in QCD could be
formally decomposed similar to the decomposition (7.1) for the
full gluon propagator. Evidently, this should be done for
quantities, which explicitly depend on the gluon momenta, i.e.,
proper vertices. The decomposition does not make any sense for the
coupling constant, quark and ghost propagators, since they do not
depend on the gluon momentum. This is also true for the quark- and
ghost-gluon proper vertices, since they depend on the quark and
ghost momenta, which are completely independent from the gluon
momentum, playing the role of the momentum transfer in these
vertices. The only proper vertices which makes sense to decompose
are the pure gluon vertices, since they crucially depend on all
the gluon momenta involved. However, we define the INP phase in
QCD in more general terms, which includes the decomposition of the
full gluon propagator only as follows:

(i) It is always transverse, i.e., it depends only on physical
degrees of freedom of gauge bosons.

(ii) Before the IR renormalization, the presence of the NP IR
singularities $(q^2)^{-2-k}, \ k=0,1,2,3,...$ is only possible.

(iii). After the IR renormalization, the INP part of the full
gluon propagator is fully saturated by the simplest NP IR
singularity, and all other NP IR singularities will be
additionally suppressed in the $\epsilon \rightarrow 0^+$ limit.

(iv). There is an inevitable dependence on the mass gap
$\Delta^2$, so that when it formally goes to zero, then the INP
phase vanishes, while the PT phase survives.

Evidently, this definition implies that the INP part of any
multi-loop skeleton diagram in QCD should contain only the INP
parts of all the corresponding gluon propagators. At the same
time, the PT part of any multi-loop skeleton diagram always
remains of arbitrary gauge. It may even contain the terms, where
the NP IR singularities are present along with the PT IR ones as
well (the so-called general PT term, see discussion below).
 The difference between the NP IR singularities $(q^2)^{-2-k}$
and the PT IR singularity $(q^2)^{-1}$ is that the latter is not
defined by its own Laurent expansion (6.1) that is dimensionally
regularized like the former ones. That is why it does not require
the IR renormalization program itself.

\subsection{A few comments}

We already know that when there is no explicit integration over
the gluon momentum, then the full gluon propagator is reduced to
the PT one. In other words, in this case $D=Z_d^{PT}(\epsilon)
\bar D^{PT}= D^{PT}$, by definition, where $Z_d^{PT}(\epsilon)$ is
the auxiliary IRMR constant of the PT part of the full gluon
propagator. However, from the general solution $Z_d=1$ ($D= \bar
D$) it follows that $Z_d^{PT} = Z_d =1$ and hence $D^{PT} = \bar
D^{PT}$ (all the IRMR constants, which do not depend on
$\epsilon$, become an arbitrary, but finite numerical constants,
and they can be put to unity not loosing generality, as mentioned
above). Let us remind that the gauge fixing parameter is not IR
renormalized as well. In principle, this is not surprising, since
the PT part of the full gluon propagator is automatically free
from the NP IR singularities, by definition. At the same time, the
IR finiteness of the full gluon propagator has been achieved
(constructed) by the nontrivial IR renormalization procedure. It
includes the IR renormalization of the required mass gap by taking
into account the distribution nature of the NP IR singularities
(see above).

From QCD sum rules it is well known that AF is stopped by
power-type terms reflecting the growth of the coupling in the IR.
Approaching the deep IR region from above, the IR sensitive
contributions were parameterized in terms of a few quantities
(gluon and quark condensates, etc.), while direct access to NP
effects (i.e., to the deep IR region) was blocked by the IR
divergences \cite{55,56}. Our approach to NP QCD, in particular,
to its true ground state within the just formulated ZMME model is
a further step into the deep IR region (in fact, we are deeply
inside it), since DT allows one to correctly deal with the NP IR
singularities.

Evidently, the ZMME mechanism for quark confinement is nothing but
the well forgotten IRS one, which can be equivalently referred to
as a strong coupling regime \cite{1,32}. Indeed, at the very
beginning of QCD it was expressed a general idea
\cite{32,33,34,35,36} that the quantum excitations of the IR
degrees of freedom, because of the self-interaction of massless
gluons in the QCD vacuum, made it only possible to understand
confinement, DCSB and other NP effects. In other words, the
importance of the deep IR structure of the true QCD vacuum has
been emphasized as well as its relevance to quark confinement,
DCSB, etc., and the other way around. This development was stopped
by the wide-spread wrong opinion that severe IR singularities
cannot be put under control. We have explicitly shown that the
correct mathematical theory of quantum YM physical theory is the
theory of distributions (the theory of generalized functions)
\cite{24,57}, complemented by the DR method \cite{25}. They
provide a correct treatment of these severe IR singularities
without any problems. Thus we come back to the old idea but on a
new basis that is why it becomes new ("new is well forgotten
old"). In other words, we put the IRS mechanism of quark
confinement on a firm mathematical ground provided by DT.
Moreover, we also emphasize the role of the purely transverse
gauge fields in this mechanism. After the authors of Ref.
\cite{35} we can repeat that what we want eventually is not AF but
transverse IRS/ZMME.

 In the light of the above-mentioned correspondence between the pure
gluon proper vertices in the deep IR region and their point-like
counterparts, it becomes almost clear that in order to control
firmly the deep IR region within our approach, we need precisely
the point-like pure gluon vertices rather than their proper
counterparts. Since the INP part of the full gluon propagator is
finally saturated by the simplest NP IR singularity $(q^2)^{-2}$
only, this implies all the pure gluon proper vertices to become
effectively point-like ones. Otherwise, each simplest NP IR
singularity multiplied by the corresponding gluon momentum coming
from the proper vertex effectively becomes less singular, and
therefore is to be shifted to the corresponding PT parts in
accordance with our method. This is just happening in the case of
the three-gluon vertex (it depends, at least, linearly on all the
gluon momenta involved). At the same time, the four-gluon vertex
remains the source of the NP IR singularities in the gluon
propagator. The quark- and ghost-gluon proper vertices at zero
momentum transfer (i.e., at zero gluon momentum) are to be
accounted for as well. However, as mentioned above, we will always
operate with the full vertices, since summing up an infinite
number of the diagrams, for example with point-like pure gluon
vertices, we again come to their proper counterparts. The
decomposition of the full gluon propagator under DT, complemented
by the DR method, is only necessary and completely enough to
firmly control the IR region in QCD within our approach.

Let us make one thing perfectly clear. The important observation
was that due to the distribution nature of the NP IR
singularities, two different types of the IR renormalization of
the INP part of the full gluon propagator have been required. The
principal distinction between them is whether there is an explicit
integration over the gluon momentum or not, preserving,
nevertheless, the IR finiteness of the full gluon propagator
($Z_d(\epsilon) =1$). The pure gluon vertices crucially depend on
all the gluon momenta involved. So, it will be necessary to
distinguish between their IR renormalization whether there is an
explicit integration over all the gluon propagators or there is an
explicit integration not over all ones (associated in both cases
with each pure gluon vertex). The previous result $Z_3(\epsilon)=
Z_4(\epsilon)= 1$ should be preserved as well (see part II). Also
let us note that calculating, for example the one-gluon exchange
potential between heavy quarks, $d^{INP}(q^2, \Delta^2) \sim
\Delta^2 (q^2)^{-2}$ should be used, by including an infinite sum
into the mass gap. Finally this combination becomes a string
tension.

Working always in the momentum space, we are speaking about the
purely transverse singular gluon fields responsible for color
confinement in our approach. Discussing the relevant field
configurations, we always will mean the functional (configuration)
space. Speaking about relevant field configurations
(chromomagnetic, self-dual, stochastic, etc), we mean, of course,
the low-frequency modes of all of these virtual transverse fields.
Only large scale amplitudes of these fields ("large transverse
gluon fields") are to be taken into account by the INP part of the
full gluon propagators. All other frequencies are to be taken into
account by corresponding PT part of the gluon propagators. To
speak about specific field configurations that are solely
responsible for color confinement is not the case, indeed. The
low-frequency components/large scale amplitudes of all the
possible in the QCD vacuum the purely transverse virtual fields
are important for the dynamical and topological formation of such
gluon field configurations which are responsible for color
confinement and other NP effects within our approach to low-energy
QCD. For convenience, we will call them the purely transverse
severely singular gluon field configurations.

In summary, we have formulated the ZMME model of the true QCD
ground state in terms of the gluon propagator. On one hand, this
makes it possible to establish its structure in the IR region. On
the other hand, it allows one to formulate the gluon confinement
criterion in Eq. (7.5) in a manifestly gauge-invariant way. In the
same way, we have defined the INP phase in QCD in terms of the
corresponding decomposition of the full gluon propagator only. It
is worth emphasizing, that just the INP phase will be responsible
for the NP effects in QCD, such as quark confinement, DBCS, etc.,
within our approach to low-energy QCD.

\section{Solution of the gluon SD equation within our approach}

As already said, the gluon SD equation (2.5) does not contain
unknown scattering amplitudes, so it can be directly solved within
our approach. By direct solution, we mean that the INP part of the
full gluon propagator, as it is determined in Eqs. (7.3) and
(7.5), identically satisfies the corresponding INP part of the
gluon SD equation. Substituting the decomposition (7.1) into the
gluon SD equation (2.5), one obtains

\begin{equation}
D^{INP}(q) + D^{PT}(q) = \tilde{D}^0(q) + \tilde{D}^0(q) T_g[
D](q) D^{INP}(q) + \tilde{D}^0(q) T_g[D](q) D^{PT}(q).
\end{equation}
In this equation and in what follows, we omit the explicit
dependence on the Dirac indices, playing no any role in tracking
down of the IR singularities in the corresponding integrals. The
nonlinear part of the gluon self-energy $T_g[D](q)$ should be also
decomposed as follows:

\begin{equation}
T_g[D](q) = T_g[D^{INP} + D^{PT}](q)=T^{INP}_g[D^{INP}](q) +
T^{PT}_g[D](q),
\end{equation}
so that the INP part will depend on $D^{INP}$ only, while the PT
part remains arbitrary, i.e., it may depend on both $D^{INP}$ and
$D^{PT}$, as mentioned above. Separating now between the INP and
PT parts in the gluon SD equation (8.1), on account of the
decomposition (8.2), one obtains

\begin{equation}
D^{INP}(q) - \tilde{D}^0(q) T^{INP}_g[D^{INP}](q)D^{INP}(q) -
\tilde{D}^0(q) T^{PT}_g[D](q) D^{INP}(q) = -D^{PT}(q) +
\tilde{D}^0(q)+ \tilde{D}^0(q) T_g[D](q)D^{PT}(q).
\end{equation}
We already know that the INP part of the full gluon propagator
with momentum $q$ vanishes as $\epsilon$ (see Eq. (7.5)), since
there is no explicit integration over the gluon momentum $q$. So,
if the INP part of the nonlinear part of the gluon self-energy
$T^{INP}_g[D^{INP}](q)$ and the composition $T^{PT}_g[D](q)$
(i.e., $T_g[D](q)$ itself) do not produce any problems in the
$\epsilon \rightarrow 0^+$ limit, then the left-hand-side of Eq.
(8.3) identically vanishes in this limit. We will be left
therefore with the gluon SD equation for the PT part of the full
gluon propagator as follows:

\begin{equation}
D^{PT}(q) = \tilde{D}^0(q) + \tilde{D}^0(q)  T_g[D](q) D^{PT}(q) =
\tilde{D}^0(q) + \tilde{D}^0(q) \Bigl[ T^{INP}_g[D^{INP}](q) +
T^{PT}_g[D](q) \Bigr]D^{PT}(q).
\end{equation}
 The main purpose of this Sec. and appendix A is to show explicitly that
this is so, indeed.

The nonlinear gluon self-energy $\Sigma_g^{NL}(q)= T_g[D](q)$ is
the sum of the four terms due to Eq. (2.13). Let us begin with Eq.
(2.14), which describes the so-called tadpole term. Substituting
the exact decomposition (7.1) for the full gluon propagator, one
obtains

\begin{equation}
T_t = T^{INP}_t  + T^{PT}_t,
\end{equation}
where

\begin{equation}
T^{INP}_t = g^2 \int {i d^4 q_1 \over (2 \pi)^4} T^0_4
D^{INP}(q_1) = - \bar \Delta_t^2 T^0_4
\end{equation}
and the PT part is given in the appendix A. Here and everywhere
below, all the finite constants will be included into the
corresponding IR finite mass scale parameters (here into $\bar
\Delta_t^2$). For the INP part has been used the relation (7.5),
since there is an explicit integration over the gluon momentum
$q_1$. So, the tadpole term does not produce problems as $\epsilon
\rightarrow 0^+$.

On account of the decompositions (7.1), the integral (2.15)
becomes

\begin{equation}
T_1(q) = T^{INP}_1(q) + T^{PT}_1(q),
\end{equation}
where

\begin{equation}
T_1^{INP}(q) = g^2 \int {i d^4 q_1 \over (2 \pi)^4} T^0_3 (q,
-q_1, q_1-q) T_3 (-q, q_1, q -q_1) D^{INP}(q_1) D^{INP}(q -q_1),
\end{equation}
and the PT part is a sum of the three terms, which are shown in
the appendix A. Integrating over the gluon momentum $q_1$ in this
integral with the help of Eq. (7.3), one obtains

\begin{equation}
T_1^{INP}(q) = - \bar \Delta_1^2 T^0_3 (q, 0, -q) T_3 (-q, 0, q)
D^{INP}(q) \sim \epsilon, \quad \epsilon \rightarrow 0^+,
\end{equation}
since there is no explicit integration over the gluon momentum $q$
(see Eq. (7.5)).

Let us now evaluate first Eq. (2.17), since it contains three
gluon propagators in comparison with Eq. (2.16), by reminding that
in both equations $q_3=q-q_1 +q_2$. Substituting again the
decompositions (7.1), one obtains

\begin{equation}
T'_2(q) = T'^{INP}_2(q) + T'^{PT}_2(q),
\end{equation}
where

\begin{equation}
T'^{INP}_2(q) =  g^4 \int {i d^4 q_1 \over (2 \pi)^4} \int {i d^4
q_2 \over (2 \pi)^4} T^0_4  T_4(-q, q_1, -q_2, q-q_1+q_2)
D^{INP}(q_1) D^{INP}(-q_2) D^{INP}(q-q_1+q_2),
\end{equation}
and the PT part is a sum of the seven terms, which are shown in
the appendix A. Integrating over the gluon momenta $q_1$ and $q_2$
(let us note that, in general, $D(q) = D(-q)$) with the help of
Eq. (7.3), one obtains

\begin{equation}
T'^{INP}_2(q) =  \bar \Delta_2'^4 T^0_4 T_4 (-q,0,0,q) D^{INP}(q)
\sim \epsilon, \quad \epsilon \rightarrow 0^+,
\end{equation}
since in this equation there is no explicit integration over the
gluon momentum $q$ (see Eq. (7.5)).

Let us now evaluate Eq. (2.16). Again substituting the
decompositions (7.1), one obtains

\begin{equation}
T_2(q) = T^{INP}_2(q) + T^{PT}_2(q),
\end{equation}
where

\begin{eqnarray}
T^{INP}_2(q) =  g^4 \int {i d^4 q_1 \over (2 \pi)^4} \int {i d^4
q_2 \over (2 \pi)^4} &T^0_4  T_3 (-q_2, q-q_1+q_2,
q_1-q) T_3(-q, q_1, q-q_1)& \nonumber\\
&D^{INP}(q_1) D^{INP}(-q_2) D^{INP}(q-q_1+q_2) D^{INP}(q-q_1),&
\end{eqnarray}
and the PT part is a sum of the fifteen terms, which are shown in
the appendix A. Integrating over the gluon momenta $q_1$ and $q_2$
with the help of Eq. (7.3), yields

\begin{equation}
T^{INP}_2(q) = \bar \Delta_2^4 T^0_4 T_3 (0, q, -q) T_3(-q, 0, q)
D^{INP}(q) D^{INP}(q) \sim \epsilon^2, \quad \epsilon \rightarrow
0^+,
\end{equation}
since in this equation there is no explicit integration over the
gluon momentum $q$ (see Eq. (7.5)).

The INP part of the nonlinear pure gluon part is a sum of the four
terms, namely

\begin{equation}
T^{INP}_g[D^{INP}](q)  = {1 \over 2} T^{INP}_t + {1 \over 2}
T^{INP}_1(q) + {1 \over 2} T^{INP}_2(q) + {1 \over 6}
T'^{INP}_2(q),
\end{equation}
where each term is given by Eqs. (8.6), (8.9), (8.12) and (8.15).
Because of Eq. (8.6) it is finite in the $\epsilon \rightarrow
0^+$ limit,

\begin{equation}
T^{INP}_g[D^{INP}](q)  = - {1 \over 2} \bar \Delta_t^2 T^0_4
 + O(\epsilon), \quad \epsilon \rightarrow 0^+.
\end{equation}
So, the combination

\begin{equation}
\tilde{D}^0(q) T^{INP}_g[D^{INP}](q) D^{INP}(q) \sim \epsilon,
\quad \epsilon \rightarrow 0^+,
\end{equation}
indeed. At the same time, in the appendix A it is shown that the
composition $T^{PT}_g[D](q)$ does not produce any problems in the
$\epsilon \rightarrow 0^+$ limit. This means that the combination

\begin{equation}
\tilde{D}^0(q) T^{PT}_g[D](q) D^{INP}(q) \sim \epsilon, \quad
\epsilon \rightarrow 0^+,
\end{equation}
as well. Thus, we are left with Eq. (8.4) for the PT part of the
full gluon propagator only, indeed. This means that the INP part
of the full gluon propagator is completely decoupled from the rest
of the gluon SD equation, i.e., it identically and independently
satisfies the INP part of the gluon SD equation (the
left-hand-side of Eq. (8.3)). Eq. (8.4) for the PT part of the
full gluon propagator can be further simplified. On account of Eq.
(8.17), it becomes

\begin{equation}
D^{PT}(q) = \tilde{D}^0(q) + \tilde{D}^0(q) \Bigl[ - {1 \over 2}
\bar \Delta_t^2 T^0_4 + T^{PT}_g[D](q) \Bigr] D^{PT}(q).
\end{equation}
In the appendix A it is shown that it produces no explicit
problems in the $\epsilon \rightarrow 0^+$ limit. However, let us
underline once more that it is not our problem, since we are, in
principle, not responsible for the PT phase in QCD.

Concluding this Sec., let us clarify the terminology. The general
gluon SD equation (2.5) can be written down for the inverse of the
full gluon propagator as

\begin{equation}
D^{-1}(q) = [\tilde{D}^0(q)]^{-1} - T_g[D](q)=
[\tilde{D}^0(q)]^{-1} - \Sigma_g^{NL}(q),
\end{equation}
so we can call $T_g[D](q)= \Sigma_g^{NL}(q)$ as the NL part of the
gluon self-energy. If one begins with the free gluon propagator
$D^0(q)$, then the gluon SD equation looks more conventionally

\begin{equation}
D^{-1}(q) = [D^0(q)]^{-1} - \Sigma_g(q),
\end{equation}
where

\begin{equation}
\Sigma_g(q) =\Sigma_{gh}(q) + \Sigma_q(q) + \Sigma_g^{NL}(q),
\end{equation}
and $\Sigma_{gh}(q)$ and $\Sigma_q(q)$ are given by the integrals
(2.7) and (2.8), respectively.

\section{General Discussion}

Let us begin our discussion with reemphasizing a few important
points.

 The first point is that the only place where the most important problem of
theoretical particle/nuclear physics -- color confinement
(together with other NP effects) can be solved is the SD system of
dynamical equations of motion, since it contains the full
dynamical information (and even more than that) on QCD. To solve
this system means to solve QCD itself and vice versa.

 The second point is that AF clearly indicates the existence of the NP phase
(with its own characteristic scale parameter) in the full QCD.
Apparently, this reflects the fact that QCD as a whole is UV
stable, and thus IR unstable theory (i.e, it has no IR stable
fixed point, indeed) \cite{1,32}. In other words, QCD as a whole
is the PT UV and NP IR divergent theory, but, nevertheless, it is
IR renormalizable as well.

The third point is that the only place where the NP dynamics can
be introduced is the deep IR region, since the PT structure of QCD
is controlled by AF.

The fourth point is that the deep IR region is dominated
(saturated) by the self-interaction of massless gluons in the true
QCD vacuum (no Schwinger-Higgs phase transition in QCD, so the
$SU(3)$ color gauge symmetry is exact and gluons remain massless
\cite{1}). The above-mentioned NL interaction leads to the ZMME in
the gluon propagation, and not to its effective/dynamical mass.

The fifth point is that at the microscopic level we identify
unambiguously the fundamental four-gluon interaction as the main
dynamical source which naturally leads to the enhancement of the
zero momentum degrees of freedom in the gluon propagation
(self-energy). So, severe IR singularities are introduced not by
hand. The corresponding purely transverse severely singular gluon
field configurations are intrinsically peculiar to the true QCD
ground state.

The sixth point is that at level of the gluon SD equation the main
source of the above-mentioned IR instability of QCD is the
two-loop skeleton term, which contains the four-gluon vertices
only. Precisely its additional IR singularities when all the gluon
momenta involved go to zero gives birth to the ZMME effect in the
true QCD vacuum.

The seventh point is that any deviation in the behavior of the
full gluon propagator from the free one in the IR requires an
automatic introduction of a characteristic mass scale parameter
responsible for the nontrivial dynamics in the IR domain, the
so-called mass gap. It is worth emphasizing that it cannot be
interpreted as the effective/dynamical gluon mass, which always
remains massless in our approach.

 The eight point is that a mass
gap appears on dynamical ground. It gains contributions from all
powers of the coupling constant squared. So, its determination
goes beyond PT, i.e., its physical meaning is essentially NP.
However, the coupling constant squared itself plays no any role in
the presence of a mass gap.

There is no doubt left that the singular configurations of gluon
fields play a permanently dominant role in the large scale quantum
structure of the QCD vacuum. Whether they can be "seen" (detected)
by other methods or approaches is not so important, since we hope
that our general consideration, to which we were restricted here,
has been convincing enough. However, there already exist a lot of
direct and indirect evidences in favor of the $(q^2)^{-2}$
behavior of the full gluon propagator in the IR domain. Let us
mention only a few of them, which are the most important in our
opinion:

a). After the pioneering papers of Mandelstam in the covariant
(Landau) gauge \cite{10} and Baker, Ball and Zachariasen in the
axial gauge \cite{58}, the consistency of the $(q^2)^{-2}$ IR
singular asymptotics of the full gluon propagator in different
gauges with the direct solution of the gluon SD equation has been
repeatedly confirmed (see, for example Refs.
\cite{11,12,59,60,61,62} and references therein). {\bf It is worth
emphasizing, however, that Eq. (7.2) is an exact result, and thus
it expresses not only the deep IR asymptotics of the full gluon
propagator.}

b). The cluster property of the Wightman functions in QCD fails,
and this allows such a singular behavior for the full gluon
propagator in the deep IR domain \cite{63}.

c). Such singular behavior of the full gluon propagator in the IR
domain leads to the area law for heavy (static) quarks (indicative
of confinement) within the Wilson loop approach \cite{64}.

d). Moreover, let us underline that without the $(q^2)^{-2}$ IR
singular component in the decomposition of the full gluon
propagator in the continuous theory, it is impossible to "see"
linearly rising potential between heavy quarks by lattice QCD
simulations \cite{65}, not involving some extra (besides gluons
and quarks) degrees of freedom. Evidently, the above-mentioned
smooth gluon propagator cannot provide a linear rising potential
between heavy quarks "seen" by lattice simulations.

e). There exists also direct lattice evidence that the zero
momentum modes are enhanced in the full gluon propagator (and
hence in the effective coupling), indeed \cite{66} (and references
therein).

f). A NP finite-size scaling technique was used in Ref. \cite{67}
to study the evolution of the running coupling (which, in
principle, can be identified with the exact gluon form factor) in
the $SU(3)$ YM lattice theory. By using the two-loop
$\beta$-function it is shown to evolve according to PT at high
energies, while at low energies it is shown to grow. Though we do
not know the NP $\beta$-function yet, nevertheless, this growing
tendency has to be acknowledged (see also Ref. \cite{68}).

g). Within DT the $(q^2)^{-2}$ singularity is the simplest NP
power-type IR singularity in 4D QCD, while the $(q^2)^{-1}$
singularity is the simplest NP power-type IR singularity in 2D
QCD, which confines quarks \cite{20,21,22}. Though the QCD vacuum
is much more complicated medium than its 2D model, nevertheless,
the above-mentioned analogy is promising even in the case of the
NP dynamics of light quarks.

h). Some classical models of the QCD vacuum also invoke a
$(q^2)^{-2}$ behavior of the gluon fields in the IR domain. For
example, it appears in the QCD vacuum as a condensation of the
color-magnetic monopoles (QCD vacuum is a chromomagnetic
superconductor) proposed by Nambu, Mandelstam and 't Hooft and
developed by Nair and Rosenzweig (see Ref. \cite{69} and
references therein), as well as in the classical mechanism of the
confining medium \cite{70} and in the effective theory for the QCD
vacuum proposed in Ref. \cite{71}.

i). It is also required to derive the linearly rising potential
between heavy quarks within the recently proposed renormalization
group flow equations approach \cite{72}.

j). It has been shown that the IR singular behavior (7.2) directly
leads to quark confinement (in a flavor independent way) and SBCS
\cite{51,73} (and references therein, see also part III of our
approach).

\subsection{Subtractions}

Let us continue our discussion recalling that many important
quantities in QCD, such as gluon and quark condensates,
topological susceptibility, the Bag constant, etc., are defined
beyond PT only \cite{50,55,74}. This means that they are
determined by such $S$-matrix elements (correlation functions)
from which all types of the PT contributions should be, by
definition, subtracted. Within the 2D covariant gauge QCD we have
already described all types of the necessary subtractions to be
done in order to define correctly such a truly NP quantity as the
quark condensate \cite{21}. Let us emphasize that such type of
subtractions are inevitable also for the sake of self-consistency.
In low-energy QCD there exist relations between different
correlation functions, for example the famous Witten-Veneziano
(WV) and Gell-Mann-Oakes-Renner (GMOR) formulae. The former
\cite{75,76} relates the pion decay constant and the mass of the
$\eta'$ meson to the topological susceptibility. The latter
\cite{55,77} relates the chiral quark condensate to the pion decay
constant and its mass. Defining the topological susceptibility and
the quark condensate by the subtraction of all types of the PT
contributions, it would be not self-consistent to retain them in
the correlation functions determining the pion decay constant and
in the expressions for the pion and $\eta'$ meson masses.

Anyway, our theory for low-energy QCD, which we call INP QCD, will
be precisely defined by the subtraction of all types of the PT
contributions (the first subtraction has been already done in Eq.
(4.1)). At the fundamental quark-gluon (microscopic) level, the
first step is to identify the terms in the corresponding SD
equations and ST identities, which should survive after the
necessary subtractions. Evidently, from the above it follows that
those terms will remain only which depend on the INP parts of the
full gluon propagators. At the hadronic (macroscopic) level, the
second step is to integrate out quark and gluon degrees of freedom
("hadronization"). At this stage, the subtraction aimed at fixing
the point at which the PT parts of the corresponding integrals
should be subtracted. In part III we will show that in the quark
sector the subtraction point should coincide with the constant of
integration of the corresponding SD equation for the INP quark
propagator. For the YM sector in part V we will determine the
subtraction point by the minimization of the INP vacuum energy
density (the effective potential in the absence of the external
sources \cite{78,79,80}) for pure gluon fields.

Our approach to QCD can be symbolically decomposed into
 two parts (consisting of three terms), namely

\begin{eqnarray}
QCD &=& INP \times INP \times .... \times INP \ + \ INP \times PT
\times INP \times .... \times PT \ + \ PT \times PT \times ....
\times PT \nonumber\\
&=& (INP + GPT) \ QCD.
\end{eqnarray}
Evidently, the first term contains only the INP parts of the full
gluon propagators, which appear in any multi-loop skeleton
diagrams. In the second term the INP parts can be present along
with the PT ones, and finally the third term depends only on the
corresponding PT parts. The last two terms can be combined into
the so-called general PT (GPT) term. It is easy to understand that
we can always render the first INP part to be IR finite (the
powers of the mass gap, which scales as $\epsilon$, is equal or
higher than the powers of the NP IR singularities, which scale as
$1 / \epsilon$). This is not the case for the second term, where
powers of $\epsilon$ coming from the IR renormalization of a mass
gap may not match the powers of $1/\epsilon$, coming from the
corresponding Laurent expansions that are dimensionally
regularized (this will be explicitly shown in part II, for
convenience). How the PT part should be rendered UV finite is a
well known procedure, and it is not our problem. So, the most
dangerous term is the second one, and its NP IR singularities
cannot be removed by the IR renormalization of a mass gap only. To
render it IR finite, some additional condition on the IR
renormalization of the four-gluon proper vertex should be imposed
(see part II).

The general idea behind all the subtractions is to completely
decouple INP QCD from QCD as a whole, i.e., to proceed from QCD to
INP QCD by the subtraction of the GPT part in the symbolical Eq.
(9.1) as follows:

\begin{equation}
QCD \Longrightarrow  INP \ QCD = QCD - GPT \ QCD.
\end{equation}
In this theory all numerical results will depend on the mass gap
and can be expressed in terms of the finite integrals, which, in
their turn, will depend only on the constants of integration of
the corresponding equations of motion (and current quark masses in
the general, nonchiral case). It is worth emphasizing that the
subtraction procedure is not only physically well-motivated, as it
follows from above. In part III we will show that it is also well
justified mathematically from the DT point of view. Evidently, the
basic key in this program is the existence of the NP phase in QCD.
In part II of our approach, it will be proven that INP QCD and QCD
as a whole are IR renormalizable theories, making thus possible to
explain and solve the color confinement phenomenon as well as all
other NP effects. It is clear now why color confinement was so
difficult to explain and to solve before the proof of the IR
remormalizability of QCD. In other words, there is a close
relationship between this proof and the solution of the
confinement problem and vice versa. It is worth emphasizing that
by subtracting an infinity from another infinity in order to get a
finite result (see Eq. (9.2)), Lorentz invariance will be not
violated. Such kind of the subtractions is a standard procedure in
any method of renormalization (to render the theory free from the
PT UV or NP IR divergences).

 Of course, the QCD Green's functions cannot be gauge-invariant
because, by definition, all fields are not. This implicit gauge
dependence (of the quark propagator as well as all other Green's
functions) always exist and cannot, in principle, be eliminated by
any means. This is a general feature of all gauge theories such as
QED and QCD. Unfortunately, in gauge theories the main problem is
not the above-mentioned unavoidable implicit gauge dependence, but
the explicit dependence of the Green's functions on the gauge
fixing parameter. As explained in the Introduction, in QED this is
not so important. It becomes crucially important in QCD due to its
non-Abelian character. In order to make gauge bosons transverse we
need ghosts. It is well known how the mechanism of the
cancellation of unphysical (longitudinal) degrees of freedom of
gauge bosons by ghosts works in PT \cite{17,31}. Though this is a
general feature, beyond PT it is technically not well known. It
seems to us that distinguishing between transverse and
longitudinal components of the gauge bosons on a dynamical ground,
we found the way in the correct direction. Subtracting finally the
PT contributions, which are always of an arbitrary covariant
gauge, while the INP contributions are always transverse, by
construction, we were able to formulate INP QCD in a manifestly
gauge-invariant way. However, this does not mean that we need no
ghosts. The quark-ghost sector contains a very important piece of
information on quark degrees of freedom themselves. Precisely this
information should be self-consistently taken into account (see
part II).

There exists also an additional but very serious argument in favor
of the inevitability of the above-discussed subtractions of the PT
contributions at all levels in order to fix the gauge of INP QCD.
In his pioneering paper \cite{81} Gribov has investigated the
quantization problem of non-Abelian gauge theories using the
functional integral representation of the generating functional
for non-Abelian gauge fields. It has been explicitly shown that
the standard Fadeev-Popov (FP) prescription fails to fix the gauge
uniquely and therefore should be modified, i.e., it is not enough
to eliminate arbitrary degrees of freedom from the theory. In
other words, there is an ambiguity in the gauge-fixing of
non-Abelian gauge fields (the so-called Gribov ambiguity
(uncertainty), which results in Gribov copies and vice versa). To
resolve this problem Gribov has explicitly demonstrated that the
above-mentioned modification reduces simply to an additional
limitation on the integration range in the functional space of
non-Abelian gauge fields, which consists in integrating only over
the fields for which the FP determinant is positive \cite{81}
(introducing thus the so-called Gribov horizon in the functional
space, see also Ref. \cite{82}). As emphasized by Gribov, this
affects the IR singularities of PT and results in a linear
increase of the charge interaction at large distances. The INP
part of QCD is a manifestly gauge-invariant, by construction (it
depends only on the transverse (physical) degrees of freedom of
gauge bosons). All problems with the gauge fixing discovered by
Gribov in the functional space should be attributed to the GPT
part of QCD within our approach. Subtracting further the GPT part
in order to proceed to INP QCD in accordance with our general
method (see Eq. (9.2)), we thus will make it free from the
gauge-fixing ambiguity in the momentum space. We fix unambiguously
the gauge (transverse) of INP QCD. Finally this will lead to the
existence of something like Gribov horizon but in the momentum
space. We would like to emphasize that the general proposal to
subtract all type of the PT contributions is our solution to the
problem of Gribov copies. These subtractions become necessary
(inevitable) and important in order to make INP QCD free from this
problem, which otherwise will plague the dynamics of any
essentially NL gauge systems \cite{82}.

A few additional remarks on the subtraction of the PT
contributions are in order. Let us remind that in lattice QCD
\cite{83} such a kind of an equivalent procedure also exists. In
order to prepare an ensemble of lattice configurations for the
calculation of any NP quantity or to investigate some NP
phenomena, the excitations and fluctuations of gluon fields of the
PT origin and magnitude should be "washed out" from the vacuum.
This goal is usually to be achieved by using "Perfect Actions"
\cite{84}, "cooling" \cite{85}, "cycling" \cite{86}, etc. (see
also Refs. \cite{42,43,44,45} and references therein). Evidently,
in lattice QCD this is very similar to our method in continuous
QCD in order to proceed to INP QCD. However, there exists also a
principle difference. As underlined above, in continuous QCD by
the subtraction of the PT contributions at all levels, any NP
quantity becomes free of the PT "contaminations" for ever, i.e.,
it remains truly NP. In lattice QCD, one should go to the
continuum limit at the final stage, which means the removal of an
UV cut-off. So, all numerical lattice results for any NP quantity,
somehow become again inevitably "contaminated" by the finite PT
contributions (and thus will be plagued by the above-mentioned
Gribov uncertainties). There is no way to escape this fundamental
difficulty. That is why the criterion of the quark
confinement--area law--derived by Wilson in lattice gauge theory
\cite{64} becomes not adequate for the continuous theory \cite{87}
(Lorentz invariance is restored but confinement is lost
\cite{30,83}). It is worth emphasizing that area law as the quark
confinement criterion makes sense only for heavy quarks. The
confinement of light quarks and gluons is the $terra \ incognita$
of lattice QCD. As underlined in the Introduction, that is why the
color confinement problem of quarks and gluons can be solved only
 within the SD system of the dynamical equations approach to
continuous QCD. The removal of an IR cut-off (the infinite volume
limit) should be done exactly after the continuum limit, and not
before it. The chiral limit should be taken last, if necessary
(however, care is needed because of the chiral log problem). To
discuss lattice QCD some other difficulties such as its
essentially PT character as a specific regularization scheme and
the absence of a mass gap in the below-discussed sense is beyond
the scope of the present investigation.

In QCD sum rules the corresponding subtraction should be also done
in order to calculate, for example such truly NP quantity as the
gluon condensate. While in our approach we should subtract finally
the INP part of the full gluon propagator integrated out over the
PT region, in QCD sum rules one needs to subtract the PT solution
of the full gluon propagator integrated out over the deep IR
region, where it certainly fails (see discussion given by Shifman
in Ref. \cite{43}). The necessarily of the subtraction of the PT
part of the effective coupling constant (integrated out) in order
to correctly calculate the gluon condensate by analytic method has
been explicitly shown in recent paper \cite{88} as well.

Let us emphasize that in the present investigation the only
subtraction which has been done so far was the above-mentioned
subtraction in Eq. (4.1), defining thus the INP phase in QCD. In
order to define INP QCD itself all other subtractions will be done
further later on in the next parts of our approach to low-energy
QCD. Here, however, it is instructive to emphasize that the
dependence on the mass gap in the full gluon propagator (2.1)
should be regular one. Evidently, only in this case it can be
considered as responsible for the NP phase in the QCD true vacuum.
Introducing it explicitly into the full gluon form factor, the
subtraction (4.1) can be equivalently written down as follows:

\begin{equation}
d^{NP}(q^2; \Delta^2)= d(q^2; \Delta^2) - d(q^2; \Delta^2=0),
\end{equation}
by obviously identifying $d(q^2; \Delta^2=0) \equiv d^{PT}(q^2)$.
It is interesting to remember that within our approach the NP
phase in QCD dies linearly when the corresponding mass gap goes to
zero, while AF dies much more slowly as $1/ \ln$ when
$\Lambda^2_{QCD}$ goes formally to zero. Concluding, let us note
that in some cases the mass of a particular particle can be
treated as the mass gap. For example, in the dual Abelian Higgs
model \cite{89,90} the mass of the dual boson plays the role of a
scale responsible for the NP dynamics. The truly NP gluon form
factor defined in Eq. (9.3) retains, nevertheless, the IR
singularity of the free gluon propagator, i.e., there is no INP
phase in this model. As a result its vacuum with string and
without string contributions is unstable against quantum
corrections \cite{90}.

\subsection{ A Mass Gap}

That the NP structure of the QCD ground state can be reflected by
the deep IR structure of the full gluon propagator is one of the
important features of our formulation of the INP phase in QCD.
That this theory requires an automatical introduction of a
characteristic mass scale parameter, responsible for the NP
dynamics, the so-called mass gap, is another relevant feature of
it. This is especially important, since there is none in the QCD
Lagrangian. As mentioned above, in the presence of a finite mass
gap the coupling constant becomes to play no any role. Since a
mass gap appears on a dynamical ground, this is also a direct
evidence for the "dimensional transmutation" \cite{1,37}, which
occurs whenever a massless theory acquires masses dynamically. It
is a general feature of spontaneous symmetry breaking in field
theories.

The relation of our mass gap to the mass gap introduced by Jaffe
and Witten (JW) \cite{91} still remains to be understood. Firstly,
we naively identified the JW mass gap $\Delta_{JW}$ with our
$\Lambda_{NP}$ \cite{92}, where $\Lambda_{NP}$ is the final
version of the mass gap $\bar \Delta$ (see part II). However,
things are apparently not so simple. In QCD there also exists
$\Lambda_{PT}$ ($\equiv \Lambda_{QCD}$), which is responsible for
the nontrivial PT dynamics. The relation between these three
scales can be symbolically reproduced as follows:

\begin{equation}
 \Lambda_{NP} \longleftarrow^{\infty \leftarrow \alpha_s}_{0
\leftarrow M_{IR}} \ \Delta_{JW} \ { }^{\alpha_s \rightarrow
0}_{M_{UV} \rightarrow \infty} \longrightarrow  \ \Lambda_{PT},
\end{equation}
where $\alpha_s$ is obviously the fine structure coupling constant
of strong interactions, while $M_{UV}$ and $M_{IR}$ are the UV and
IR cut-offs, respectively. The right-hand-side limit is well known
as the weak coupling regime, while the left-hand-side can be
regarded as the strong coupling regime. We know how to take the
former \cite{1}, and we don't know how to take the latter one yet.
However, there is no doubt that the final goal of this limit,
namely, the mass gap $\Lambda_{NP}$ exists, and should be the
renormalization group invariant in the same way as $\Lambda_{QCD}$
($\equiv \Lambda_{PT}$) is. Taking the weak coupling limit, a
dependence on the number of flavors appears. Apparently, the
strong coupling limit is also flavor dependent. At the same time,
it is well understood that the fundamental mass gap should come
from the quantum YM theory \cite{91}, i.e., it should be, in
principle, by definition, flavor independent. This means that the
JW mass gap $\Delta_{JW}$ is the fundamental mass, determining the
scale of QCD as a whole. The mass gap $\Lambda_{NP}$ and
$\Lambda_{PT}$ are responsible for the NP and PT dynamics,
determining thus the deviation of the full gluon propagator from
the free one in the IR and UV regions, respectively. They are two
different "faces" of the same fundamental mass gap $\Delta_{JW}$
("two sides of the same coin"). However, it is necessary to
emphasize that the asymptotic scale parameter $\Lambda_{PT}$
cannot be considered as the mass gap in the sense understood by
JW, while $\Lambda_{NP}$ looks like very similar to $\Delta_{JW}$.
In the INP phase of QCD there is no spectrum in the interval $(0,
\Lambda_{NP})$, indeed. It is possible to say that in the INP
phase of QCD our mass gap $\Lambda_{NP}$ plays the same role which
in QCD as a whole belongs to the JW mass gap $\Delta_{JW}$.

 The JW theorem \cite{91} is formulated as follows:

\vspace{3mm}

{\bf Yang-Mills Existence And Mass Gap:} Prove that for any
compact simple gauge group $G$, quantum Yang-Mills theory on $R^4$
exists and has a mass gap $\Delta > 0$,

\vspace{3mm}

then one of our main results obtained in this paper can be
formulated in terms of the similar theorem, namely

\vspace{3mm}

{\bf I. Yang-Mills Existence And Mass Gap:} If four-dimensional
quantum Yang-Mills theory with compact simple gauge group
$G=SU(3)$ exists then it should have a mass gap $\Delta > 0$.

\vspace{3mm}

In fact, we have proven the second (physical) part of the JW
theorem. The first (mathematical) part is beyond of reach
(apparently, it requires a sophisticated mathematics, indeed). Let
us outline the main general points of our investigation/proof.

1. We assume that non-Abelian quantum YM theory exists.

2. It is IR unstable theory (the ZMME effect).

3. At the microscopic level, the main dynamical source of this
instability is the NL fundamental four-gluon interaction.

4. Precisely it leads to the NP (severe) IR singularities in the
full gluon propagator.

5. In its turn, this automatically requires the existence of a
mass gap, so it appears on a dynamical ground.

We have done even more than that. The rest of our main results can
be also formulated as a following theorem:

\vspace{3mm}

{\bf II. Yang-Mills Existence And Mass Gap:} If four-dimensional
quantum Yang-Mills theory with compact simple gauge group
$G=SU(3)$ exists and exhibits a mass gap $\Delta > 0$ then it
confines gluons.

\vspace{3mm}

Let us outline the main general points of this part of our
investigation/proof.

6. The distribution nature of the NP IR singularities is to be
underlined.

7. The use of the corresponding Laurent expansion that is
dimensionally regularized in order to control the above-mentioned
NP IR singularities.

8. The IR renormalization properties of the mass gap only lead the
confinement of gluons.

9. It will explain the quark confinement and SBCS as well.

Of course, our investigation can be given in mathematically
standard way (to formulate theorems, the corresponding lemmas,
etc.), however, this is not the case here, indeed. At the same
time, we would like to emphasize here that our general conclusion
that the mass gap in non-Abelian YM theory arises from the quartic
potential $(A \wedge A)^2$ in the action is in complete agreement
with that expressed by Feynman \cite{93} (see also Ref.
\cite{91}).

The ultimate goal of QCD is to calculate all physical parameters,
for example hadron masses, as pure numbers times the
characteristic scale, the mass gap \cite{30}. Thus, in general,
should be $ \Lambda_{NP} = \mu \times \Delta_{JW}$ and
$\Lambda_{PT} = \nu \times \Delta_{JW}$, where $\mu$ and $\nu$ are
some pure numbers. Obviously, from the above-mentioned relations
one can deduce that $\Lambda_{NP} = (\mu / \nu) \times
\Lambda_{PT}$. However, such kind of relations do not mean that
both quantities are dependent from each other. Such kind of
relations, for example between hadron masses, mean only that we
know how to calculate them from first principles in terms of the
fundamental mass gap. Let us note in advance that in a pure YM
theory there are no good physical numbers (physical observables
which can be directly experimentally measured). So, there is a
problem how to formulate a well-defined scale-setting scheme in
order to determine any mass gap: the JW mass gap $\Delta_{JW}$ or
our mass gap $\Lambda_{NP}$. Evidently, it should be done in a
more sophisticated way in the framework of QCD itself
\cite{51,94}.

In the next Subsec. it makes sense to discuss some technical
aspects of our approach to low-energy QCD.

\subsection{A technical outlook}

Let us begin with some tentative remarks on a possible IR
renormalizability of INP QCD and QCD itself, which will be proven
in part II. There is no doubt that it is due to the DT fundamental
result, which requires that any NP IR singularity with respect to
momentum in terms of $\epsilon$ should always be $1 / \epsilon$.
And this does not depend on how the IR regularization parameter
$\epsilon$ has been introduced in a way compatible with DT itself.
On the other hand, this fundamental result relates the IR
regularization to the number of space-time dimensions (see
discussion in Sec. 2 and Refs. \cite{24,57}), i.e., the so-called
"compactification" \cite{91}. It is easy to imagine that otherwise
none of the IRMR programs would be possible. In other words, we
know the mathematical theory which has to be used - the theory of
distributions \cite{24,57} (apparently, for the first time the
distribution nature of the Green's functions in quantum field
theory has been recognized and used in Ref. \cite{95}). It
provides the basis for the adequate mathematical investigation of
a global character of the NP IR divergences. Each one-loop
skeleton diagram diverges as $1/ \epsilon$. Moreover, each
independent loop part of the multi-loop skeleton diagrams will
diverge as $1/ \epsilon$ as well (see part II). On the other hand,
the UV divergences have a local character, and thus should be
investigated term by term in the coupling constant squared.
Bearing in mind that INP QCD becomes trivially an UV finite theory
(after subtractions of the PT contributions at all levels, which
contain an UV divergent tails, see the symbolical Eq. (9.2)) and
in order not to complicate notations, that is why we have started
from the PT unrenormalized Green's functions.

In principle, none of the regularization schemes (how to introduce
the IR regularization parameter in order to parameterize the NP IR
divergences and thus to put them under control) should be
introduced by hand. First of all, it should be well defined.
Secondly, it should be compatible with DT \cite{24}. The DR scheme
\cite{25} is well defined, and here we have shown how it should be
introduced into DT (complemented by the number of subtractions, if
necessary). Though the so-called $\pm i\epsilon$ regularization is
formally equivalent to the regularization used in our paper (see
again Ref. \cite{24}), nevertheless, it is rather inconvenient for
practical use. Especially this is true for the gauge-field
propagators, which are substantially modified due to the response
of the vacuum (the $\pm i\epsilon$ prescription is designated for
and is applicable only to the theories with PT vacua, indeed
\cite{39,82}). Other regularization schemes are also available,
for example such as analytical regularization used in Ref.
\cite{9} or the so-called Speer's regularization \cite{96}.
However, they should be compatible with DT, as emphasized above.
Anyway, not the regularization is important but DT itself. Whether
the theory is IR multiplicative renormalizable or not depends on
neither the regularization nor the gauge. Due to the chosen
regularization scheme or the gauge only the details of the
corresponding IRMR program can be simplified. In other words, if
the theory is proven to be IR or UV renormalizable in one gauge,
it is IR or UV renormalizable in any other gauge. This is true for
the regularization schemes as well.

It is worth emphasizing the difference between QED and QCD. In
former theory an electron/quark loop insertion into the photon
propagation with momentum $q$ is only possible. As mentioned above
in Subsec. D of Sec. 2, the estimate similar to the estimate
(2.37) could be formally performed. However, the summation of an
infinite series of the most singular terms in the IR will yield
final zero, i.e., the mass gap will not survive in QED. That is
why an electron loop insertion into the photon propagation with
momentum $q$ requires its proportionality to $q^2$ in the
numerator. In the iteration solution for the photon propagator it
cancels one of $q^2$ in the denominator, and the photon always
propagates like the free one, i.e., as $1/q^2$ in the IR (even
summing up all insertions). The reason is that in QED the cluster
property of the Wightman functions forbids a more singular
behavior of the full photon propagator in the IR than the behavior
of its free photon counterpart. So, in QED the IR singularity of
the full photon propagator (2.1) is as much singular as $1/q^2$,
i.e., it is always the PT IR singularity. In QCD, contrary to QED,
the cluster property of the above-mentioned Wightman functions
fails due to the self-interactions of the massless gluons (the
so-called Strocchi theorem \cite{69}). This allows a behavior of
the full gluon propagator more singular in the IR than the
behavior of its free counterpart. That is why the mass gap finally
survives after the summation of an infinite series of the most
singular terms in the IR. This is precisely the principal
dynamical distinction between these two theories. Thus, the
estimate like the estimates, which have derived in Subsec. C of
Sec. 2, do not exist in QED, which is Abelian gauge theory (no
direct interactions between photons), while in QCD they exist,
since it is non-Abelian gauge theory (direct interactions between
gluons do exist).

 There also exists a principal difference in their IR
renormalization procedures. Since in QED the IR singularity cannot
be as much singular as $1/q^4$ (which is the simplest NP IR
singularity possible in 4D gauge theory), Laurent expansion, that
is dimensionally regularized for the purpose of the
renormalization, becomes useless. Thus the PT IR singularity can
be regularized even by hand, introducing, for example, the
gluon/photon "mass", which goes to zero at the end of the
computations. In other words, the PT IR singularity needs only the
regularization in one place, i.e., all other Green's functions are
not affected by the dependence on the regularization "mass". This
is the principal difference between the PT and the NP IR
singularities. The latter ones additionally need also the
nontrivial renormalization program to be done for the whole
theory, as it was done for the YM sector in this paper.

Let us underline also that due to the distribution nature of the
NP IR singularities, any solution to the gluon SD equation found
in closed form (for example, some combination of special
functions), has to be presented further by the series of the
Laurent expansion, anyway. This makes it possible to put the
corresponding integrals and equations in the deep IR domain under
firm mathematical control, provided by correctly applying DT,
complemented by the DR method, to each term of the expansion. As
we already know, the rest of the Laurent expansion (which begins
from the PT IR singularity) is to be shifted to the PT part of the
full gluon propagator. Thus, the Laurent expansion (6.7) is a
general one. It shows any solution to the INP part of the full
gluon propagator as an infinite sum of all possible NP IR
singularities. The next step is to perform the IR renormalization
program (a scaling analysis with respect to the IR regularization
parameter). A crucial role in this analysis belongs to the mass
gap. This finally makes it possible to fix the IR structure of the
full gluon propagator, presented by its INP part in Eq. (7.2).
Thus, we have found a solution to the full gluon propagator (7.1)
up to its unimportant PT part, not solving directly the gluon SD
equation.

The coupling constant in QCD is an effective one. In fact, it is
"running", i.e., it is a function of the gluon momentum. The
solution of the gluon SD equation (2.5) for the full gluon form
factor $d(q^2)$ (which is dimensionless) can be identified with
the QCD effective coupling. Because of AF we know its behavior in
the deep UV limit. It depends on the asymptotic scale parameter
$\Lambda_{PT}$ \cite{1,17,18,27,31}. The behavior of the effective
coupling in the deep IR limit has been established here (see also
our preliminary publication \cite{97}). It depends on its own
scale parameter, the mass gap $\bar \Delta^2$. In both cases the
presence of the corresponding masses indicates the scaling
violation. So, for the calculation of any numerical value of the
QCD effective coupling, it is necessary to chose the scale, at
which it should be done. In this case, it is difficult to assign a
physical meaning to any of it numerical values (see also remarks
in Refs. \cite{30,98}), for example to its critical value, at
which some NP phenomena may occur. It is much more relevant to
speak about the scale at which some physical phenomena become
important or even occur. Just this is happening in INP QCD, the
numerical results of which depend on the mass gap being the
renormalization group invariant. It gains the contributions from
all powers of the QCD strong coupling constant squared. Neither
any value of it nor it itself plays any role in our approach.

This is also in complete agreement with our definition of the INP
phase in QCD. We distinguish between the two phase in QCD by the
character of the IR singularities and the presence of the mass
gap. We do not use the coupling constant for this purpose. We
already know that the INP part can be present by an infinite
series in the coupling constant squared. At the same time, the PT
part can be responsible for the gluon field configurations, which,
in principle, cannot be even described by an infinite series in
the coupling constant, for example something like "quantum"
instantons. Apparently, the standard definition of the
nonperturbativity based on the impossibility to expand in the
coupling constant should be abandoned. The primary feature of the
nonperturbativity seems to be the dependence on the corresponding
mass gap.

In the functional space we should distinguish between the purely
transverse severely singular gluon field configurations, which
decrease more slowly than $1/r$ (in fact, they increase at large
distances $r$, at least linearly, the so-called "large gluon
fields" \cite{81} (see also Ref. \cite{82})), and the gluon field
configurations of an arbitrary covariant gauge, which decrease at
least as $1/r$. Of course, this separation corresponds to the
difference between the NP (severe) and the PT IR singularities
defined above in the momentum space. However, the former ones are
prevented from growing up to infinity. The above-described IR
renormalization renders the amplitudes of these virtual fields
finite, but large enough to provide quark confinement by shifting
quarks from the mass shell (see part III). Moreover, the
amplitudes of the purely transverse severely singular actual gluon
field configurations (no explicit integration over the
corresponding gluon momenta, i.e., configurations with external
gluon legs or equivalently configurations with soft-gluon
emissions) are suppressed providing thus the gluon confinement,
see Eq. (7.5). This is our response to the Gribov's $dilemma$
formulated in Ref. \cite{82} as follows: "the solution of the
confinement problem lies not in the understanding of the
interaction of "large gluon fields" but instead in the
understanding of how the QCD dynamics can be arranged as to
prevent the non-Abelian fields from growing real big". Contrary to
this, the interaction of just "large gluon fields" leads to color
confinement within our approach. Only two different cases of the
purely transverse severely singular actual and virtual gluon field
configurations should be carefully distinguished due to the
distribution nature of the same NP IR singularities for both
cases. The Gribov's $dilemma/mystery$ can be also formulated as
follows: if indeed the NL interaction of "large gluon fields" is
responsible for color confinement (and precisely this statement is
to be deduced from his paper \cite{81}, though finally Gribov
pursued another approach \cite{39,82}), then why we cannot "see"
them. This $dilemma/mystery$ is nothing else but the first
formulation of the gluon confinement problem (see also Ref.
\cite{93}). Based on his paper, this problem in some detail has
been discussed by Zwanziger in Ref. \cite{99}, where the vanishing
of zero-momentum lattice Landau and Coulomb gauge gluon
propagators have been investigated. Contrary to this case, first
we have shown that the gluon propagator can be only severely
singular in the IR. Secondly, this does not prevent to formulate
the gluon confinement criterion in a manifestly gauge-invariant
way, and thus to resolve finally the gluon confinement mystery.
The problem of Gribov copies in lattice QCD has been recently
addressed via the computation of the gluon propagator in the
Landau gauge \cite{100}.

\section{Conclusions}

Emphasizing the highly nontrivial structure of the true QCD ground
state in the deep IR region, one can conclude:

1). The self-interaction of massless gluons (i.e., the NL
gluodynamics) is responsible for the large scale structure of the
true QCD vacuum.

2). It is saturated by the so-called purely transverse severely
singular gluon field configurations.

3). Precisely these field configurations are behind the ZMME
effect in the true QCD vacuum, which is to be taken into account
by the deep IR structure of the full gluon propagator.

4). The full gluon propagator thus is inevitably more singular in
the IR than its free counterpart.

5). This requires the existence of a mass gap, which is
responsible for the NP dynamics in the QCD vacuum. It appears on
dynamical ground due to the self-interaction of massless gluons
only. It cannot be interpreted as the effective/dynamical gluon
mass.

6). We define the NP and the PT IR singularities as more severe
than and as much singular as $1 / q^2$, respectively, which is the
power-type, exact IR singularity of the free gluon propagator. In
the functional space the former ones correspond to the purely
transverse gluon field configurations which decrease more slowly
than $1/r$ at large distances $r$. The latter ones correspond to
the gluon field configurations of an arbitrary covariant gauge
which decrease as $1/r$, at least.

7). The main dynamical source of the NP (severe) IR singularities
in the full gluon propagator is the two-loop skeleton term of the
corresponding SD equation, which contains the four-gluon vertices
only (Eq. (2.17)).

8). We decompose algebraically (i.e., exactly) the full gluon
propagator as a sum of its INP and PT parts. We additionally
distinguish between them dynamically by the different character of
the IR singularities in each part.

9). We have established the deep IR structure of the full gluon
propagator, reproduced by its INP part, as an infinite sum over
all possible NP IR singularities in the expansion (5.1) or
equivalently (5.4).

10). The next step is to regularize them correctly, i.e., to use
the Laurent expansion (6.1) that is dimensionally regularized with
respect to the IR regularization parameter $\epsilon$.

11). We emphasize once more that the IR renormalization program is
based on an important observation that the NP IR singularities
$(q^2)^{-2-k}$, being distributions, always scale as $1 /
\epsilon$, not depending on the power of the singularity $k$,
i.e., $(q^2)^{-2-k} \sim \ 1/ \epsilon$. It is easy to understand
that otherwise none of the IR renormalization program in the INP
phase of QCD  and in QCD itself would be possible.

12). The IR renormalization of the initial mass gap is only needed
in order to fix uniquely and exactly the IR structure of the full
gluon propagator. It is saturated by the simplest NP IR
singularity, the famous $(q^2)^{-2}$, see Eq. (7.2).

13). This also means that the mass gap after summing up an
infinite number of the corresponding diagrams survives, indeed. It
gains contributions from all orders of PT in the coupling constant
squared, which remains IR finite from the very beginning as well
as the gauge fixing parameter.

14). The smooth in the IR the full gluon propagator should be
ruled out. Only its PT part can be rendered finite at zero due to
the special gauge choice (Landau gauge).

15). As a functions ghost and quark degrees of freedom contribute
into the PT part of the full gluon propagator within our approach.
As integrated out in all orders of linear PT (i.e., numerically)
they contribute into its INP part as well.

16). On this basis, we have formulated the ZMME model of the true
QCD ground state. Due to the distribution nature of the NP IR
singularities, two different types in the IR renormalization of
the INP part of the full gluon propagator are required,
preserving, nevertheless, its IR finiteness ($Z_d(\epsilon)=1$).

17). In its turn, this allows one to rigorously prove that color
gluons can never be isolated. The gauge-invariant, analytical
formulation of the gluon confinement criterion is given in Eq.
(7.5).

18). In a manifestly gauge-invariant way, we define the INP phase
in QCD at the fundamental gluon level. The corresponding
decomposition of the full gluon propagator is only needed in order
to firmly control the IR region in QCD within our approach (only
it contains explicitly the mass gap).

19). We have shown that the INP part of the full gluon propagator
identically satisfies the corresponding part of the gluon SD
equation.

 Our general conclusions are:

I. The NP structure of the true QCD ground state is to be
described by the IR structure of the full gluon propagator.

II. It is an infinite sum over all possible NP IR singularities.
Due to their distribution nature any solution to the gluon SD
equation has to be always present in the form of the corresponding
Laurent expansion. It makes it possible to control both the whole
expansion as well as each term of it by the correct use of DT.

III. The mass gap responsible for the NP dynamics in the true QCD
ground state is required. This is important, since there is none
in the QCD Lagrangian.

IV. The mass gap arises from the quartic gluon potential (since it
survives when all the gluon momenta involved go to zero, while the
triple gluon potential does not). It just makes the full gluon
propagator so singular in the IR.

V. In the presence of a mass gap the QCD coupling constant plays
no any role.

VI. Complemented by the DR method, DT puts the treatment of the NP
IR singularities on a firm mathematical ground. So, there is no
place for theoretical uncertainties. The wide-spread opinion that
they cannot be controlled is not justified.

VII. We emphasize the importance of the general relation between
the IR regularization and the number of space-time dimensions
("compactification"), which is crucial for the general IR
renormalization program.

VIII. We would like to stress the importance of the consideration
in the Euclidean momentum space, which is automatically free from
the kinematical (unphysical) singularities in the gauge-field
propagators.

IX. All this makes it possible to fix uniquely the IR structure of
the full gluon propagator in QCD, not solving directly the
corresponding SD equation itself. Thus, we have exactly
established the interaction between quarks (concerning its pure
gluon (i.e., NL) contribution up to its unimportant PT part).

X. This somehow astonishingly radical result has been achieved at
the expense of the PT part of the full gluon propagator. It
remains of arbitrary covariant gauge and its functional dependence
cannot be determined. However, as explained above, we are not
responsible for the PT phase in QCD.

XI. Collective motion of all the purely transverse $virtual$ gluon
field configurations with low-frequency components/large scale
amplitudes (the purely transverse severely singular virtual gluon
field configurations) is solely responsible for the color
confinement phenomenon within our approach.

XII. The amplitudes of all the purely transverse severely singular
$actual$ gluon field configurations are totally suppressed,
leading thus to the confinement of gluons (no transverse gluons at
large distances).

XIII. The difference between the above-mentioned $virtual$ and
$actual$ gluon field configurations at the microscopic fundamental
level should be traced back to the two different types in the IR
renormalization of the INP part of the full gluon propagator.

XIV. In its turn, the above-mentioned difference is determined by
the IR renormalization properties of the mass gap only, taking
into account the distribution nature of the NP IR singularities,
which inevitably appear in the full gluon propagator due to the NL
character of the gluodynamics.

XV. We unambiguously identify the main source of the IR
instability of QCD. It is the four-gluon vertex at the Lagrangian
level, and the two-loop skeleton term, which contains only the
four-gluon vertices, at the level of the gluon SD equation.
Precisely this interaction exhibits an additional IR singularities
(and hence the mass gap) in the corresponding loop integrals when
all the gluon momenta involved go to zero.

XVI. Color confinement is an IR renormalization effect within our
approach. If AF is mainly determined by the three-gluon
interactions, then color confinement is mainly due to the
four-gluon interactions.

XVII. We would like to emphasize once more a self-consistency and
manifestly gauge invariance of our approach to low-energy QCD.

Our main results can be summarized similar to the JW theorem as
follows:

\vspace{3mm}

{\bf Theorem: If four-dimensional quantum Yang-Mills theory with
compact simple gauge group

\hspace{18mm} $G=SU(3)$ exists then it exhibits a mass gap and
confines gluons}.

\vspace{3mm}

Concluding finally, let us make a few remarks. How to correctly
formulate the quark confinement problem was more or less clear
from the very beginning of QCD. As mentioned above, the
confinement of heavy quarks can be understood in terms of the
linear rising potential between them. The confinement of light
quarks can be formulated as the absence of the pole-type
singularities in the quark Green's functions
\cite{73,101,102,103,104}, which can be generalized on heavy
quarks as well. At the same time, how to correctly formulate the
gluon confinement problem was not clear, and up to these days the
confinement of gluons remained rather mysterious. It seems to us
that in the present investigation we have correctly formulated the
gluon confinement problem, which allowed us to solve it as well.
In other words, the above-formulated theorem is proven here (as
rigorously as possible in theoretical physics). In its turn, this
will allow us to solve the quark confinement problem (the
gauge-invariant formulation of the quark confinement criterion
will be given in part III of our approach. For preliminary
discussion see our paper \cite{20}). A brief description of the
most important results obtained in this paper has been already
published in Refs. \cite{96,105}. Though we are not interested in
the PT part of the full gluon propagator, nevertheless it is
important for high-energy QCD (for a new analytic approach to PT
QCD see Ref. \cite{106}).

\begin{acknowledgments}

The author would like to thank S. Adler, H. Georgi, K. Nishijima,
V.I. Zakharov, H. Fried, M. Faber, A. Ivanov, V.A. Rubakov (and
members of his seminar at INR, RAS), D.V. Shirkov, A.A. Slavnov,
A.T. Filippov, G.V. Efimov, V.P. Gusynin, Gy. Pocsik, P. Forg\'{a}cs,
K. Toth, B. Lukacs, P. Levai, T. Biro, J. Revai, T. Csorgo and
especially Gy. Kluge and J. Nyiri for useful correspondence,
discussions, remarks, support and help. We are also grateful to H.
Ejiri and H. Toki for support and collaboration and useful
discussions at the first stage of this investigation. A financial
support from HAS-JINR Scientific Collaboration Fund is to be also
acknowledged.

\end{acknowledgments}

\appendix
\section{The PT part of the gluon SD equation}

In this appendix we evaluate the gluon SD equation (8.19) for the
PT part of the full gluon propagator. The PT part of the so-called
tadpole term, which is given in Eq. (2.14), is

\begin{equation}
T^{PT}_t = g^2 \int {i d^4 q_1 \over (2 \pi)^4} T^0_4 D^{PT}(q_1),
\end{equation}
so it is finite in the $\epsilon \rightarrow 0^+$ limit. Moreover,
in dimensional regularization with $D^{PT}=D^0$ (but not with
$D^{PT}=\tilde{D}^0$) it yields simply zero \cite{17}. That is a
reason why in the general iteration solution of the gluon SD
equation (2.4) the tadpole contribution can be generally
discarded, which means the omission of the term depending on $\bar
\Delta^2_t$ in Eq. (8.20) as well. So, the tadpole term itself is
not important, as emphasized above.

 The PT part of the three-gluon vertex contribution into
nonlinear gluon self-energy comes from Eq. (2.15) and it is a sum
of the three terms

\begin{equation}
T^{PT}_1(q) = \sum_{n=1}^3 T_1^{(n)}(q),
\end{equation}
where

\begin{equation}
T_1^{(1)}(q) =  g^2 \int {i d^4 q_1 \over (2 \pi)^4} T^0_3 (q,
-q_1, q_1-q) T_3 (-q, q_1, q -q_1) D^{INP}(q_1) D^{PT}(q -q_1),
\end{equation}

\begin{equation}
T_1^{(2)}(q) = g^2 \int {i d^4 q_1 \over (2 \pi)^4} T^0_3 (q,
-q_1, q_1-q) T_3 (-q, q_1, q -q_1) D^{PT}(q_1) D^{INP}(q -q_1),
\end{equation}

\begin{equation}
T_1^{(3)}(q) =  g^2 \int {i d^4 q_1 \over (2 \pi)^4} T^0_3 (q,
-q_1, q_1-q) T_3 (-q, q_1, q -q_1) D^{PT}(q_1) D^{PT}(q -q_1).
\end{equation}
Integrating over the gluon momentum $q_1$ in Eq. (A3) and over the
gluon momentum $q-q_1$ in Eq. (A4) with the help of Eq. (7.3), one
finally obtains

\begin{eqnarray}
T_1^{PT}(q) = &-& \bar \Delta_1^2 \Bigl[ T^0_3 (q, 0, -q) T_3 (-q,
0, q) +
T^0_3 (q, -q,0) T_3 (-q, q, 0) \Bigr] D^{PT}(q) \nonumber\\
&+&  g^2 \int {i d^4 q_1 \over (2 \pi)^4} T^0_3 (q, -q_1, q_1-q)
T_3 (-q, q_1, q -q_1) D^{PT}(q_1) D^{PT}(q -q_1).
\end{eqnarray}
One can conclude that one-loop skeleton PT contribution into the
gluon self-energy is under control in the $\epsilon \rightarrow
0^+$ limit, i.e., formally it exists as $\epsilon$ goes to zero.

The simplest two-loop contribution into the gluon self-energy is
determined by Eq. (2.17). Its PT part is a sum of the seven terms

\begin{equation}
T'^{PT}_2(q) =  \sum_{n=1}^7 T'^{(n)}_2(q),
\end{equation}
where

\begin{equation}
T'^{(1)}_2(q) = g^4 \int {i d^4 q_1 \over (2 \pi)^4} \int {i d^4
q_2 \over (2 \pi)^4} T^0_4  T_4 (-q, q_1, -q_2, q-q_1+q_2)
D^{INP}(q_1) D^{PT}(-q_2)D^{INP}(q-q_1+q_2),
\end{equation}

\begin{equation}
T'^{(2)}_2(q) = g^4 \int {i d^4 q_1 \over (2 \pi)^4} \int {i d^4
q_2 \over (2 \pi)^4} T^0_4 T_4 (-q, q_1, -q_2, q-q_1+q_2)
D^{PT}(q_1) D^{INP}(-q_2) D^{INP}(q-q_1+q_2),
\end{equation}

\begin{equation}
T'^{(3)}_2(q) =  g^4 \int {i d^4 q_1 \over (2 \pi)^4} \int {i d^4
q_2 \over (2 \pi)^4} T^0_4 T_4 (-q, q_1, -q_2, q-q_1+q_2)
D^{PT}(q_1) D^{PT}(-q_2) D^{INP}(q-q_1+q_2),
\end{equation}

\begin{equation}
T'^{(4)}_2(q) = g^4 \int {i d^4 q_1 \over (2 \pi)^4} \int {i d^4
q_2 \over (2 \pi)^4} T^0_4 T_4 (-q, q_1, -q_2, q-q_1+q_2)
D^{INP}(q_1) D^{INP}(-q_2) D^{PT}(q-q_1+q_2),
\end{equation}

\begin{equation}
T'^{(5)}_2(q) = g^4 \int {i d^4 q_1 \over (2 \pi)^4} \int {i d^4
q_2 \over (2 \pi)^4} T^0_4  T_4 (-q, q_1, -q_2, q-q_1+q_2)
D^{INP}(q_1) D^{PT}(-q_2) D^{PT}(q-q_1+q_2),
\end{equation}

\begin{equation}
T'^{(6)}_2(q) =  g^4 \int {i d^4 q_1 \over (2 \pi)^4} \int {i d^4
q_2 \over (2 \pi)^4} T^0_4 T_4 (-q, q_1, -q_2, q-q_1+q_2)
D^{PT}(q_1) D^{INP}(-q_2) D^{PT}(q-q_1+q_2),
\end{equation}

\begin{equation}
T'^{(7)}_2(q) = g^4 \int {i d^4 q_1 \over (2 \pi)^4} \int {i d^4
q_2 \over (2 \pi)^4} T^0_4 T_4 (-q, q_1, -q_2, q-q_1+q_2)
D^{PT}(q_1) D^{PT}(-q_2) D^{PT}(q-q_1+q_2).
\end{equation}

 Integrating over the gluon momentum $q_1$ in Eq. (A8) and
further over the gluon momentum $q_2=-q$, one obtains

\begin{equation}
T'^{(1)}_2(q) = \bar \Delta_2'^4 T^0_4 T_4 (-q, 0, q, 0)
D^{PT}(q).
\end{equation}
Integrating over the gluon momentum $-q_2$ in Eq. (A9) and further
over the gluon momentum $q_1=q$, one gets

\begin{equation}
T'^{(2)}_2(q) = \bar \Delta_2'^4 T^0_4 T_4 (-q, q, 0, 0)
D^{PT}(q).
\end{equation}
Integrating over the gluon momentum $q_2 = q_1 -q$ in Eq. (A10),
one arrives at

\begin{equation}
T'^{(3)}_2(q) = - g^4 \bar \Delta_2'^2 \int {i d^4 q_1 \over (2
\pi)^4} T^0_4 T_4 (-q, q_1, q-q_1, 0) D^{PT}(q_1) D^{PT}(q-q_1).
\end{equation}
Integrating over the gluon momenta $q_1$ and $-q_2$ in Eq. (A11),
one obtains

\begin{equation}
T'^{(4)}_2(q) = \bar \Delta_2'^4 T^0_4 T_4 (-q, 0, 0, q)
D^{PT}(q).
\end{equation}
Integrating over the gluon momentum $q_1$ in Eq. (A12), leads to

\begin{equation}
T'^{(5)}_2(q) = - g^4 \bar \Delta_2'^2 \int {i d^4 q_2 \over (2
\pi)^4} T^0_4 T_4 (-q, 0, -q_2, q+q_2) D^{PT}(-q_2) D^{PT}(q+q_2).
\end{equation}
Integrating over the gluon momentum $-q_2$ in Eq. (A13), one gets

\begin{equation}
T'^{(6)}_2(q) = - g^4 \bar \Delta_2'^2  \int {i d^4 q_1 \over (2
\pi)^4} T^0_4 T_4 (-q, q_1, 0, q-q_1) D^{PT}(q_1) D^{PT}(q-q_1),
\end{equation}
and the pure PT contribution (A14) remains unchanged. It is
convenient replace $q_2 \rightarrow - q_1$ in Eq. (A19), then the
total PT contribution becomes

\begin{eqnarray}
T'^{PT}_2(q) &=& \bar \Delta_2'^4 \Bigl[ T^0_4 T_4 (-q, 0, q, 0) +
T^0_4 T_4 (-q, q, 0, 0) + T^0_4 T_4 (-q, 0, 0, q) \Bigr]
D^{PT}(q) \nonumber\\
&-& \bar \Delta_2'^2 \int {i d^4 q_1 \over (2 \pi)^4} \Bigl[ T^0_4
T_4 (-q, q_1, q-q_1, 0) + T^0_4 T_4 (-q, 0, q_1, q-q_1) + T^0_4
T_4 (-q, q_1, 0, q-q_1) \Bigr] D^{PT}(q_1)D^{PT}(q-q_1) \nonumber\\
&+&  g^4 \int {i d^4 q_1 \over (2 \pi)^4} \int {i d^4 q_2 \over (2
\pi)^4} T^0_4 T_4 (-q, q_1, -q_2, q-q_1+q_2) D^{PT}(q_1)
D^{PT}(-q_2) D^{PT}(q-q_1+q_2).
\end{eqnarray}
Again one can conclude that two-loop skeleton contribution into
the gluon self-energy due to four-gluon vertices only is under
control in the $\epsilon \rightarrow 0^+$ limit. Here the coupling
constant $g^4$ is included into the corresponding mass gap, for
convenience apart from the last term, of course.

In connection with the last integral in Eq. (A21) it is worth
making a few remarks. There is no explicit integration over
variable $q$, but there are an explicit integrations over
variables $q_1$ and $q_2$. In order to estimate the behavior of
this integral in the deep IR domain (setting $T_4 = T_4^0$ and
using the free gluon propagators in the Feynman gauge instead of
the corresponding $D^{PT}$), one can utilize the method which has
been used in Subsec. B of Sec. 2. Omitting some algebra, to
leading order one obtains $T'^{(7)}_2(q) \sim (\Delta^4 / q^2)$.
We already know, however, that the mass gap scales as $\epsilon$
as it goes to zero, i.e., $\Delta^2 = \epsilon \bar \Delta^2$. So,
this integral in the deep IR region vanishes as $\epsilon^2$,
i.e., $T'^{(7)}_2(q) \sim \epsilon^2$ (compare with Eq. (8.12)).
Thus, in the integrals which contain only the PT parts of the full
gluon propagators, the integration is to be effectively taken from
some finite value to infinity (for all the loop variables), in
accordance with the decomposition (2.24). Let us emphasize once
more that the PT parts of the corresponding integrals, however,
are not important for us. In accordance with our method they have
to be subtracted, anyway.

 The next two-gluon contribution into the gluon self-energy is given in
Eq. (2.16), so its PT part is a sum of the fifteen terms

\begin{equation}
T^{PT}_2(q) =  \sum_{n=1}^{15} T^{(n)}_2(q),
\end{equation}
where

\begin{eqnarray}
T^{(1)}_2(q) =  g^4 \int {i d^4 q_1 \over (2 \pi)^4} \int {i d^4
q_2 \over (2 \pi)^4} &T^0_4  T_3 (-q_2, q-q_1+q_2,
q_1-q) T_3(-q, q_1, q-q_1)& \nonumber\\
&D^{INP}(q_1) D^{INP}(-q_2) D^{INP}(q-q_1+q_2) D^{PT}(q-q_1),&
\end{eqnarray}

\begin{eqnarray}
T^{(2)}_2(q) = g^4 \int {i d^4 q_1 \over (2 \pi)^4} \int {i d^4
q_2 \over (2 \pi)^4} &T^0_4 T_3 (-q_2,
q-q_1+q_2, q_1-q) T_3(-q, q_1, q-q_1)& \nonumber\\
&D^{INP}(q_1) D^{INP}(-q_2)D^{PT}(q-q_1+q_2) D^{INP}(q-q_1),&
\end{eqnarray}

\begin{eqnarray}
T^{(3)}_2(q) =  g^4 \int {i d^4 q_1 \over (2 \pi)^4} \int {i d^4
q_2 \over (2 \pi)^4} &T^0_4 T_3 (-q_2, q-q_1+q_2,
q_1-q)  T_3(-q, q_1, q-q_1)& \nonumber\\
&D^{INP}(q_1) D^{INP}(-q_2) D^{PT}(q-q_1+q_2) D^{PT}(q-q_1),&
\end{eqnarray}

\begin{eqnarray}
T^{(4)}_2(q) =  g^4 \int {i d^4 q_1 \over (2 \pi)^4} \int {i d^4
q_2 \over (2 \pi)^4} &T^0_4 T_3 (-q_2,
q-q_1+q_2, q_1-q) T_3(-q, q_1, q-q_1)& \nonumber\\
&D^{INP}(q_1) D^{PT}(-q_2) D^{INP}(q-q_1+q_2) D^{INP}(q-q_1),&
\end{eqnarray}

\begin{eqnarray}
T^{(5)}_2(q) =  g^4 \int {i d^4 q_1 \over (2 \pi)^4} \int {i d^4
q_2 \over (2 \pi)^4} &T^0_4 T_3 (-q_2,
q-q_1+q_2, q_1-q) T_3(-q, q_1, q-q_1)& \nonumber\\
&D^{INP}(q_1) D^{PT}(-q_2) D^{INP}(q-q_1+q_2) D^{PT}(q-q_1),&
\end{eqnarray}

\begin{eqnarray}
T^{(6)}_2(q) =  g^4 \int {i d^4 q_1 \over (2 \pi)^4} \int {i d^4
q_2 \over (2 \pi)^4} &T^0_4  T_3 (-q_2,
q-q_1+q_2, q_1-q) T_3(-q, q_1, q-q_1)& \nonumber\\
&D^{INP}(q_1) D^{PT}(-q_2) D^{PT}(q-q_1+q_2) D^{INP}(q-q_1),&
\end{eqnarray}

\begin{eqnarray}
T^{(7)}_2(q) = g^4 \int {i d^4 q_1 \over (2 \pi)^4} \int {i d^4
q_2 \over (2 \pi)^4} &T^0_4 T_3 (-q_2,
q-q_1+q_2, q_1-q) T_3(-q, q_1, q-q_1)& \nonumber\\
&D^{INP}(q_1) D^{PT}(-q_2)  D^{PT}(q-q_1+q_2) D^{PT}(q-q_1),&
\end{eqnarray}

\begin{eqnarray}
T^{(8)}_2(q) = g^4 \int {i d^4 q_1 \over (2 \pi)^4} \int {i d^4
q_2 \over (2 \pi)^4} &T^0_4 T_3 (-q_2,
q-q_1+q_2, q_1-q) T_3(-q, q_1, q-q_1)& \nonumber\\
&D^{PT}(q_1) D^{INP}(-q_2) D^{INP}(q-q_1+q_2) D^{INP}(q-q_1),&
\end{eqnarray}

\begin{eqnarray}
T^{(9)}_2(q) = g^4 \int {i d^4 q_1 \over (2 \pi)^4} \int {i d^4
q_2 \over (2 \pi)^4} &T^0_4 T_3 (-q_2,
q-q_1+q_2, q_1-q) T_3(-q, q_1, q-q_1)& \nonumber\\
&D^{PT}(q_1) D^{INP}(-q_2) D^{INP}(q-q_1+q_2) D^{PT}(q-q_1),&
\end{eqnarray}

\begin{eqnarray}
T^{(10)}_2(q) = g^4 \int {i d^4 q_1 \over (2 \pi)^4} \int {i d^4
q_2 \over (2 \pi)^4} &T^0_4 T_3 (-q_2,
q-q_1+q_2, q_1-q) T_3(-q, q_1, q-q_1)& \nonumber\\
&D^{PT}(q_1) D^{INP}(-q_2) D^{PT}(q-q_1+q_2) D^{INP}(q-q_1),&
\end{eqnarray}

\begin{eqnarray}
T^{(11)}_2(q) =  g^4 \int {i d^4 q_1 \over (2 \pi)^4} \int {i d^4
q_2 \over (2 \pi)^4} &T^0_4 T_3 (-q_2,
q-q_1+q_2, q_1-q) T_3(-q, q_1, q-q_1)& \nonumber\\
&D^{PT}(q_1) D^{INP}(-q_2) D^{PT}(q-q_1+q_2) D^{PT}(q-q_1),&
\end{eqnarray}

\begin{eqnarray}
T^{(12)}_2(q) =  g^4 \int {i d^4 q_1 \over (2 \pi)^4} \int {i d^4
q_2 \over (2 \pi)^4} &T^0_4 T_3 (-q_2,
q-q_1+q_2, q_1-q) T_3(-q, q_1, q-q_1)& \nonumber\\
&D^{PT}(q_1) D^{PT}(-q_2) D^{INP}(q-q_1+q_2) D^{INP}(q-q_1),&
\end{eqnarray}

\begin{eqnarray}
T^{(13)}_2(q) =  g^4 \int {i d^4 q_1 \over (2 \pi)^4} \int {i d^4
q_2 \over (2 \pi)^4} &T^0_4 T_3 (-q_2,
q-q_1+q_2, q_1-q) T_3(-q, q_1, q-q_1)& \nonumber\\
&D^{PT}(q_1) D^{PT}(-q_2) D^{INP}(q-q_1+q_2) D^{PT}(q-q_1),&
\end{eqnarray}

\begin{eqnarray}
T^{(14)}_2(q) = g^4 \int {i d^4 q_1 \over (2 \pi)^4} \int {i d^4
q_2 \over (2 \pi)^4} &T^0_4 T_3 (-q_2,
q-q_1+q_2, q_1-q) T_3(-q, q_1, q-q_1)& \nonumber\\
&D^{PT}(q_1) D^{PT}(-q_2) D^{PT}(q-q_1+q_2) D^{INP}(q-q_1),&
\end{eqnarray}

\begin{eqnarray}
T^{(15)}_2(q) =  g^4 \int {i d^4 q_1 \over (2 \pi)^4} \int {i d^4
q_2 \over (2 \pi)^4} &T^0_4  T_3 (-q_2,
q-q_1+q_2, q_1-q) T_3(-q, q_1, q-q_1)& \nonumber\\
&D^{PT}(q_1) D^{PT}(-q_2) D^{PT}(q-q_1+q_2) D^{PT}(q-q_1).&
\end{eqnarray}

Integrating over the gluon momenta $q_1$ and $-q_2$ in Eqs. (A23),
(A24) and (A25), one obtains

\begin{equation}
T^{(1)}_2(q) =  \bar \Delta_2^4 T^0_4 T_3 (0, q, -q) T_3(-q, 0, q)
D^{INP}(q) D^{PT}(q) \sim  \epsilon, \quad \epsilon \rightarrow
0^+,
\end{equation}

\begin{equation}
T^{(2)}_2(q) =  \bar \Delta_2^4 T^0_4 T_3 (0, q, -q) T_3(-q, 0, q)
D^{PT}(q) D^{INP}(q) \sim  \epsilon, \quad \epsilon \rightarrow
0^+.
\end{equation}
Here and below such kind of contributions vanish because of Eq.
(7.5), since there is no explicit integration over the gluon
momentum $q$.

\begin{equation}
T^{(3)}_2(q) = \bar \Delta_2^4 T^0_4 T_3 (0, q, -q) T_3(-q, 0, q)
D^{PT}(q) D^{PT}(q).
\end{equation}

Integrating first over the gluon momentum $q_1$ and then over the
gluon momentum $q_2=-q$ in Eq. (A26), one gets

\begin{equation}
T^{(4)}_2(q) = \bar \Delta_2^4 T^0_4 T_3 (q, 0, -q) T_3(-q, 0, q)
D^{PT}(q) D^{INP}(q) \sim  \epsilon, \quad \epsilon \rightarrow
0^+.
\end{equation}

Integrating first over the gluon momentum $q_1$ and then over the
gluon momentum $q_2=-q$ in Eq. (A27), one arrives at

\begin{equation}
T^{(5)}_2(q) =  \bar \Delta_2^4 T^0_4  T_3 (q, 0, -q) T_3(-q, 0,
q) D^{PT}(q) D^{PT}(q).
\end{equation}

Integrating over the gluon momentum $q_1$ in Eq. (A28), leads to

\begin{equation}
T^{(6)}_2(q) =  - \bar \Delta_2^2 \int {i d^4 q_2 \over (2 \pi)^4}
T^0_4  T_3 (-q_2, q+q_2, -q) T_3(-q, 0, q) D^{PT}(-q_2)
D^{PT}(q+q_2) D^{INP}(q) \sim  \epsilon, \quad \epsilon
\rightarrow 0^+.
\end{equation}

Integrating over the gluon momentum $q_1$ in Eq. (A29), one
obtains

\begin{equation}
T^{(7)}_2(q) = - \bar \Delta_2^2 \int {i d^4 q_2 \over (2 \pi)^4}
T^0_4 T_3 (-q_2, q+q_2, -q) T_3(-q, 0, q) D^{PT}(-q_2)
D^{PT}(q+q_2) D^{PT}(q).
\end{equation}

Integrating over the gluon momentum $-q_2$ in Eq. (A30), one gets

\begin{equation}
T^{(8)}_2(q) = - \bar \Delta_2^2 \int {i d^4 q_1 \over (2 \pi)^4}
T^0_4 T_3 (0, q-q_1, q_1-q) T_3(-q, q_1, q-q_1) D^{PT}(q_1)
D^{INP}(q-q_1) D^{INP}(q-q_1) \sim \epsilon, \quad \epsilon
\rightarrow 0^+.
\end{equation}
In connection with this integral, let us note that its behavior
with respect to $\epsilon$ comes from $D^{INP}(q-q_1)
D^{INP}(q-q_1) = \Delta^2 [(q-q_1)^2]^{-2} \Delta^2
[(q-q_1)^2]^{-2} = \epsilon^2 \bar \Delta_2^4[(q-q_1)^2]^{-4}$.
However, the singularity in the integrand function is not so
severe, since both three-gluon vertices depend on the momentum
$q-q_1$ linearly. Due to symmetric integration, in fact, the
singularity is $\sim [(q-q_1)^2]^{-3}$, which in terms of
$\epsilon$ produces $1 / \epsilon$ pole with residue being the
first derivative of the $\delta$-function (see Eq. (6.1)). If one
ignores the structure of the corresponding three-gluon vertices,
nevertheless, the singularity $\sim [(q-q_1)^2]^{-4}$ produces $1/
\epsilon$ pole as well. So, the integral (A45) vanishes as
$\epsilon$. Let us note that this and other integrals above and
below reproduce the case when the number of loop integrations does
not coincide with the number of the gluon propagators. The
derivatives of the $\delta$-function should appear, and not the
product of the $\delta$-functions at the same point, which has no
mathematical meaning, as underlined above.

Integrating over the gluon momentum $-q_2$ in Eq. (A31), one
obtains

\begin{equation}
T^{(9)}_2(q) = - \bar \Delta_2^2 \int {i d^4 q_1 \over (2 \pi)^4}
T^0_4 T_3 (0, q-q_1, q_1-q) T_3(-q, q_1, q-q_1) D^{PT}(q_1)
D^{INP}(q-q_1) D^{PT}(q-q_1) =0.
\end{equation}
In connection with this integral, let us note that we cannot
integrate directly over the gluon momentum $q_1=q$, since
dependence on it appears in both gluon propagators, namely,
$D^{INP}(q-q_1) D^{PT}(q-q_1) = \Delta^2 [(q-q_1)^2]^{-2}
[(q-q_1)^2]^{-1} = \epsilon \bar \Delta_2^2[(q-q_1)^2]^{-3}$.
However, the singularity in the integrand function is not so
severe, since both three-gluon vertices depend on the momentum
$q-q_1$ linearly. Due to symmetric integration, in fact, the
singularity is $\sim [(q-q_1)^2]^{-2}$, which in terms of
$\epsilon$ produces $1 / \epsilon$ pole, which residue is
$\delta^4(q-q_1)$. Since $T_3(0,0,0) = 0$, one obtains the
above-displayed result.

Integrating over the gluon momentum $-q_2$ in Eq. (A32), one
arrives at

\begin{equation}
T^{(10)}_2(q) = - \bar \Delta_2^2 \int {i d^4 q_1 \over (2 \pi)^4}
T^0_4 T_3 (0, q-q_1, q_1-q) T_3(-q, q_1, q-q_1) D^{PT}(q_1)
D^{PT}(q-q_1) D^{INP}(q-q_1) = 0
\end{equation}
because of the same reason as the previous integral.

Integrating over the gluon momentum $-q_2$ in Eq. (A33), one
obtains

\begin{equation}
T^{(11)}_2(q) = - \bar \Delta^2_2 \int {i d^4 q_1 \over (2 \pi)^4}
T^0_4 T_3 (0, q-q_1, q_1-q) T_3(-q, q_1, q-q_1) D^{PT}(q_1)
D^{PT}(q-q_1) D^{PT}(q-q_1).
\end{equation}
Similar to the integral (A46), we cannot integrate directly over
the gluon momentum $q_1=q$, since dependence on it appears in both
gluon propagators, namely, $D^{PT}(q-q_1) D^{PT}(q-q_1) =
[(q-q_1)^2]^{-2}$. However, the singularity in the integrand
function is not so severe, since both three-gluon vertices depend
on the momentum $q-q_1$ linearly. Due to symmetric integration, in
fact, the singularity is $\sim [(q-q_1)^2]^{-1}$, which is not the
NP IR singularity at all. So, this integral is finite in the
$\epsilon \rightarrow 0^+$ limit.

Integrating over the gluon momentum $q_1=q$ in Eq. (A34), one gets

\begin{equation}
T^{(12)}_2(q) = - \bar \Delta_2^2 \int {i d^4 q_2 \over (2 \pi)^4}
T^0_4 T_3 (-q_2, q_2, 0) T_3(-q, q, 0) D^{PT}(q) D^{PT}(-q_2)
D^{INP}(q_2) = 0
\end{equation}
Again we cannot integrate directly over the gluon momentum $q_2$,
since dependence on it appears in both gluon propagators, namely,
$ D^{INP}(q_2)D^{PT}(-q_2) = \Delta^2 [(q_2)^2]^{-2}
[(q_2)^2]^{-1} = \epsilon \bar \Delta_2^2[(q_2)^2]^{-3}$. However,
the singularity in the integrand function is not so severe, since
the three-gluon vertex depends on the momentum $q_2$ linearly. Due
to symmetric integration, in fact, the singularity is $\sim
[(q_2)^2]^{-2}$, which in terms of $\epsilon$ produces $1 /
\epsilon$ pole, which residue is $\delta^4(q_2)$. Since
$T_3(0,0,0) = 0$, one obtains the above-displayed result.

Integrating over gluon momentum $q_1=q+q_2$ in Eq. (A35), one
obtains

\begin{equation}
T^{(13)}_2(q) = - \bar \Delta_2^2 \int {i d^4 q_2 \over (2 \pi)^4}
T^0_4  T_3 (-q_2, 0, q_2) T_3(-q, q+q_2, -q_2) D^{PT}(q+q_2)
D^{PT}(-q_2) D^{PT}(-q_2).
\end{equation}
We cannot integrate directly over the gluon momentum $-q_2$, since
dependence on it appears in both gluon propagators, namely,
$D^{PT}(-q_2) D^{PT}(-q_2) = [(q_2)^2]^{-2}$. However, the
singularity in the integrand function is not so severe, since the
three-gluon vertex depends on the momentum $-q_2$ linearly. Due to
symmetric integration, in fact, the singularity is $\sim
[-q_2)^2]^{-1}$, which is not the NP IR singularity at all. So,
this integral is finite in the $\epsilon \rightarrow 0^+$ limit.

Integrating over the gluon momentum $q_1=q$ in Eq. (A36), we get

\begin{equation}
T^{(14)}_2(q) =  \bar \Delta_2^2 \int {i d^4 q_2 \over (2 \pi)^4}
T^0_4  T_3 (-q_2, q_2, 0) T_3(-q, q, 0) D^{PT}(q) D^{PT}(-q_2)
D^{PT}(q_2),
\end{equation}
and it is finite in the $\epsilon \rightarrow 0^+$ limit in the
same way as the previous integral. The last integral (A37) for
$T_2^{(15)}(q)$ remains the same.

Summing up nonzero contributions, one finally obtains

\begin{eqnarray}
T_2(q) &=& \bar \Delta_2^4 T^0_4 T_3 (0, q, -q) T_3(-q,
0, q) D^{PT}(q) D^{PT}(q) \nonumber\\
&+& \bar \Delta_2^4 T^0_4 T_3 (q, 0, -q) T_3(-q,
0, q) D^{PT}(q) D^{PT}(q) \nonumber\\
&-& \bar \Delta_2^2 \int {i d^4 q_2 \over (2 \pi)^4} T^0_4 T_3
(-q_2, q+q_2, -q) T_3(-q, 0, q)
D^{PT}(-q_2) D^{PT}(q+q_2) D^{PT}(q) \nonumber\\
&-&  \bar \Delta^2_2 \int {i d^4 q_1 \over (2 \pi)^4} T^0_4 T_3
(0, q-q_1, q_1-q) T_3(-q, q_1, q-q_1)
D^{PT}(q_1) D^{PT}(q-q_1) D^{PT}(q-q_1) \nonumber\\
&-&  \bar \Delta_2^2 \int {i d^4 q_2 \over (2 \pi)^4} T^0_4 T_3
(-q_2, 0, q_2) T_3(-q, q+q_2, -q_2)
D^{PT}(q+q_2) D^{PT}(-q_2) D^{PT}(-q_2) \nonumber\\
&-& \bar \Delta_2^2 \int {i d^4 q_2 \over (2 \pi)^4} T^0_4 T_3
(-q_2, q_2, 0) T_3(-q, q, 0) D^{PT}(q)
D^{PT}(-q_2) D^{PT}(q_2) \nonumber\\
&+&  g^4 \int {i d^4 q_1 \over (2 \pi)^4} \int {i d^4 q_2 \over (2
\pi)^4} T^0_4 T_3 (-q_2, q-q_1+q_2,
q_1-q) T_3(-q, q_1, q-q_1) \nonumber\\
 &\times& D^{PT}(q_1) D^{PT}(-q_2)
D^{PT}(q-q_1+q_2) D^{PT}(q-q_1).
\end{eqnarray}
Again one can conclude that two-loop skeleton contribution into
the gluon self-energy due to four- and three-gluon vertices does
not produce explicitly any problems in the $\epsilon \rightarrow
0^+$ limit.

 The PT part of nonlinear pure gluon part is a sum of
four terms, namely

\begin{equation}
T^{PT}_g[D](q)  = {1 \over 2} T^{PT}_t + {1 \over 2}  T^{PT}_1(q)
+ {1 \over 2} T^{PT}_2(q) + {1 \over 6} T'^{PT}_2(q),
\end{equation}
where each term is given by Eqs. (A1), (A6), (A21) and (A52). The
total sum $T^{PT}_g[D](q)$ is finite in the $\epsilon \rightarrow
0^+$ limit. This ends the investigation of the gluon SD equation
(8.20) for its PT part. The main goal of this appendix to show
that the PT part does not explicitly produce any problems in the
$\epsilon \rightarrow 0^+$ limit has been achieved, indeed.

\end{document}